\documentclass[letterpaper,11pt,fleqn]{article}
\pdfoutput=1
\usepackage{jheppub}

\usepackage{graphicx}
\usepackage{bm,amsmath,amssymb}
\usepackage[mathscr]{eucal}

\usepackage{wasysym}
\usepackage{slashed}
\usepackage{graphicx}
\usepackage{pdfcomment}
\long\def\comment#1{ }
\newcommand{\eqn}[1]{Eq.~(\ref{#1})}
\newcommand{\order}[1]{\mcal{O}{(#1)}}
\newcommand{\mcal}{\mathcal}
\newcommand{\rmd}{{\rm d}}
\newcommand{\rme}{{\rm e}}

\newcommand{\F}{\mcal{F}}
\newcommand{\abar}{\bar{\alpha}}
\newcommand{\tth}{t_{\rm th}}
\newcommand{\tbr}{t_{\rm br}}
\newcommand{\trel}{t_{\rm rel}}
\newcommand{\tf}{t_{\rm form}}
\newcommand{\mfp}{\lambda_{\rm mfp}}

\newcommand{\bp}{\bm{p}}
\newcommand{\bx}{\bm{x}}

\newcommand{\be}{\begin{equation}}
\newcommand{\ee}{\end{equation}}
\newcommand{\bea}{\begin{eqnarray}}
\newcommand{\eea}{\end{eqnarray}}
\newcommand{\f}{\frac}

\newcommand{\nn}{\nonumber}

\newcommand{\bra}{\langle}
\newcommand{\ket}{\rangle}

\title{\Large Thermalization of mini-jets in a quark-gluon plasma} 

\author{Edmond Iancu}

\author{and Bin Wu}

\affiliation{Institut de Physique Th\'{e}orique, CEA Saclay, UMR 3681,
F-91191 Gif-sur-Yvette, France}

\emailAdd{Edmond.Iancu@cea.fr}
\emailAdd{Bin.Wu@cea.fr}

\abstract{We complete the physical picture for the evolution of a high-energy jet 
propagating through a weakly-coupled quark-gluon plasma
by investigating the thermalization of the soft  components of the jet.  
We argue that the following scenario should hold:
the leading particle emits a significant number of mini-jets which promptly evolve via
quasi-democratic branchings and thus degrade into a myriad of soft gluons,
with energies of the order of the medium temperature $T$.
Via elastic collisions with the medium constituents,  
these soft gluons relax to local thermal equilibrium with the plasma
over a time scale which is considerably shorter than the typical lifetime of the mini-jet.
The thermalized gluons form a tail which lags behind the hard components of the jet.
We support this scenario, first, via parametric arguments and, next, by studying
a simplified kinetic equation, which  describes the jet
dynamics in longitudinal phase-space.
We solve the kinetic equation using both (semi-)analytical and numerical methods.
In particular, we obtain the first exact, analytic, solutions to the ultrarelativistic
Fokker-Planck equation in one-dimensional phase-space.
Our results confirm the physical picture aforementioned
and demonstrate the quenching of the jet via multiple branching followed by the 
thermalization of the soft gluons in the cascades.
}


\begin{document}
\maketitle

\section{Introduction}
\label{sec:intro}

It is by now well established that, within weakly-coupled QCD at least, the energy loss by an energetic
parton and/or the associated jet propagating through a dense medium, such as a quark-gluon plasma,
is dominated by {\em medium-induced radiation}, that is, the additional radiation triggered by the
interactions between the partons from the jet and the medium constituents 
\cite{Gyulassy:1993hr,Baier:1996kr,Baier:1996sk,Zakharov:1996fv,Zakharov:1997uu,Baier:1998kq,Wiedemann:2000za,Wiedemann:2000tf,Arnold:2001ba,Arnold:2001ms,Arnold:2002ja} (see also the review papers
\cite{Baier:2000mf,CasalderreySolana:2007zz,Mehtar-Tani:2013pia}). This picture has led to
a rather successful phenomenology, based on, or at least inspired by, calculations in perturbative QCD,
which has allowed one to understand many interesting observables at RHIC and the LHC, like the
nuclear modification factor or the suppression of di-hadron azimuthal correlations in 
ultrarelativistic nucleus-nucleus collisions \cite{d'Enterria:2009am,Majumder:2010qh,Burke:2013yra,Roland:2014jsa}.

More recently, the experimental studies of the phenomenon
known as `di-jet asymmetry' in Pb+Pb collisions at the LHC have demonstrated that a substantial fraction 
of the energy loss by an energetic jet is carried by relatively soft hadrons propagating 
at large angles with respect to the jet axis \cite{Aad:2010bu,Chatrchyan:2011sx,Chatrchyan:2012nia,Aad:2012vca,Chatrchyan:2013kwa,Chatrchyan:2014ava,Aad:2014wha,Gulhan:2014}. This pattern too can be understood, at least qualitatively, 
within the pQCD picture for medium-induced radiation, which predicts the formation of well-developed 
gluon cascades, or `mini-jets',  via multiple branching \cite{Blaizot:2012fh,Blaizot:2013hx,Blaizot:2013vha,Fister:2014zxa,Kurkela:2014tla,Blaizot:2014ula,Apolinario:2014csa,Blaizot:2014rla,Blaizot:2015jea} (see also 
Refs.~\cite{Baier:2000sb,Baier:2001yt,Jeon:2003gi,Schenke:2009gb} for earlier, related, studies and the recent review
paper \cite{Blaizot:2015lma}). 
Within these cascades, the energy is efficiently 
transmitted, via quasi-democratic branchings, from the `leading particle' --- the parton that has 
initiated the (mini)jet --- to a large number of comparatively soft gluons, which can be easily deviated 
towards large angles by rescattering in the medium. The theoretical description of medium-induced 
multiple branching started being developed only recently and in its current formulation 
it leaves unanswered a number of important questions.

The basic question is, what is the microscopic mechanism responsible for the energy loss ? Of course, one already
knows that, from the perspective of the leading particle, the dominant mechanism at work is radiative energy loss
and that the energy carried by the primary radiation is efficiently transmitted to softer and softer particles, 
via successive branchings in the gluon cascades. But what is the physical mechanism which {\em stops} these
cascades, and at which energy scale (i.e., what is the {\em low-energy end} of the cascade) ? 
How does this mechanism influence the dynamics of the
branchings at {\em higher} energy scales ? And {\em where} does the energy go, 
when it flows out of the cascade ? 


In order to better appreciate these questions, it is useful to briefly recall our current understanding of
the medium-induced jet evolution via multiple  branching (see Sect.~\ref{sec:phys} below for details).
The gluon cascades which are relevant for us here are those generated by iterating gluon emissions
of the BDMPSZ  type\footnote{The acronym `BDMPSZ' stands for Baier, Dokshitzer, 
Mueller, Peign\'e, Schiff, and Zakharov.}  \cite{Baier:1996kr,Baier:1996sk,Zakharov:1996fv,Zakharov:1997uu}.
The BDMPSZ mechanism governs the emission of relatively hard gluons, with 
energies $\omega \gg T$, which undergo multiple scattering in the surrounding medium.
Here, $T$ is the characteristic energy scale of the medium, say, the temperature for the case of a quark-gluon 
plasma in thermal equilibrium, or, more generally, the average $p_T$ of the background hadrons.
The distinguished feature of this mechanism for our present purposes is the fact that it favors 
`quasi-democratic branchings', that is, $1\to 2$ gluon splittings where the daughter 
gluons carry comparable fractions of the energy of their parent gluon. 
Such branchings occur fast and are extremely efficient in redistributing the energy among the 
branching products: they lead to turbulent cascades, in which the energy flows 
from one parton generation to the next one, without accumulating at intermediate steps.

In most theoretical analyses so far, on has assumed, for simplicity, that 
the branching dynamics remains unmodified down to arbitrarily low energies. This allowed for
elegant and physically transparent solutions \cite{Blaizot:2013hx,Fister:2014zxa,Blaizot:2015jea}, 
which exhibit {\em wave turbulence} with a characteristic scaling spectrum 
 --- the analog of the Kolmogorov--Zakharov spectrum  \cite{KST,Nazarenko}  
 for the medium-induced cascades --- and have interesting consequences for the energy loss by the
jet (see Sect.~\ref{sec:phys} below). But such an `ideal' cascade leads also to
unphysical results: after a finite interval of time 
(the `branching time' $\tbr(E)$, to be specified in Sect.~\ref{sec:phys})
the original energy $E$ of the leading particle gets transmitted to quanta which are arbitrarily soft 
($\omega\to 0$) and hence can propagate at arbitrarily large angles. 
While this peculiar `final state' is clearly unacceptable, 
it is important to stress that the phenomenon of wave turbulence is in fact more general
and could very well coexist with a physically acceptable final state: the scaling spectrum survives
unchanged if one stops the branching process at some finite energy scale $p_*\ll E$,
by introducing there a  `perfect sink'.  The `perfect sink' --- a concept familiar in the theory
of turbulence  \cite{KST,Nazarenko}   
--- is, by definition, a mechanism which is capable to absorb the energy flux
generated by the cascade at $p_*$, without influencing the branching dynamics at higher
energies $\omega\gg p_*$.

On physical grounds, there is an obvious candidate for such a `sink' : the surrounding medium.
The soft gluons from the cascade with $\omega\sim T$ are expected to thermalize via collisions
in the medium and thus deposit their energy inside the medium. This motivated proposals in
the literature to terminate the cascade at the medium scale $T$, e.g. by enforcing an 
`infrared' cutoff $p_*\sim T$ on the branching dynamics \cite{Fister:2014zxa,Blaizot:2014ula}.
But such previous arguments were insufficiently developed; it was not clear, e.g., what is the actual 
thermalization mechanism, why should this inhibit the branching process, 
and whether this should act as a `perfect sink', or, on the contrary, wash out the wave turbulence. 
[Numerical simulations using an ad-hoc infrared cutoff $p_*$  \cite{Blaizot:2014ula}
observed a strong distortion of the scaling spectrum, due to the accumulation of gluons in the bins 
above $p_*$. With increasing time, this pile-up extends up to high energies $\omega\gg p_*$ 
(see also the discussion in Sect.~\ref{sec:spectrum} below).]
  
It is our main purpose in this paper to clarify these and related questions, via
a dedicated theoretical analysis. Specifically, assuming the medium to be a weakly-coupled quark-gluon
plasma with temperature $T$, we shall study the possibility that the soft components of the jet,
with energies $\omega\sim T$, thermalize via elastic constituents with the quarks and gluons from
the plasma. To that aim, we shall consider a special kinetic equation, which emerges via
specific approximations from more general (but also more difficult to solve) equations existing
in the literature \cite{Baier:2000sb,Arnold:2002zm},  and which will be argued to capture the 
interesting dynamics to parametric accuracy at least. This equation, to be introduced
in Sect.~\ref{sec:kin}, describes the evolution of the
gluon distribution $f(t,z,p_z)$ created by the jet in the longitudinal phase-space, with the $z$ 
axis referring to the direction of propagation of the leading particle. 

The kinetic equation includes two types of collision terms:
an inelastic one, describing multiple branching with the BDMPSZ splitting rate, and an elastic one,
which describes $2\to 2$ collisions with the medium constituents in the Fokker-Planck approximation
\cite{Lifshitz:1981}. The latter is {\em a priori} suitable for a probe particle, or a dilute system of such particles,
which can be distinguished from the thermal bath (like a heavy quark \cite{Moore:2004tg,Rapp:2009my}),
but can also be applied to the gluons from the jet with relatively large momenta $p_z\simeq\omega\gg T$
\cite{Ghiglieri:2015zma}.
Indeed, the occupation numbers for the gluons generated via multiple branching
remain very small down to $\omega= T$, as we shall see (cf. the discussion in Sect.~\ref{sec:branch}). 

We shall further argue that, as a result of thermalization, the
branching process effectively terminates at the medium scale $\omega\sim T$ and we shall mimic this
by inserting an `infrared' cutoff $p_*\sim T$ in the splitting rate. The physical mechanism beyond
this cutoff will be clarified too --- this is related to the non-linear effects in the {\em total} gluon distribution, 
i.e. the distribution produced by the medium plus the jet  (see the discussion in Sect.~\ref{sec:kin})
---, but a proper treatment of this mechanism would require working 
with the non-linear kinetic equation obeyed by the full distribution, a task which is extremely hard in practice.
As we shall see in our numerical simulations, the physics around this cutoff is smeared by elastic
collisions and thermalization, so in practice we do not expect strong artifacts related to $p_*$.

Finally, the restriction to the longitudinal dynamics is needed to simplify the problem (in particular,
in view of numerical calculations) but it is also physically motivated. The longitudinal momenta remain
much larger than the respective transverse components ($p_z\gg p_\perp$) so long as $\omega\gg T$,
that is, during most stages of the dynamics, where they control the relevant time scales.
This approximation fails, strictly speaking, in the approach towards thermal equilibrium,
but even in that case it captures the correct time dependence to parametric accuracy. 
(In fact, a similar approximation has been used in all the previous studies of the in-medium 
cascade, including those which have explicitly considered the dependence upon transverse momenta
\cite{Blaizot:2013vha,Kurkela:2014tla,Blaizot:2014ula,Blaizot:2014rla}.) Furthermore, the gluon distribution
along the longitudinal axis $z$ is the most interesting one in view of a study of thermalization, 
since this is strongly inhomogeneous to start with: in the absence of collisions, all the gluons
in the jet, even the softest ones, would propagate along the light-cone at $z=t$, together with the
leading particle. 


In view of the above approximations, we expect that a thermalized distribution emerging from
our kinetic equation should look like a `tail' lying well behind the front of the jet, i.e. at $|z|\ll t$,
where it is quasi-homogeneous, and which in longitudinal momentum features the classical thermal
distribution for massless particles in one spatial dimension, that is, the Maxwell-Boltzmann distribution
$\rme^{-|p_z|/T}$. This is indeed what we shall observe in our solutions.

Specifically, we shall study the kinetic equation via a combination of (semi)analytic and numerical methods.
For the Fokker-Planck dynamics alone, we will be able to find exact, analytic, solutions --- in particular, the
solution corresponding to a steady source along the light-cone (in Sect.~\ref{sec:steady})
and the exact Green's function in longitudinal phase-space
(see Sect.~\ref{sec:Green}). To our knowledge, these are
the first examples of exact solutions for a relativistic Fokker-Planck equation. By combining this 
Green's function with known, analytic, solutions for the (ideal) branching process 
\cite{Blaizot:2013hx,Fister:2014zxa,Blaizot:2015jea},
we have been able to give a relatively simple, semi-analytic, analysis of the full dynamics, 
under the assumption that the medium acts as a perfect sink at the scale $p_*=T$.
(Under this assumption, it is indeed well justified to treat the branching part of the dynamics
as a source which injects gluons at the scale $p_*$ at a rate known from the previous analyses of
the turbulent cascade; see the discussion in Sect.~\ref{sec:analytic}.)

In order to test the `perfect sink' assumption and also to have more a complete study which
includes the interplay between branchings and elastic collisions at energies $\omega > p_*$,
we present in Sect.~\ref{sec:numerics} a detailed numerical analysis of the kinetic equation,
with infrared cutoff $p_*=T$ in the branching integral.
The numerical solutions turn out to be qualitatively similar to the semi-analytic ones in 
Sect.~\ref{sec:analytic}, but they bring additional clarifications,
in particular, on the role of the infrared cutoff $p_*$ and, related to that, on the
limitations of the  `perfect sink' approximation.

The overall physical scenario which emerges from these explicit solutions is in agreement with
the general picture anticipated in Sect.~\ref{sec:phys}, but it is more precise and also more complete
than the latter. It can be summarized as follows  (see Sects.~\ref{sec:phys} and
\ref{sec:numerics} for details):

A `leading particle' (LP) with initial energy $E\gg T$ which crosses the medium along a distance $L$ 
radiates abundantly (i.e., with a probability of order one) 
relatively soft gluons with energies $\omega\lesssim \omega_{\rm br}(L)$.
Here, $ \omega_{\rm br}(L) \sim \alpha_s^2\hat q L^2$, with $\hat q\sim
\alpha_s^2 T^3\ln(1/\alpha_s)$ the jet quenching parameter, is the characteristic energy scale 
for democratic branchings: gluons with smaller energies $\omega\lesssim \omega_{\rm br}(L)$
can undergo a democratic branching within a time $\Delta t\lesssim L$, where those with higher
energies $\omega\gg \omega_{\rm br}(L)$ cannot. In the experimental conditions at the LHC,
one typically has $T \ll \omega_{\rm br}(L) \ll E$, so the LP belongs to the second category above,
whereas the `primary gluons' emitted by it belong to the first one. Accordingly,
each of these `primary gluons' generates a gluon cascade (`mini-jet') via successive democratic branchings.
Each such a cascade ends at the thermal scale $T$, meaning that the energy $\omega$ of a primary gluon
gets distributed among a large number $\omega/T\gg 1$ of soft gluons. These gluons
undergo elastic collisions with the medium constituents and thus relax to a thermal distribution 
in momentum after a time of order $\trel\sim T^2/\hat q$. 
At the same time they separate in $z$ from the harder partons (which keep propagating along the
light-cone) and thus form a tail at $|z| < t$, which lags behind the front. Therefore the
momentum distribution looks very different near the front of the jet ($z\simeq t$), where the
spectrum shows an approximate scaling behavior, as expected for `ideal' branching, 
and in the tail at $z<t$, where the distribution is nearly thermal. 

\begin{figure}[t]
\begin{center}
\includegraphics[width=0.48\textwidth]{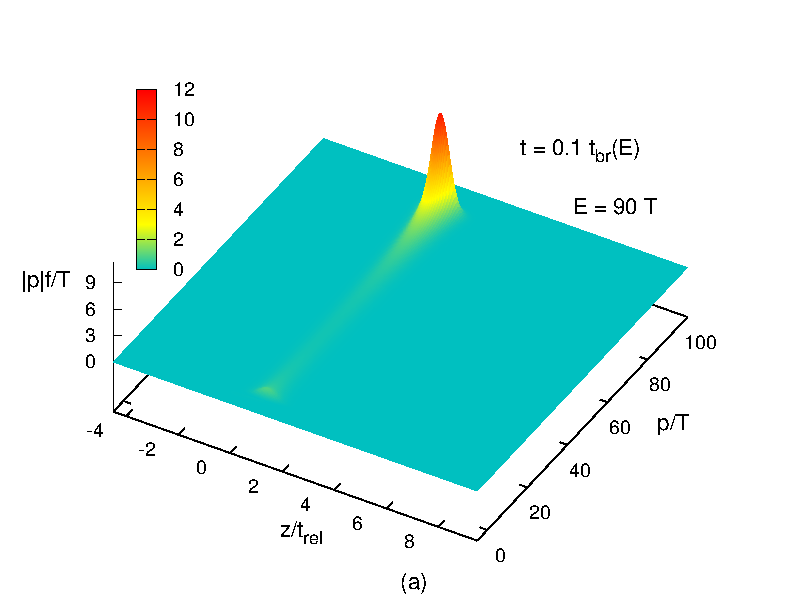}
\includegraphics[width=0.48\textwidth]{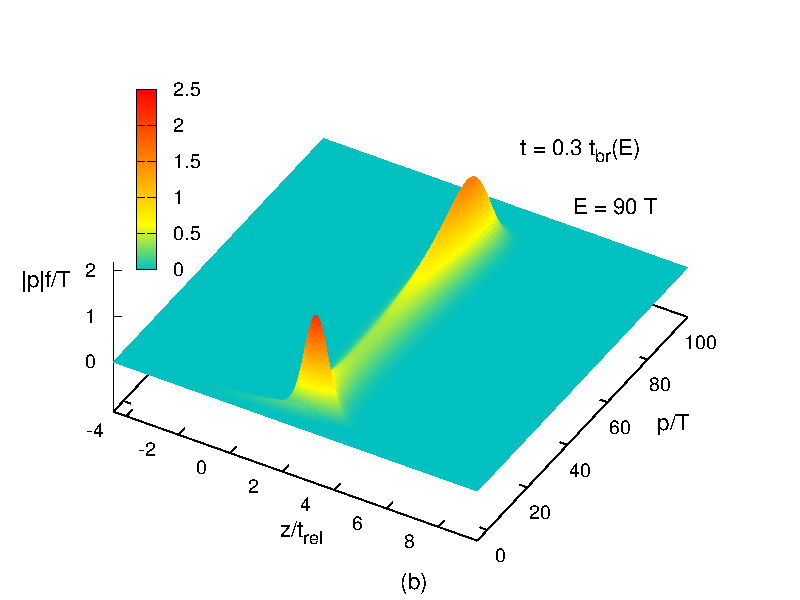}
\caption{The phase-space energy density $|p|f/T$ produced by an energetic
jet with $E=90\,T$ at two successive times: (a) an early time $t=0.1\tbr(E)$, when the jet is almost
unquenched; 
(b) a larger time $t=0.3\tbr(E)$, when the jet is partially quenched.
The secondary peak visible around $p\sim T$ in plot (b) represents the energy lost towards 
the medium via thermalization.
The reference scale $\tbr(E)$ is the characteristic time scale for the evolution of
the jet via democratic branching, to be explained in Sect.~\ref{sec:branch} (this is given by \eqn{tbr}).}
\label{fig:intro}
\end{center}
\end{figure}

The energy carried away by the gluons in the thermalized tail is naturally
interpreted as the {\em energy lost by the jet to the medium}. 
This is controlled by the hardest `mini-jets', those with energies  $\omega\sim \omega_{\rm br}(L)$,
and therefore it scales with the medium size like $L^2$. More precisely, the $L^2$--scaling
is the respective prediction of the ideal branching process, 
where the energy loss is computed as the energy carried by the turbulent flow  
\cite{Blaizot:2013hx,Fister:2014zxa}. For the full dynamics including elastic collisions
and thermalization, we numerically find that the energy deposited by the jet inside the medium 
is somewhat smaller than this `ideal' prediction, albeit comparable to it. This reduction reflects the
fact that the medium is not a `perfect sink', more precisely the fact that the soft gluons 
cannot thermalize  {\em instantaneously} (see the discussion in Sect.~\ref{sec:eloss}). 

The above considerations are illustrated by the plots  in Fig.~\ref{fig:intro} which are in fact extracted from
our numerical results in Sect.~\ref{sec:numerics}. These plots show the (normalized) energy distribution 
$(|p|/T)f(t,z,p)$ (with $p\equiv p_z$) produced by an incoming leading particle with $E=90\,T$ at two
successive times, an early time, when the jet is almost unquenched, and a later time, when the jet is partially
quenched. In the `late time' plot in Fig.~\ref{fig:intro} (b), one clearly sees the accumulation of particles
at soft momenta $p\lesssim T$, due to branchings and elastic collisions.

Let us finally stress that the above picture strictly holds for a very energetic jet, with initial
energy $E\gg \omega_{\rm br}(L)$, which is only `slightly quenched' --- meaning that the LP
survives in the final state and the energy
lost towards the medium is only a small fraction of the initial energy $E$. In that case, the final distribution,
as visible in Fig.~\ref{fig:intro} (b), may be viewed as the superposition of the LP  peak together with
the distributions separately created by all the `mini-jets'. But the individual `mini-jets' have lower energies
$\omega\lesssim \omega_{\rm br}(L)$, so they are fully quenched by the medium --- their whole
energy gets transmitted to the thermalized tail and the mini-jets disappear in the medium.
The distribution created by a single mini-jet will be discussed too in Sect.~\ref{sec:numerics}.


\section{The physical picture}\label{sec:phys}
In this section, we summarize the physical picture underlying the in-medium evolution
of a jet generated by a high-energy parton propagating through a weakly-coupled quark-gluon plasma.
This picture largely reflects the current understanding of this problem, as emerging from the literature,
but includes some additional arguments which will be physically motivated, together 
with some expectations to be subsequently confirmed by the new analysis in this work. 
We start with a brief review of recent studies of the medium-induced gluon cascade
\cite{Blaizot:2012fh,Blaizot:2013hx,Blaizot:2013vha,Fister:2014zxa},
which recognized the importance of multiple branchings, but did not address
the important problem of the  thermalization of the soft components of the jet. 
Then we discuss the interplay between multiple branching
and the elastic collisions responsible for thermalization.

\subsection{Multiple branching and the medium-induced gluon cascade}
\label{sec:branch}

An energetic `probe' parton (gluon or quark) which propagates through a dense QCD medium, such as
a quark-gluon plasma, undergoes elastic collisions with the constituents of the medium,
leading to `collisional' energy loss --- on the average, the energy transferred from the 
probe to the medium is larger than in the opposite direction --- and 
to the broadening of the probe distribution in (longitudinal and transverse) momentum, 
due to the random nature of the `kicks'. Besides, the collisions trigger gluon emissions by the
probe, leading to additional, `radiative', energy loss, which in practice dominates over the
collisional one, in spite of the fact that the emission probability is suppressed by a factor of $\alpha_s$.
This is possible because the coherence effects inherent in a quantum emission lead to a stronger 
dependence of the energy loss upon the medium size --- the radiative component rises, roughly,
like $L^2$ (with $L$ the distance traveled by the probe through the medium), whereas the 
collisional component rises only like $L$. Notwithstanding, the elastic collisions 
will play an important role for the subsequent discussion --- as we shall see, it is 
responsible for  the in-medium deposition of the energy lost via radiation in a typical event.

The precise mechanism responsible for medium-induced radiation
depends upon the ratio between the gluon `formation time' $\tf\sim \omega/p_\perp^2$
(the typical duration of the emission, as fixed by the uncertainty principle) 
and the mean free path $\mfp\sim 1/[\alpha_s T\ln(1/\alpha_s)]$ 
between two successive small-angle collisions. Here, $\omega$ is the energy of the emitted gluon 
and $p_\perp$ is its transverse momentum, as acquired during the formation time,
via collisions. For a single collision, one typically has $p_\perp^2\sim m_D^2
\sim \alpha_s T^2$, with $m_D$ the Debye mass. For a series of independent 
collisions occurring during a time interval $\Delta t\gg
\mfp$, one has $p_\perp^2\sim \hat q \Delta t$, where $\hat q \simeq m_D^2/\mfp \sim \alpha^2 T^3\ln(1/\alpha_s)$
is the `jet quenching parameter' --- a transport coefficient which characterizes momentum diffusion.
Using these estimates, one finds that, with increasing $\omega$,
one interpolates between a single-scattering (or `Bethe-Heitler') regime at low energies
$\omega \lesssim T$, where $\tf\sim \omega/m_D^2\lesssim \mfp$, and a multiple-scattering
(or `LPM', from Landau, Pomeranchuk, and Migdal) regime at high energies $\omega \gg T$, where 
$\tf\sim \sqrt{\omega/\hat q}\gg \mfp$. In the LPM regime, there is a large number of
collisions which coherently contribute to a single emission. 

In what follows, we shall consider the LPM regime alone. Indeed, we shall later argue that 
the branching process terminates around $\omega\sim T$, 
hence the phase-space for Bethe-Heitler radiation is comparatively small.
The calculation of the gluon branching rate in the LPM regime to leading order in pQCD has
been first given by Baier, Dokshitzer, Mueller, Peign\'e, and Schiff \cite{Baier:1996kr,Baier:1996sk,Baier:1998kq}, and independently by Zakharov \cite{Zakharov:1996fv,Zakharov:1997uu}.  One has thus obtained the 
following result for the differential probability per unit time and per unit $x$ for the collinear
splitting\footnote{The splitting is effectively collinear since the transverse momentum squared acquired
by the gluon during the formation time $\tf$ is much smaller than the respective quantity acquired after 
formation, namely during the lifetime $\tbr\sim \tf/\abar$ of the gluon until its next splitting (see below).}
 of a gluon with energy $\omega$ into two daughter gluons with 
energy fractions $x$ and  $1-x$, with $0< x <1$:
 \be\label{Pdef}
 \frac{\rmd^2 \mcal{I}_{\rm br}}{\rmd x\,\rmd t}\,=\,\frac{\alpha_s}{2\pi}\,
 \frac{P_{g\to g}(x)}{\tf(x, \omega)}\,.
  \ee
In this equation, $P_{g\to g}(x)=N_c [1-x(1-x)]^2/x(1-x)$, 
with $N_c$ the number of colors,  is the leading order gluon--gluon splitting function
of the DGLAP equation and  $\tf(x, \omega)$
is a more precise estimate for the formation time, which involves the average
energy $x(1-x)\omega$ of the two daughter gluons:
 \be\label{tform}
 \tf(x, \omega)\equiv \sqrt{\frac{x(1-x)\omega}
 {\hat q_{\text{eff}}(x)}}\,,\qquad 
 \hat q_{\text{eff}}(x)\equiv \hat q \left[1-x(1-x)\right]\,.\ee
Using \eqn{Pdef}, one can evaluate the probability for 
a branching to occur during a given interval $\Delta t$\,:
\be\label{Pdeltat}
 \Delta \mcal{P}\,\simeq\,x\frac{\rmd^2 \mcal{I}_{\rm br}}{\rmd x\,\rmd t}\,\Delta t\,
 \sim\,\abar \,\sqrt{\frac {\hat q}{x(1-x) \omega}}\,
 \Delta t\,,\ee
where $\abar \equiv \alpha_s N_c/\pi$. This probability becomes of order one,
meaning that multiple branching is important during $\Delta t$, provided
\be\label{omegas}
x(1-x)\omega \,\sim\,\abar^2\hat q\Delta t^2\,\equiv\,\omega_{\rm br}(\Delta t)\,.\ee
This condition can be satisfied by two types of emissions:

\texttt{(a)} very asymmetric splittings, for which either $x\ll1$ or $1-x\ll 1$, whereas the energy
$\omega$ of the parent gluon can be relatively hard (for instance, $\omega\gg \omega_{\rm br}(\Delta t)$)\,;

\texttt{(b)}  `quasi-democratic' branchings, where the two daughter gluons carry 
comparable fractions of the total energy, $x\sim 1-x\sim \order{1}$, but the parent gluon is
relatively soft: $\omega\sim \omega_{\rm br}(\Delta t)$.

Reversing the argument for case \texttt{(b)} above, we conclude that it takes a time 
$\Delta t\sim \tbr(\omega)$, with
\be\label{tbr}
\tbr(\omega)\equiv \frac{1}{\abar}\sqrt{\frac{\omega}
 {\hat q}}\,,\ee
for a gluon with energy $\omega$ to undergo a `quasi-democratic' branching 
\cite{Baier:2000sb,Kurkela:2011ti,Blaizot:2013hx}. This duration
$\tbr(\omega)$ should be compared to the medium size $L$ which is available
to that parton:

 \comment{The BDMPSZ rate \eqref{Pdef} is formally of order $\alpha_s$, but it is 
enhanced for very asymmetric splittings ($x\ll1$ or $1-x\ll 1$) and also for sufficiently soft
parent gluons (relatively small $\omega$). As a result, there is no suppression for multiple
branching at weak coupling, as we now explain. }

\texttt{(i)} If $\tbr(\omega)\gg L$, then the parton with energy $\omega$
can emit {\em abundantly} --- i.e. with probability of $\order{1}$ ---  only relatively
soft gluons with $x\ll 1$, in such a way that $x\omega \lesssim
\omega_{\rm br}(L)=\abar^2\hat q L^2$.
Accordingly, the original parton will `survive the medium' : it will be recognizable in the final
state due to the fact that its final energy will be considerably higher than for all the other
gluons, as produced by radiation.

\texttt{(ii)} If $\tbr(\omega)\lesssim L$, then the parton with energy $\omega$ will
 `disappear inside the medium ' --- it will undergo a quasi-democratic branching before
it exits the medium and thus it will be replaced by a pair of softer gluons, which can
 `democratically' split again. Eventually, the
original parton will leave behind it a gluon cascade generated via successive, 
quasi-democratic splittings. Note that each new gluon generation in that cascade has
a lower energy and hence a shorter lifetime than the previous ones. 
Accordingly, the overall lifetime of the cascade is of the order of the branching time 
$\tbr(\omega)$ of the initial gluon.

Case \texttt{(i)} is the interesting situation for the leading particle (LP) which initiates a
typical jet measured in Pb+Pb collisions at the LHC. Indeed, denoting the
energy of this LP by $E$, one generally has $E\ge 100$~GeV, whereas the characteristic
energy for multiple branching is much smaller: 
$\omega_{\rm br}(L)=\abar^2\hat q L^2\simeq 12$\,GeV 
for a medium with $\hat q=1$GeV$^2$/fm and $L=5$\,fm (we used $\abar=0.3$).
The above estimate also shows that, for the interesting values of $L$, the branching scale 
$\omega_{\rm br}(L)$ is much harder than the medium temperature $T\simeq 0.5$~GeV;
hence, one typically has $T\ll \omega_{\rm br}(L)\ll E$.

Case \texttt{(ii)} applies to the {\em typical} primary gluons --- the gluons which are directly radiated by the LP
in a typical event ---, which have relatively soft energies $\omega\lesssim \omega_{\rm br}(L)$ and
thus have the time to develop gluon cascades (`mini-jets') via quasi-democratic branchings.
Harder primary emissions, with energies up to\footnote{This upper limit $\omega_c$ on the energy of 
medium-induced gluon
emissions follows from the condition that the gluon formation time $\tf(\omega)$ be at most
as large as $L$.} $\omega_c\equiv \hat q L^2$, are possible as well (provided $E>\omega_c$,
of course), but these are rare events which occur with a probability of $\order{\abar}$ and
do not generate mini-jets. Such hard but rare emissions will not be explicitly 
considered in what follows, since they do not contribute to the energy loss by the jet towards the medium.
(But they are important for the average energy loss by the LP 
\cite{Baier:1996kr,Baier:1996sk,Baier:1998kq,Zakharov:1996fv,Zakharov:1997uu}.)

The quasi-democratic nature of the splittings has important consequences for the energy
flow across a mini-jet: it leads to {\em wave turbulence} \cite{Blaizot:2013hx,Fister:2014zxa}.
Via successive branchings, the energy flows from one gluon generation to the next one,
without accumulating at any intermediate value of $\omega$.  The precise mathematical
condition for wave turbulence is that the energy flux --- the rate for energy
flow along the cascade --- should be independent of  
$\omega$. This condition is indeed satisfied for the gluon cascade
generated by a primary gluon with energy $\omega_0\lesssim \omega_{\rm br}(L)$,
at least at sufficiently small values $\omega\ll \omega_0$   \cite{Blaizot:2013hx,Fister:2014zxa}.
(See the discussion in Sect.~\ref{sec:source} for more details.)

If this branching dynamics was to remain unmodified down to arbitrarily small values 
of $\omega$, then the whole energy would end up into a `condensate'
at $\omega=0$ \cite{Blaizot:2013hx}. In reality though, we expect the gluon cascade
to terminate at the thermal scale $T$, because the very soft quanta with 
$\omega\lesssim T$ can efficiently thermalize via elastic collisions (see the discussion
in the next subsection). If the medium acts as a {\em perfect sink} at the lower end of
the cascade --- in the sense of absorbing all the quanta with $\omega\lesssim T$ without
modifying the dynamics of branching at higher energies $\omega\gg T$ --- then the whole
energy carried by the turbulent flow is eventually transmitted to the medium. Under this
assumption, the total energy loss by the jet towards the medium is obtained as
\cite{Blaizot:2013hx,Fister:2014zxa}
\be\label{DEflow}
{\Delta E}_{\rm flow}\,\simeq\,\frac{\upsilon}{2}  \, \omega_{\rm br}(L) \,=\,
\frac{\upsilon }{2}\, \abar^2 \hat q L^2
\,,\ee
with  $\upsilon\simeq 4.96$. 
This result, which is independent of the initial energy $E$ of the LP and grows
with the medium size like $L^2$, truly applies (under the `perfect sink' assumption) for a
very energetic jet with $E\gg \omega_{\rm br}(L)$. In the opposite limit where
$E\lesssim \omega_{\rm br}(L)$ (the case of a mini-jet), one rather has
${\Delta E}_{\rm flow}\simeq E$ (see \eqn{Econd} for a more general expression). 
\eqn{DEflow} admits a natural physical
interpretation \cite{Fister:2014zxa}: the LP radiates an average number $\upsilon/2 $ of primary gluons 
with energies of order  $\omega_{\rm br}(L)$, which then transmit their whole energy to the medium, 
via democratic branchings followed by the thermalization of the soft gluons ($\omega\sim T$)
at the lower end of the cascades.
We shall later discover that the assumption that the medium acts as a `perfect sink' has some
limitations in practice, yet it can be used for qualitative considerations and parametric estimates.

Another important assumption that was implicitly postulated by previous analyses of multiple
branching \cite{Blaizot:2013hx,Fister:2014zxa} is that the branching dynamics is {\em linear},
meaning that the gluons from the cascade can split, but not also recombine with each other.
This assumption is correct provided the partons cascade are sufficiently dilute. The precise
condition is that $f(t,\bx,\bp)\ll 1$, where $f(t,\bx,\bp)$ is the {\em gluon phase-space occupation number}
(below, $N_g$ the total number of gluons in the jet),
\begin{align}\label{f3D}
f(t,\bx,\bp)\,\equiv\,\f{(2\pi)^3}{2(N_c^2-1)}\,\f{\rmd N_g}{\rmd^3\bx \rmd^3\bp}\,.
\end{align}
This quantity has not been explicitly computed in the previous studies, but it is straightforward to
obtain an order-of-magnitude estimate for it via physical considerations.

Consider a typical mini-jet, as generated by a primary gluon with initial energy 
$\omega_0\lesssim \omega_{\rm br}(L)$. Over a time interval of order $\tbr(\omega_0)$, 
this whole energy gets redistributed, via multiple branching, 
among a large number $N_g \simeq \omega_0/T$ of soft quanta with energies $\omega\sim T$.
Their occupancy can therefore be estimated as
\be\label{fT}
f(T)\,\sim\,\frac{1}{N_c^2}\,\f{\omega_0/T}{p_z\Delta z\,\Delta p_\perp^2 \Delta x_\perp^2}\,,
\ee
where $p_z\sim \omega\sim T$,  $\Delta p_\perp^2$ is the transverse momentum 
squared acquired by a gluon via rescattering in the medium, 
$\Delta z$ is the longitudinal extent of the distribution,
and $\Delta x_\perp^2$ is the corresponding spread in the transverse plane. 

The transverse phase-space $\Delta p_\perp^2 \Delta x_\perp^2$ occupied by a gluon with energy
$\omega\sim p_z$ turns out to be independent of $\omega$.  
Indeed, during a time $\Delta t$, a gluon accumulates  a transverse momentum broadening 
$\Delta p_\perp^2\simeq\hat q\Delta t$, leading to an uncertainty
\begin{align}
\Delta x_\perp^2\,\simeq\,\f{\Delta p_\perp^2}{p_z^2}\,\Delta t^2
\,\simeq\,\f{\hat q \Delta t^3}{\omega^2}
\end{align}
 in its transverse location. Taking $\Delta t$ of the order of the gluon lifetime, $\Delta t\sim \tbr(\omega)$, one finds 
  \be\label{perpPS}
 \Delta p_\perp^2\, \Delta x_\perp^2\,\simeq\,\left(\frac{\hat q\, \tbr^2(\omega)}{\omega}\right)^2
 \sim\,\f{1}{\abar^4}\,.
 \ee
This is independent of $\omega$, as anticipated, and parametrically large. \eqn{perpPS} is truly a lower limit,
since the soft gluons can also inherit part of the transverse momentum of their harder parents.
The above argument also shows that, so long as $\omega\gg T$ --- as is
the case for the gluons which control the splitting process --- the transverse momenta
remain much smaller than the longitudinal ones: $p_\perp\ll p_z\simeq\omega$.

Consider now the longitudinal distribution of the soft quanta within the mini-jet. As already discussed,
soft gluons are emitted promptly, so they can be produced anywhere along the cascade.
After being emitted, they efficiently lose energy and randomize their
direction of motion, via elastic collisions.
Accordingly, they separate from each other and also from the harder ($p\gg T$) 
partons in the mini-jet over a time interval $\sim\trel$, which is small compared to 
the overall lifetime $\sim \tbr(\omega_0)$ of the cascade. Hence,
on the average, these soft gluons should be quasi-uniformly distributed along $z$, 
within a distance $\Delta z\sim \tbr(\omega_0)$. Using this,
together with \eqref{fT} and \eqref{perpPS}, we finally deduce
\be\label{fpav}
f(T)\,\sim\,\frac{\abar^4}{N_c^2}\,\f{\omega_0}{T^2\tbr(\omega_0)}\,.
\ee
This number increases with $\omega_0$, so it is
interesting to evaluate it for the largest possible value $\omega_0\sim \omega_{\rm br}(L)$ ---
the one which also controls the energy loss by the overall jet (cf. \eqn{DEflow}). In that case,
$\tbr(\omega_0)\simeq L$, so one finds
\be\label{fTav}
f(T)\,\sim\,\frac{\abar^6}{N_c^2}\,\f{\hat q L}{T^2}\,.
\ee
The corresponding estimate for the jet as a whole can be simply obtained by multiplying this result
by a number $\sim{\upsilon}$, cf. \eqn{DEflow}. The occupation number \eqref{fTav}
is parametrically small at weak coupling and furthermore suppressed at large $N_c$. It is easy to check
that the condition $f(T)\ll 1$ is always very well satisfied in practice. 

The estimate in \eqn{fTav} should be more properly viewed as a {\em lower} limit on $f$\,: 
as we shall explain in Sect.~\ref{sec:source}, the soft gluons with $\omega\sim T$ are
more abundantly produced during the late stages of the cascade, at times $t\sim \tbr(\omega_0)$,
so their distribution in $z$ is not really homogeneous. (This will be also confirmed by the numerical 
simulations in Sect.~\ref{sec:numerics}; see in particular Fig.~\ref{fig:fEvol}.) 
An {\em upper} limit on the longitudinal occupancy
is however easily obtained by assuming the smallest possible longitudinal spread for the soft gluons,
namely their lifetime $\tbr(T)$. With $\Delta z\sim \tbr(T)$ and $\omega_0\sim \omega_{\rm br}(L)$,
\eqn{fT} implies (recall that $\tbr(T)\sim 1/\abar^2 T$),
\be\label{fTrare}
f(T)\,\sim\,\frac{\abar^4}{N_c^2}\,\f{\omega_{\rm br}(L)}{T^2\tbr(T)}\,
\sim\,\frac{\abar^8}{N_c^2}\,\f{\hat q L^2}{T}\,,
\ee
which is still much smaller than one in all practical situations of interest, as one can easily check.
 
For what follows, one should keep in mind that the previous estimates for $f$
refer exclusively to the soft gluons generated by the jet via multiple branching.
When $\omega\sim T$, these gluons add to those from the background medium,
whose occupation numbers are given by the usual, Bose-Einstein,
thermal distribution and hence are of order one. So the present
results also show that the effect of the jet on the occupancy of gluons with $\omega\lesssim
T$ represents only a small perturbation. This in particular implies that the jet cannot produce 
`hot spots' in the medium (it cannot significantly increase the local energy density): 
the energy density carried by the soft jet constituents with $\omega\sim T$ is obtained
by multiplying $f(T)$ by $N_c^2 T^4$ and hence remains much smaller than the respective quantity
for the thermal gluons, i.e. $\varepsilon \sim N_c^2 T^4$, so long as $f(T)\ll 1$.

\subsection{Elastic collisions and thermalization}
\label{sec:therm}

In the discussion in the previous subsection,
we have implicitly assumed that the only effect of the elastic collisions between the gluons
from the jet and the medium constituents is to trigger new branchings. 
In reality though, such interactions can also transfer energy and
momentum between the colliding particles, leading to energy loss and momentum broadening
for the jet constituents. As we shall see, these effects remain negligible so long
as the gluons in the cascade have relatively large energies $\omega\gg T$, but they become
a leading-order effect, and the driving force towards thermalization, when $\omega\sim T$.

To propose a theoretical description for these interactions, it is essential to recall,
from the previous discussion, that the gluon system produced via multiple branching is dilute.
Hence, it is appropriate to study the effects of elastic collisions on individual gluons from the jet.
So long as the gluon under study has a relatively large energy $\omega \gg T$, it can be
unambiguously distinguished from the thermal gluons, so it is possible to describe its
dynamics by using the same methods as for other energetic probes, like a relativistic
heavy quark (see e.g.  \cite{Lifshitz:1981,Moore:2004tg,Rapp:2009my,Ghiglieri:2015zma}). 
In this subsection we shall use a Langevin description
because of its formal simplicity. Later on, we shall employ the equivalent method of the 
Fokker-Planck equation for more elaborate studies. Strictly speaking, both descriptions
will eventually fail when the energy of the `probe' gluon decreases down to $\omega\sim T$, 
but even in that case they remain qualitatively correct, in that they capture 
the correct time scale for thermalization to parametric accuracy.

For more clarity, let us first assume that the branching dynamics is switched off,
meaning that the energetic gluon suffers only elastic collisions in the plasma.
The Langevin equation which encompasses the effects of these collisions reads as follows:
\be\label{Langevin}
\frac{\rmd p^i}{\rmd t}\,=\,-\eta v^i+\xi^i\,,\qquad\langle \xi^i(t)\xi^j(t')\rangle=\frac{\hat q}{2}\,
\delta^{ij}\delta(t-t')\,,
\ee
where $v^i=p^i/p$, with $i=1,2,3$, is the particle velocity, $\eta$ is a friction coefficient,
and $\xi^i$ is a stochastic force (the `noise'). Microscopically, the total force in the r.h.s.
of \eqn{Langevin} represents the Lorentz force 
generated by random, quasi-classical, fields in the plasma --- the color fields of the thermal particles, 
which are slightly disturbed out-of-equilibrium by their scattering with the external particle.
The `drag force' $f^i=-\eta v^i$ describes the average effect of this microscopic force,
which is the energy transfer from the probe to the medium, whereas the noise term $\xi^i$ 
represents its random component leading to momentum broadening. The average over the noise reflects
the thermal average over the microscopic sources of this force. The facts that the noise correlator is local
in time (`white noise') and also isotropic are non-trivial and reflect some approximations. The first property 
is true because we follow the dynamics on time scales much larger than the
typical correlation time for the sources. The isotropy is `accidental' (at least in the
relativistic context at hand), in the sense that it is specific to the 
lowest-order approximation at weak coupling and to the
ultrarelativistic limit for the external particle\footnote{In general, $(\hat q/2)\delta^{ij}$ in the r.h.s. of
the noise correlator gets replaced by $\hat q^{ij} =\hat q_\ell\hat v^i\hat v^j + (\hat q/2)(\delta^{ij}
-\hat v^i\hat v^j)$, where the longitudinal ($\hat q_\ell$) and transverse ($\hat q$) momentum
diffusion coefficients are different from each other. But for a massless energetic particle and in the
lowest, leading-logarithmic, approximation, it so happens that $\hat q_\ell=\hat q/2$;
see e.g. \cite{Moore:2004tg,Ghiglieri:2015zma,Blaizot:2014jna}.} \cite{Moore:2004tg}.

By using a properly discretized version of the stochastic
equation \eqref{Langevin} (see e.g.  \cite{Moore:2004tg}), 
one can deduce the following evolution equation for the average of the particle momentum squared:
\begin{align}\label{avp}
\frac{\rmd \langle p^2\rangle}{\rmd t}\,=\,-2\eta\langle p\rangle+\frac{3}{2}\hat q\,.
\end{align}
For this equation to be consistent with the approach towards the
thermal distribution\footnote{The probe gluon is here treated as a 
classical particle, hence its momentum distribution in thermal
equilibrium is the relativistic version of the classical Maxwell-Boltzmann distribution, also known as
the Maxwell-J\"uttner distribution.}
 $f_p\propto \rme^{-p/T}$ (which in turn implies
$\langle p\rangle=3T$ and $\langle p^2\rangle=12T^2$), one needs to fulfill the Einstein relation
$\hat q = 4T\eta$ between diffusion and drag. This is indeed guaranteed 
by the fluctuation-dissipation theorem for thermal correlations.

Let us now assume that the energetic gluon enters the medium at $t=0$ with a large momentum
$p_0\gg T$ oriented along the $z$ axis ($i=3$) : $v_z\equiv v^3=1$. At early stages, the longitudinal momentum 
remains large, $p_z\gg T$, and the effects of fluctuations are unimportant: $p_\perp\ll p_z$ and $v_z\simeq 1$.
During these stages, one can take
the average in \eqn{Langevin} to deduce $\rmd \langle p_z\rangle/{\rmd t}
\simeq -\eta$ and therefore $\langle p_z(t)\rangle\simeq p_0 -\eta t$. This shows that
the particle loses most of its energy, from the initial value $p_0\gg T$ down to a value $p\sim T$ 
where diffusion effects start to be important, over an interval $\Delta t\simeq p_0/\eta =(p_0/T)\trel$, with
\be\label{trel}
\trel\equiv \f{4T^2}{\hat{q}}\,\sim\, \frac{1}{\abar^2 T\ln(1/\abar)}\,.
\ee
From that moment on, the particle approaches the thermal distribution
quite fast, over a time $\Delta t\sim\trel$, under the combined effect of drag and diffusion.
This can be understood from the fact that its momentum broadening increases with time  like
$\langle p^2\rangle\simeq (3/2)\hat q\Delta t$, cf. \eqn{avp}, and hence it becomes of $\order{T^2}$
after a time $\Delta t\sim T^2/\hat q\sim \trel$. Clearly, when $p_0\gg T$, 
 the total duration of the thermalization process is controlled by the
first period --- the energy loss via drag --- and is of order $\tth(p_0)$, with
\begin{align}\label{tth}
\tth(p)\,\equiv\,\frac{p}{T}\,\trel \,=\,\f{4pT}{\hat{q}}\,.
\end{align}

Let us now switch on the branching dynamics, on top of the elastic collisions. 
From the previous subsection, we know that the incoming gluon
with energy $p_0\gg T$ has a lifetime $\Delta t\sim \tbr(p_0)$ before it undergoes
a first democratic branching. By comparing Eqs.~\eqref{tbr} and \eqref{tth}, it is clear that 
$\tbr(p_0)\ll\tth(p_0)$ so long as $p_0\gg T$. Indeed,
\begin{align}\label{tbrtth}
\f{\tbr(p)}{\tth(p)}\,=\,\frac{1}{4\abar}\sqrt{\f{\hat q}{p T^2}}\,\sim\,\sqrt{\f{T}{p}}\,,
\end{align}
where we have used $\hat q\sim \abar^2 T^3$ for the weakly-coupled QGP. This implies that
the incoming gluon disappears via branching before having the time to lose a substantial fraction of 
its original energy via drag. A similar conclusion applies to all the successive generations within
the ensuing gluon cascade, so long as the respective momenta are 
hard, $p\gg T$\,: the elastic collisions cannot significantly modify the kinematics of the hard 
gluons during the time interval between two successive democratic branchings. 


However, the situation changes in the later stages of the cascade when, as a result of successive 
branchings, the gluons have been degraded to lower energies $p\sim T$. Then, the various time-scales previously 
introduced become degenerate (at least, parametrically),
\begin{align}\label{scales}
\tbr(T)\,\sim\,\tth(T)\,\sim\,\trel\,\sim\,\frac{1}{\abar^2 T \ln(1/\abar)}\,,\end{align}
meaning that the various processes start to compete with each other. Before they have the time to branch
again, the gluons with $p\sim T$ can lose a substantial fraction of their energy towards the medium
and also suffer a considerable broadening of their momentum distribution. Such (drag and diffusion)
processes will naturally drive the soft gluons towards a thermal distribution in momentum which, within the 
present approximations, appears to be the classical, Maxwell-Boltzmann, distribution.
In reality though, our approximations fail to properly describe the final equilibrium state, for the
reasons already explained. The gluons from the jet which approach thermal equilibrium cannot
be distinguished anymore from the gluons in the plasma, so a proper theoretical description 
of the late stages should rather follow the {\em complete} gluon distribution. 
For the latter, the effects of the quantum statistics are essential 
(since the occupation numbers are of order one) and the final distribution in equilibrium must be of the
Bose-Einstein type. When this equilibrium distribution is finally
reached, the branching process naturally stops, because of the compensation between 
splittings and recombination (see the discussion in Sect.~\ref{sec:kin} below). 
Thus the cascade effectively ends at a scale $\sim T$.

This discussion shows that the characteristic time scale for the thermalization of a mini-jet
--- the duration of the  overall process
which starts with the emission of a primary gluon with energy $p_0\gg T$ and ends up with
the thermalization of its soft branching products --- is controlled by the branching part of the dynamics
and hence is of order $\tbr(p_0)$. As discussed around \eqn{tbrtth}, this time scale $\tbr(p_0)$ is much smaller 
than $\tth(p_0)$  --- the would-be thermalization time for a gluon with the same initial energy in the absence of 
branchings. This is so since the branchings are more efficient than the elastic collisions in degrading
the energy of the relatively hard ($p\gg T$) constituents of the jet. 

In the previous considerations, we have focused on the relaxation of the {\em energy-momentum} 
distribution, but we ignored the {\em spatial} distribution of the gluons inside the jet.
In particular, we have discarded the fact that  this distribution is highly inhomogeneous to start with ---
the leading particle and, more generally, all the energetic constituents of the jet with $p\gg T$ 
propagate at the speed of light and are concentrated near the light-cone ($z=t$) ---,
a feature which could prevent, or at least delay, thermalization. Also, we have implicitly assumed that
the dynamics responsible for thermalization at the low-energy end of the cascade ($p\sim T$) does not
affect the branching dynamics at higher energies $p\gg T$. In order to address such complex issues,
test our various assumptions, and thus firmly establish the physical picture that we previously exposed, 
we need more explicit calculations. 
A suitable formalism in that sense will be described in the next section.

\section{The kinetic equation for the longitudinal dynamics}
\label{sec:kin}

In what follows we shall propose a relatively simple kinetic equation which captures the general 
dynamics exposed in the previous section, in the sense that it respects the various
parametric estimates and thus is expected to reproduce the correct physical picture, 
and which allows for explicit studies, via semi-analytic and numerical techniques.

The starting point is a kinetic equation 
introduced in Ref.~\cite{Baier:2000sb} and thoroughly derived in 
Refs.~\cite{Arnold:2002zm} (see also Refs.~\cite{Blaizot:2012fh,Blaizot:2013vha} for
a careful analysis of the quantum branching process, which justifies treating the successive branchings
as independent from each other), with the schematic structure
\bea\label{generaleq}
\left(\frac{\partial}{\partial t}+{\bm v}\cdot \nabla_{\bm x} \right)f(t,\bx,\bp)\, =\, \mcal{C}_{\rm el}[f]+
\mcal{C}_{\rm br}[f]\,.\eea
Here, $f$ is the gluon occupation number defined in \eqn{f3D}, which {\em a priori} refers to 
gluons from both the jet and the surrounding medium\footnote{To \eqn{generaleq}, one should 
add corresponding equations for the quark and the antiquark occupation numbers, but here 
these are not necessary, since the (anti)quarks will only appear as constituents of the thermal medium.}, 
$\mcal{C}_{\rm el}[f]$ is a collision integral encoding the effects of
$2\to 2$ elastic collisions, whereas $\mcal{C}_{\rm br}[f]$ encodes the inelastic processes, like
$2\to 3$ collisions, which lead to (collinear) branchings. The most general structure of these collision
terms can be found in \cite{Arnold:2002zm,Arnold:2003zc}, while a simpler form of $\mcal{C}_{\rm br}[f]$,
which is sufficient for our present purposes, is presented in \cite{Baier:2000sb}. These general collision
terms are non-linear with respect to the gluon occupation number, in such a way to respect the 
detailed balance principle and the quantum statistics. Accordingly, the 
thermal Bose-Einstein distribution is a fixed point for both $\mcal{C}_{\rm el}[f]$ and $\mcal{C}_{\rm br}[f]$.

The general equation \eqref{generaleq} is difficult to solve in
practice, because of the non-linear effects alluded to above and also because of the generally
complicated structure of the two collision integrals, which involve multi-dimensional momentum
integrations together with complicated kernels (see \cite{Arnold:2002zm,Ghiglieri:2015zma} for details).
Numerical solutions have been presented for special limits of this equation
\cite{Schenke:2009gb,Kurkela:2014tea,Kurkela:2015qoa}, which however do not cover the
present physical situation.
Here, however, we shall follow a different strategy: motivated by the physical discussion in the 
previous sections, we shall propose a simplified version of the kinetic equation, which
allows for efficient numerical studies and even for piecewise analytic solutions.
Similar approximations have been already
used in the literature, separately for the elastic collisions and for the branching dynamics,
but here we shall combine them for the first time in a study of the jet thermalization.

First, we would like to write
an equation for the gluons from the jet alone, with the surrounding medium treated as a thermal bath which
influences the jet dynamics but is not significantly disturbed by the latter. 
The omission of the back-reaction is indeed well justified, given that the
gluonic system produced via branching remains dilute ($f(p)\ll 1$) down to the thermal
scale $p\sim T$, as we have seen. By the same token, the kinetic equation can be {\em linearized}
w.r.t. the  occupation number $f$ for the gluons in the jet.

The distinction between the gluons in the jet and those in the medium
is strictly possible for the energetic quanta with $p\gg T$, which
control most stages of the branching process, but it becomes ambiguous at 
later stages, notably during the thermalization process, where all the gluons have $p\lesssim T$. 
Notwithstanding, the discussion in the previous sections shows that the dynamics is rather smoothly changing
around $p\sim T$, where the relevant time scales become commensurable with each other,
cf. \eqn{scales}. Hence, there should be no danger (to parametric accuracy, at least)
with extrapolating down to $p\sim T$ an equation which is strictly valid for $p\gg T$. 
With that in mind, one can deduce rather simple approximations to the collision integrals
in \eqn{generaleq}. 

Consider the elastic collisions first, which preserve the number
of particles. Due to the
infrared singularity of the Coulomb exchanges, the dominant contribution to the elastic collision
integral comes from the small angle scatterings, i.e. from collisions where the momentum transfer 
between the colliding particles --- here, a gluon from the jet and another one from the thermal bath --- 
is much smaller than the individual momenta of these particles. More precisely, this property holds
to leading logarithmic accuracy, since the momentum exchanges within the range $m_D\ll q\ll T$
produce a Coulomb logarithm $\ln(T^2/m_D^2)\sim \ln(1/\alpha_s)$ in the relevant
transport coefficient (see below). Within this approximation, $\mcal{C}_{\rm el}[f]$ can be
replaced with the Fokker-Planck dynamics, 
i.e. the sum of diffusion and drag \cite{Lifshitz:1981,Moore:2004tg,Rapp:2009my,Ghiglieri:2015zma} 
\bea\label{CFP}
\mcal{C}_{\rm el}[f]\,\simeq\,
\f{1}{4}\,\hat{q}\,\nabla_{\bp}\cdot\left[ \left( \nabla_{\bp} + \f{{\bm v}}{T}\right) f\right]\,.
\eea
In writing this expression, we have exploited the isotropy of the diffusion tensor and 
the Einstein relation between drag and diffusion, as already
discussed in relation with \eqn{Langevin}. In fact, there is a one-to-one correspondence
between the Langevin dynamics described by  \eqn{Langevin} and the  Fokker-Planck dynamics 
encoded in the r.h.s. of \eqn{CFP}. In particular, it is easy to check that the above collision term 
admits the classical thermal distribution (the Maxwell-Boltzmann distribution for a massless particle)
as a fixed point: $\mcal{C}_{\rm el}[f_{\rm eq}]=0$ for $f_{\rm eq}(\bp)=\kappa \rme^{-p/T}$, where $\kappa$
is independent of $\bp$, but otherwise arbitrary. Clearly, one cannot expect gluons to obey a classical
distribution in thermal equilibrium, yet one can rely on this Fokker-Planck dynamics to qualitatively study the
{\em approach} towards equilibrium for the gluons from the jet, which have low occupancy.
The relevant mechanism at work (elastic collisions with the thermal particles) and the characteristic time scales
(namely $\tth(p)$ for energy loss, cf. \eqn{tth}, and $\trel$ for momentum broadening, cf. \eqn{trel}) are
correctly captured by \eqn{CFP}, to parametric accuracy at least. 

For a dilute quark-gluon plasma (QGP) and to the leading 
logarithmic accuracy of interest, $\hat{q}$ is given by \cite{Moore:2004tg,Huang:2014iwa,Ghiglieri:2015zma,Blaizot:2014jna}
\bea\label{qhat}
\hat{q}&=&8\pi\alpha_s^2 N_c \ln\left(\f{\bra k_{\rm max}^2\ket}{m_D^2}\right)\int \frac{\rmd^3{\bm p}}{(2\pi)^3} \left[N_c f_{_{\rm BE}}(1+f_{_{\rm BE}} ) +N_f f_{_{\rm FD}}( 1-f_{_{\rm FD}} )\right]\nn\\
&=&\alpha_s N_c T m_D^2   \ln\left(\f{\bra k_{\rm max}^2\ket}{m_D^2}\right),
\eea
where $m_D$ is the leading-order result for the Debye screening mass, that is,
\be
m_D^2\,=\,\f{2\pi}{3} \alpha_s T^2\left(2 N_c + N_f \right)\,,\ee
$N_f$ is the number of active quark flavors, $\bra k_{\rm max}^2\ket\sim T^2$ is the maximal 
momentum transfer squared between the probe gluon and the medium constituents, 
and we have included contributions from both
thermal gluons and thermal quarks, with respective occupation numbers
\be
f_{_{\rm BE}}(p) = \f{1}{e^{\f{p}{T}}-1}\,,\qquad f_{_{\rm FD}}(p) = \f{1}{e^{\f{p}{T}}+1}\,.
\ee

For the inelastic collision integral $\mcal{C}_{\rm br}[f]$, one can use the corresponding approximation
in Ref.~\cite{Baier:2000sb}, which describes medium-induced gluon branching in the LPM regime,
with the BDMPSZ branching rate shown in \eqn{Pdef}. The original equation in \cite{Baier:2000sb} involves
both splitting ($1\to 2$) and recombination ($2\to 1$) processes, but only the splitting terms survive
after linearizing w.r.t. the gluon occupation number. The ensuing expression reads
(see also Refs.~\cite{Blaizot:2012fh,Blaizot:2013vha})
\bea\label{Cbr}
\mcal{C}_{\rm br}[f]\,\simeq\,
\f{1}{\tbr(p)}\int_0^1 \rmd x\,  {\cal K}(x)\left[\f{1}{x^{\f{5}{2}}}f\left(t,\bx,{\f{{\bm p}}{x}}\right) - \f{1}{2} f(t,\bx,{{\bm p}}) \right]
\eea
with $\tbr(p)$ as defined in \eqn{tbr} and\footnote{Strictly speaking, the jet quenching
parameter $\hat q$ which enters the expression \eqref{tbr} for the branching time is not exactly the 
same as that occurring in the Fokker-Planck term, \eqn{CFP}, because of the difference between
the respective transverse scales $\bra k_{\rm max}^2\ket$: one has $\bra k_{\rm max}^2\ket\sim T^2$
for the Fokker-Planck dynamics and, roughly, $\bra k_{\rm max}^2\ket\sim \hat q\tbr\gg T^2$ for
the branching of sufficiently energetic gluons 
(see the discussion in \cite{Arnold:2008iy,Arnold:2008vd}). Here however we shall ignore
this subtle difference, which is anyway small, due to the weak, logarithmic, dependence upon 
$\bra k_{\rm max}^2\ket$, cf. \eqn{qhat}.}
\be\label{Kbr}
{\cal K}(x)\equiv\frac{[1-x(1-x)]^{\frac{5}{2}}}{[x(1-x)]^{\frac{3}{2}}}\,.
\ee
 We recall that $x$ and $1-x$ represent the longitudinal
momentum fractions of the daughter gluons.
The two terms in the r.h.s. of \eqn{Cbr} are recognized as the gain and loss terms associated with a collinear splitting:
in the gain term, a gluon with 3-momentum $\bp$ is produced via the splitting of a parent gluon with
momentum $\bp/x$\,; in the loss term, a gluon with momentum $\bp$ disappears because it splits.
The integrand has singularities at $x=0$ and $x=1$, which however cancel between the gain and loss terms,
and the integral is well defined. One can easily check that the branching integral \eqref{Kbr} preserves 
the total energy: $\int \rmd^3{\bm p}\,|\bp|\,\mcal{C}_{\rm br}[f](\bp)=0$.

Notice that the r.h.s. of \eqn{Cbr} does not vanish when $f$ approaches the thermal distribution
$f_{\rm eq}(\bp)\propto \rme^{-p/T}$.
(The only fixed point of this particular collision integral is the turbulent spectrum $f\propto 1/p^{7/2}$
\cite{Baier:2000sb,Kurkela:2011ti,Blaizot:2013hx}.)  However,
from the previous discussion, we also know that the branching term in \eqn{Cbr}
is strictly correct only when $p\gg T$. On one hand, the mechanism which triggers radiation
changes around $p\sim T$, from multiple scattering to single scattering, so the correct branching rate 
at lower momenta should be of the Bethe-Heitler type. On the other hand, the gluons from 
the jet having $p\lesssim T$ cannot be distinguished from the thermal gluons; so, in this
soft region of the phase-space, the collision integrals should involve the {\em total} occupation number 
(medium plus jet), including the associated non-linear effects. This total occupation number 
efficiently relaxes to the Bose-Einstein distribution.
When this happens, the non-linear terms --- which are omitted in \eqn{Cbr}, but would be
present in a more general version of $\mcal{C}_{\rm br}$ valid at soft momenta
\cite{Baier:2000sb,Arnold:2002zm}  --- will also stop the branching process.

However, this physical mechanism for stopping the branchings is not included 
in our linear equation, which applies to the gluon distribution created by the jet {\em alone}.
To cope with that while keeping the formalism as simple as possible, 
we shall cut off by hand the branching process at some arbitrary scale $p_*\sim T$.
In practice, we shall
enforce the condition $p\ge p_*$ for all the particles participating in a $1\to 2$ splitting process,
that is, we shall require $p\ge p_*$ for the daughter gluons and hence $p\ge 2p_*$ for their parent. 
This cutoff  $p_*$ should be viewed as a free
parameter of our model: the dependence of our predictions upon this scale, which as we shall
see is weak so long as $p_*$ remains of $\order{T}$, is indicative of the error that we have 
introduced by neglecting the non-linear terms in the (total) occupation number.

To summarize, our basic kinetic equation reads (with the compact notation $f_{{\bm p}}\equiv f(t,\bx,\bp)$)
\begin{align}
\label{eq:equation}
\left(\frac{\partial}{\partial t}+{\bm v}\cdot \nabla_{\bm x} \right)f_{{\bm p}} = {\f{1}{4}\hat{q}\,\nabla_{{\bm p}} \cdot\left[ \left( \nabla_{{\bm p}}  + \f{{\bm v}}{T}\right) f_{{\bm p}} \right]}+
{\f{1}{\tbr(p)}\int\limits_{r} \rmd x\,  {\cal K}(x)\left[\f{1}{x^{\f{5}{2}}}f_{\f{{\bm p}}{x}} - \f{1}{2} f_{{\bm p}} \right]},
\end{align}
where the symbol $\int\limits_{r}$ denotes the restricted integration over $x$ (see Sect.~\ref{sec:numerics} for details).
This equation should be solved with the following initial condition at $t=0$:
\be\label{fIC}
f(t=0,\bx,\bp)\,=\,\f{(2\pi)^3}{2(N_c^2-1)}\,\delta^{(3)}(\bx)\,\delta(p_z-E)\,
\delta^{(2)}(\bp_\perp)\,,
\ee
which represents the leading particle propagating along the $z$ axis with energy $E\gg T$.

\eqn{eq:equation} matches our present 
purposes: it correctly encodes the dynamics of the relatively hard constituents of the jet
with $p\gg T$ and, when extrapolated down to $p\lesssim T$, it also describes (at least to parametric accuracy)
their approach to {\em kinetic} equilibrium --- that is, the fact that the gluons from the jet individually approach
a Maxwell-Boltzmann distribution in momentum, via elastic collisions in the plasma. On the other hand, the
approach to {\em chemical} equilibrium --- the evolution of the ensemble of the complete gluon
distribution (jet+medium) 
towards the quantum Bose-Einstein distribution ---  is not encoded in this equation, but merely mimicked in a rather
crude way by the lower cutoff $p_*$ on the branching process.

Albeit considerably simpler than the original equations, \eqn{eq:equation} 
is still too complicated to be solved as it stands, including via numerical techniques.  A main source of
complication is the spatial inhomogeneity inherent in our problem, which is very strong to start with, 
cf. \eqn{fIC}, and plays an essential role in the subsequent dynamics.
In order to keep the salient features of this evolution in a numerically tractable way, we shall project
\eqn{eq:equation} along the longitudinal axis and at the same
time perform approximations based on the separation of scales $p_z\gg p_\perp$ between longitudinal 
and transverse momenta. This separation is physically realized so long as $p\gg T$, which is the regime
where \eqn{eq:equation} strictly applies, but is progressively washed out when decreasing the momenta
towards $T$. Yet, this approximation correctly keeps trace (to parametric accuracy, once again) of the
separation of time scales in the problem: indeed, as already explained, the characteristic time scales for
branching, \eqn{tbr}, and for the thermalization of hard particles, \eqn{tth}, are controlled by the
longitudinal momenta and become degenerate with each other (and with $\trel$, \eqn{trel}) only when
$p\sim T$.

Specifically, by integrating \eqn{eq:equation}  over the transverse phase-space, while at
the same time approximating $p\simeq p_z$ within the definition of $v_z$, within $\tbr(p)$,
and within the condition $p\ge p_*$, we finally obtain
\begin{align}\label{eq:equationL}
\left(\frac{\partial}{\partial t}+v_z\frac{\partial}{\partial z} \right)f_\ell(t,z,p_z) &= \f{1}{4}\hat{q}\frac{\partial}{\partial {p_z}}\left[\left(\frac{\partial}{\partial {p_z}} + \frac{v_z}{T}\right) f_\ell(t,z,p_z) \right]\nn\\*[0.3cm]
&
+\f{1}{\tbr(p_z)}\int\limits_r \rmd x\,  {\cal K}(x)\left[\f{1}{\sqrt{x}}\,
f_\ell\left(t,z,\f{p_z}{x}\right) - \frac{1}{2} f_\ell(t,z,p_z) \right],
\end{align}
where $v_z\equiv {p_z}/{|p_z|}$ and $f_\ell(t,z,p_z)$ is the longitudinal gluon distribution\footnote{Notice
 that this quantity $f_\ell$ is not an occupation number by itself
(because of the integration over the transverse phase-space in \eqn{f1D}), so in practice
one can very well have $f_\ell > 1$ and still use a linear kinetic equation, provided one can justify
that the actual occupation number is indeed small.},
\be\label{f1D}
f_\ell(t,z,p_z)\,\equiv\,\f{\rmd N_g}{\rmd z \rmd p_z}\,= \f{2(N_c^2-1)}{(2\pi)^3}\,
\int{\rmd^2\bx_\perp \rmd^2\bp_\perp}\,f(t,\bx,\bp)\,.
\ee
In \eqn{eq:equationL} it is understood that the partial derivative $\partial_{p_z}\equiv
\partial/\partial {p_z}$ commutes with $v_z$: $\partial_{p_z}(v_z f_\ell)=
v_z\partial_{p_z} f_\ell$. (This prescription follows for the limit $p_\perp\ll p_z$: starting with
$v_z=p_z/p$ with $p=\sqrt{p_z^2+p_\perp^2}$, one obtains $\partial_{p_z}v_z
=p_\perp^2/p^3\simeq p_\perp^2/p_z^3$, which is much smaller than the respective natural value 
$\sim1/p_z$.)

\eqn{eq:equationL}
is the equation that we shall explicitly study in what follows, via a combination of analytic and numerical
methods. To that aim, it is also useful to remind that the branching integral above admits
the turbulent fixed point $f_\ell\propto 1/p_z^{3/2}$, which is expected to control the shape
of the spectrum at intermediate momenta $p_*\ll p_z\ll E$.

\section{Semi-analytic studies of the kinetic equation}\label{sec:analytic}

As explained in the previous section, the two `collision terms' in the  r.h.s. of \eqn{eq:equationL}
become important in different kinematical regions, which are complementary to each other:
the high--energy region at $p\gg T$ for the branching term and, respectively, 
the low--energy region at $p\lesssim T$ for the elastic collisions. 
This distinction makes it possible to {\em separately} study their physical consequences --- 
at least, at a qualitative level. Namely, one can effectively treat the branching process 
as a {\em source} of relatively soft  gluons, which get injected into the medium at a scale 
$p_*\sim T$ and subsequently feel the effects of elastic collisions, in the form of drag and diffusion.
These considerations motivate the following, simplified, version of the kinetic equation  
\eqref{eq:equationL} :
\bea\label{eq:FP}
&&\left(\frac{\partial}{\partial t}+v\frac{\partial}{\partial z} \right)f(t,z,p) = \f{1}{4}\hat{q}\,\frac{\partial}{\partial {p}}\left[\left(\frac{\partial}{\partial {p}} + \frac{v}{T}\right) f(t,z,p) \right]+\mathcal{S}(t,z,p; p_*).
\eea
(From now on, we shall omit the subscript $\ell$ on $f$ as well as
the subscript $z$ on longitudinal momenta and velocities, to simplify notations.) 
Notice that, when using this source approximation, we implicitly assume that the medium
acts as a perfect sink: the branching dynamics at $p>p_*$ is not at all 
affected by collisions. For consistency, one must also construct the source $\mathcal{S}(t,z,p)$
by assuming an `ideal' gluon cascade, with wave turbulence, at $p>p_*$. This source
has the general structure
\bea\label{eq:Source}
\mathcal{S}(t,z,p; p_*)\,=\, \delta(t-z)\delta(p- p_*)\,\Gamma(t,p_*)
\,,\eea
where $\Gamma(t,p_*)$ is the flux of gluons at the lower end of the cascade
at time $t$. This flux can be easily inferred from the previous studies of the
`ideal' branching process \cite{Blaizot:2013hx,Fister:2014zxa} and will
be presented in Sect.~\ref{sec:source} below. Our ultimate purpose in this section is to
solve \eqn{eq:FP} with this particular source. 

In preparation to that, it will be useful to
study a couple of simpler cases, which have a physical interest by themselves and
for which we will be able to obtain exact solutions in analytic form. We shall start by
considering in Sect.~\ref{sec:steady} a steady source which propagates at the speed of light.
Then, in Sect.~\ref{sec:Green}, we shall construct the exact Green's function for
the differential operator appearing in \eqn{eq:FP} (i.e. for the ultrarelativistic Fokker--Planck 
equation in 1+1 dimensions). Finally, in Sect. \ref{sec:source}, we shall use this Green's function
to give a semi-analytic calculation of the gluon distribution produced by the physical source.


\subsection{Thermalization for a steady source}
\label{sec:steady}

In this subsection we shall present an exact solution for the case where the time-dependence 
of the injection rate $\Gamma(t,p_*)$ in the r.h.s. of  \eqn{eq:Source} can be neglected. 
This is a good approximation if one is interested in the effects of the collisions over a time 
interval $\Delta t$ which is much smaller than the characteristic time scale $\tbr(E)$ for the evolution of the source
via branchings, but much larger than $\trel$ (in order for the effects of collisions to be indeed significant); 
that is, $\trel < \Delta t\ll \tbr(E)$. During this time $\Delta t$, the source can be effectively treated as
`frozen' and the corresponding distribution at $z\le t$ is expected to depend only on $t-z$.

For convenience, we choose to normalize the injection rate as $\Gamma(t,p_*)=T$. (This brings no
loss of generality since the equation is linear.)  Also, within the context of this subsection, it is preferable 
to denote the energy of the soft gluons as $p_0$, rather than $p_*$. With these conventions, \eqn{eq:FP} 
becomes (below, a prime denotes a derivative w.r.t. $\hat p$)
\bea\label{eq:eqstat}
\left(\frac{\partial}{\partial \hat t}+v\frac{\partial}{\partial \hat z}\right)f=(f'+vf)' + \delta(\hat t- \hat z)
\delta(\hat p- \hat p_0),
\eea
in terms of dimensionless variables which measure the respective quantities in natural units,
that is, in units of $\trel$ for all the time and length scales ($\hat t=t/\trel$, $\hat z=z/\trel$) and in
units of $T$ for the various momenta ($\hat p=p/T$, etc). In what follows, we shall use 
such reduced variables in most formul\ae, to simplify the notation, but we shall
restore the physical units when discussing the
physical interpretation of the results. Also, we shall drop the hat 
on the reduced variables (e.g. $\hat p\to p$), 
as the distinction should be clear from the context.

\begin{figure}\begin{center}
\includegraphics[width=0.5\textwidth]{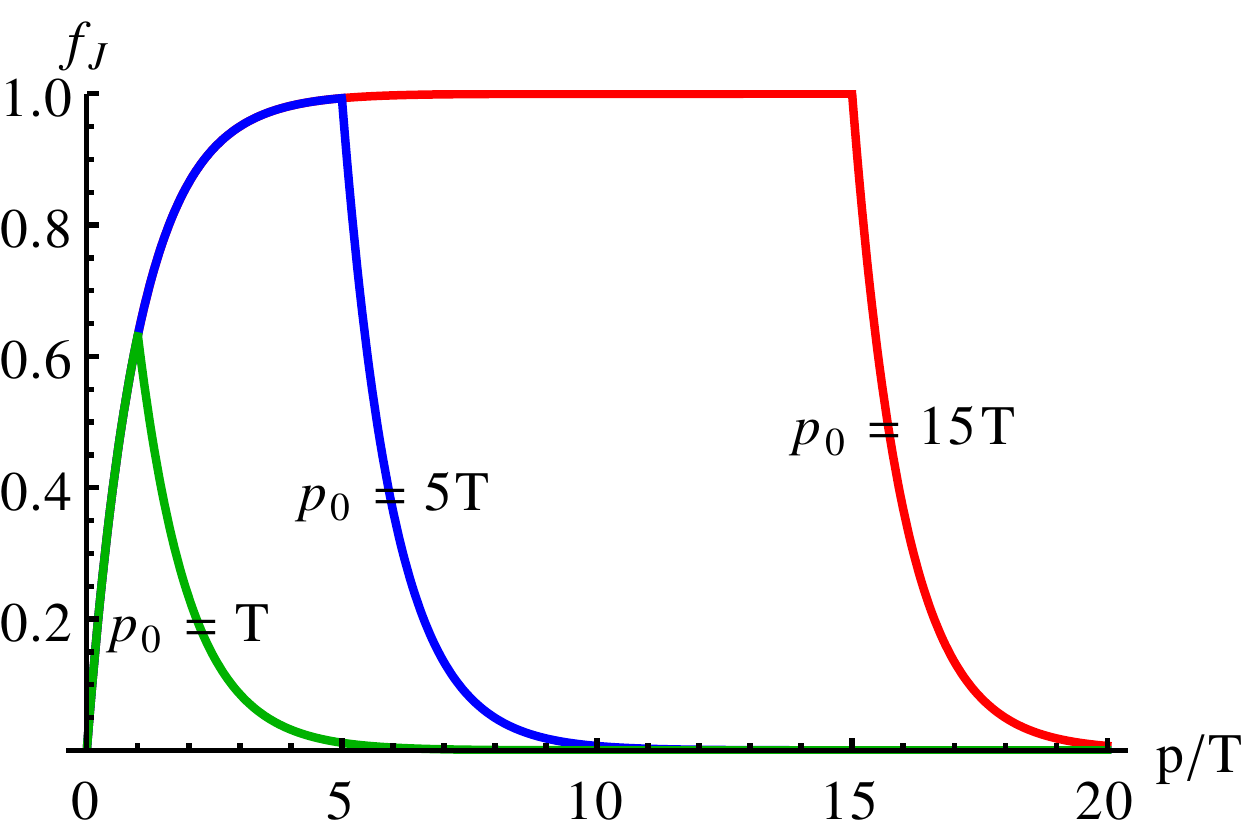}
\end{center}
\caption{The jet front function $f_{\rm J}(p,p_0)$ is displayed as a function of $p$ for $p_0/T =1, 5$ and $15$.}\label{fig:fs}
\end{figure}

We search for a stationary distribution $f(x^-, p; p_0)$ with $x^-\equiv t - z$. This function obeys
 \begin{equation}\label{eq:eqstatpm}
        \begin{cases}
        \displaystyle{0=(f'+f)'+\delta(x^-)\delta(p-p_0)   }
         &
        \text{ for\,  $p>0$,}
        \\*[0.3cm]
       \displaystyle{
       2\frac{\partial}{\partial x^-}f =(f'-f)' }&
        \text{ for\,  $p<0$}.
    \end{cases}
 \end{equation}
together with the condition for particle number conservation at $p=0$:
\begin{eqnarray}
\left(f'+f)\right|_{p=0^+}=\left(f'-f)\right|_{p=0^-}=\delta(x^-).\label{eq:numCon}
\end{eqnarray}
For $p>0$ the solution to Eq. (\ref{eq:eqstatpm}) is found in the form
\begin{eqnarray}
f(x^-, p; p_0)=f_{\rm J}(p,p_0)\delta(x^-)+C^+(x^-)\,\rme^{-p}
\end{eqnarray}
where $f_{\rm J}(p,p_0)$ is the `jet front function' (see also Fig. \ref{fig:fs})
\be\label{eq:fs}
f_{\rm J}(p,p_0)\equiv\,\rme^{-p} \left(\rme^{p_0}-1\right) \theta (p-p_0)+\left(1-\rme^{-p}\right) \theta (p_0-p)\,,
\ee
and $C^+(x^-)$ is an unknown function, to be later determined.
For $p<0$, it is convenient to use the Laplace transform of the solution, 
 $f_s(p)\equiv\int_0^\infty \rmd x^- \rme^{-sx^-}f(x^-,p;p_0)$. By taking the 
Laplace transform in the second equation \eqref{eq:eqstatpm}, one obtains
\begin{eqnarray}
f_s(p)= \tilde C^-(s)\rme^{\frac{p}{2}(\sqrt{1+8s}+1)}\qquad\text{for $p<0$}\,.
\end{eqnarray}
By imposing the conservation condition \eqref{eq:numCon} (which actually introduce
two constraints), we can determine both
 `coefficient' functions $\tilde C^-(s)$ and $\tilde C^+(s)$ (the Laplace transform of $C^+(x^-)$) :
\begin{eqnarray}
\tilde C^+(s)= \tilde C^-(s)=\frac{2}{\sqrt{8s+1}-1}
\end{eqnarray}
After also performing the inverse Laplace transformation, we finally obtain
 \begin{align}\label{eq:solstat}
f(t-z, p; p_0)&&=\left\{
\begin{array}{lr}
f_{\rm J}(p,p_0)\delta(t-z)+\left[\frac{1}{4} \text{erf}\left(\frac{\sqrt{t-z}}{2 \sqrt{2}}\right)+\frac{1}{4}+\frac{\rme^{-\frac{t-z}{8}}}{\sqrt{2 \pi } \sqrt{t-z}}\right]\rme^{-p}&\text{  for $p\geq 0$}, \\*[0.5cm]
\frac{1}{4} \rme^p \left[\text{erf}\left(\frac{2 p+t-z}{2 \sqrt{2} \sqrt{t-z}}\right)+1\right]+\frac{\rme^{-\frac{(-2 p+t-z)^2}{8 (t-z)}}}{\sqrt{2 \pi } \sqrt{t-z}}&\text{  for $p\leq 0$}.
\end{array}
\right.
 \end{align}
We have introduced here the error function
\bea\label{eq:erfc}
\text{erf}(x)\equiv \frac{2}{\sqrt{\pi }}\int _0^x\rmd t \,\rme^{-t^2}\,.
\eea

\begin{figure}
\begin{center}
 {\includegraphics[width=0.7\textwidth]{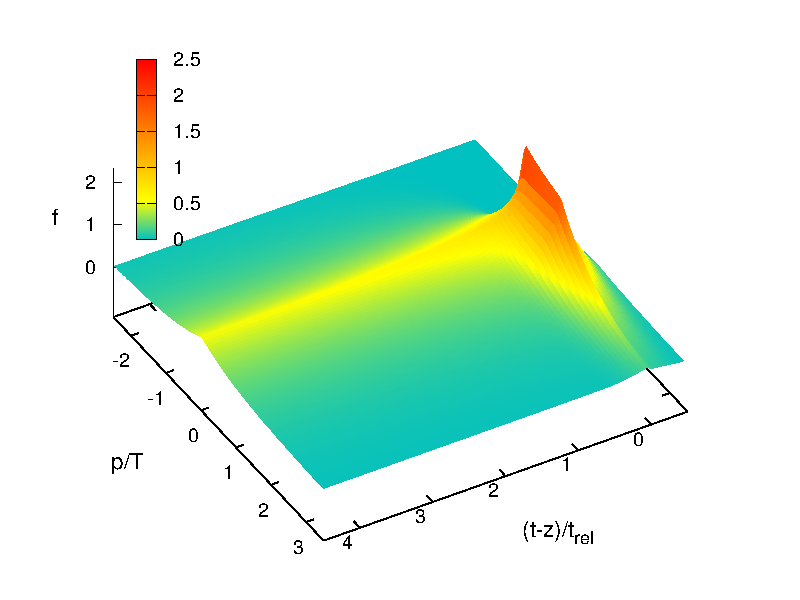}}
\end{center}
\caption{The distribution (\ref{eq:solstat}) as produced by a steady source is 
shown as a function of $t-z$ and $p$ for $p_0 = T$. This figure exhibits a front
moving along the light-cone at $z=t$ and a thermalized tail at $z\lesssim t-\trel$.
In displaying the front, the $\delta$-function in Eq. (\ref{eq:solstat}) is regulated as
$\delta_\epsilon(t-z)=\f{1}{\sqrt{\epsilon \pi}}\rme^{-\f{(t-z)^2}{\epsilon}}$ with $\epsilon=0.1$.}\label{fig:sourced}
\end{figure}

By inspection of the solution in \eqn{eq:solstat}, one can recognize a {\em front} which propagates at the speed 
of light with the profile in $p$ shown in Fig. \ref{fig:fs}, and a {\em tail} at $z < t$ which is localized
around $p=0$. The overall distribution is illustrated in Fig.~\ref{fig:sourced} for the case $p_0=T$,
which is the most interesting one for the physical problem at hand (recall that $p_0\equiv p_*$ is the
infrared end of the gluon cascade). However, in order to better appreciate the physical content of this stationary
distribution, it is useful to consider first the high-energy case $p_0\gg T$. Then, as visible in Fig. \ref{fig:fs}, the front
profile becomes a $\theta$--function with support at $0 < p < p_0$. This can be understood as follows:
a particle injected by the source at time $t_0$ with  $p_0 \gg T$ loses energy towards 
the medium at a constant rate, via drag (recall the discussion following \eqn{Langevin}); hence,
 its energy decreases with time according to 
\begin{eqnarray}
p(t)=p_0-(T/\trel)(t-t_0).\label{eq:ptcol}
\end{eqnarray}
So long as this energy remains much larger than $T$, which is indeed the case during a large interval
$t-t_0\simeq (p_0/T)\trel\gg \trel$, the diffusion effects are negligible and the distribution created by this
particle can be as well studied by neglecting the second--order 
derivative $f^{''}$ in \eqn{eq:eqstat}. The corresponding solution is easily found as
\be\label{eq:nodiff}
f(t-z, p; p_0)=\delta(t-z)\theta (p_0-p)\theta(p)
\qquad \mbox{(without diffusion)}\,,
\ee
which is indeed very similar to the `front' piece of \eqn{eq:solstat} in the case $p_0\gg T$. Hence,
the `front' is built with those particles that have been recently injected by the source, within a time interval 
$\Delta t=(p_0/T)\trel$ prior to the time $t$ of measurement, and which  have a still a relatively large
energy $p\gtrsim T$ at time $t$. All the other particles, that have been injected at earlier times
$t' < t-(p_0/T)\trel$, have been degraded by the viscous drag to energies $p\lesssim T$, where the 
diffusion effects {\em are} important. This becomes clear by inspection of the function $f_{\rm J}(p,p_0)$
for $p_0=1$ in Fig. \ref{fig:fs}.

As a consequence of the competition between diffusion and drag, the gluons with $|p|\lesssim T$ can 
have both positive and negative velocities, hence their distribution moves at a slower speed $|v| < 1$
This explains the depletion visible in $f_J$ at $p\lesssim T$  (for any $p_0$) and also the formation of the 
tail. At points sufficiently far away from the front, such that $t-z\gg t_{rel}$, the distribution reaches
thermal equilibrium, since this is the fixed point of the Fokker-Plank dynamics. Indeed, 
the solution in \eqn{eq:solstat} implies
\be\label{MB}
f(t-z, p; p_0) \,\simeq\, \f{1}{2} \,\rme^{-|p|/T}\,\quad\mbox{when}\ \ t-z\gg \trel\,.
\ee
To understand the energy balance between the jet and the medium, notice that the external source
in \eqn{eq:eqstat} inserts energy at a rate $\rmd E_s/\rmd t=p_0T$, whereas the energy carried by the 
thermalized tail $t-z\gg \trel$ increases at a rate $\rmd E_{\rm ther}/\rmd t=T^2$. (The total energy
in the front in independent of time, $E_{\rm J}=\int \rmd p p f_{\rm J}(p,p_0)\sim \trel p_0^2$, and represents
only a negligible fraction of the total energy injected by the source over large times.) If $p_0\gg T$,
then the insertion rate is much larger than the thermalization rate, meaning most of the energy is
lost via viscous drag --- meaning it is transferred via collisions to the medium constituents.
By choosing $p_0\simeq T$, i.e. by inserting the particles directly at the medium scale, we can
minimize the effects of the drag and thus recover most of the injected energy in the thermalized tail.

The qualitative features that we have discovered in this simple example are in fact generic and will
be recovered in the more general situations to be studied later on. In particular, the peculiar structure of
the distribution visible in Fig.~\ref{fig:sourced}, with a jet localized on the light-cone ($z=t$) and a 
thermalized tail well behind it ($z< t-\trel$), will also show up for a physical jet initiated by a leading particle
with energy $E\gg T$, at least for sufficiently small times $t\lesssim \tbr(E)$.

\subsection{Jet quenching in the source approximation}
\label{sec:queching}
In this subsection we shall use the source approximation, cf. Eq. (\ref{eq:FP}), in order 
to unveil generic features of the jet evolution in the presence of both branchings
and elastic collisions. The solution to \eqn{eq:FP} corresponding to a 
general source $\mathcal{S}(t,z,p)$ can be written as
\bea\label{eq:gensol}
f(t,z,p)=&& \int \rmd p_0 \rmd z_0\, f_G(t,z-z_0,p,p_0) f_0(z_0,p_0)\nn\\
&& + \int \rmd p_0 \rmd z_0 \int_{-\infty}^t \rmd t' f_G(t-t',z-z_0,p;p_0) \mathcal{S}(t',z_0,p_0),
\eea
where we have chosen the initial condition $f(0,z,p)=f_0(z, p)$ and
 $f_G$ is the appropriate Green's function, that is, the solution to the homogeneous equation
\bea
&&\left(\frac{\partial}{\partial t}+v\frac{\partial}{\partial z} \right)f_G(t,z,p) = \frac{\partial}{\partial {p}}\left[\left(\frac{\partial}{\partial {p}} + v\right) f_G(t,z,p) \right],
\eea
with initial condition $f_G(0,z,p;p_0)=\delta(z)\delta(p-p_0)$ with $p_0>0$. 
An analytic form for this Green's function will be constructed in the next subsection and then applied
to the source representing an ideal branching process, in Sect.~\ref{sec:source}.
 
\subsubsection{The Fokker-Planck Green's function}
\label{sec:Green}
The Green's function for the longitudinal Fokker-Planck equation can be constructed
via a mathematical method similar to that described in the previous subsection for the case of
a steady source. In what follows, we shall omit the details but merely show the starting point equations, 
which replace the previous equations \eqref{eq:eqstatpm} and \eqref{eq:numCon}.
After performing Laplace and Fourier transforms with respect to $t$ and $z$ respectively, we deduce
 \begin{equation}\label{eq:fsQ}
        \begin{cases}
        \displaystyle{(s+iQ)-\delta(p-p_0)=f_{sQ}''+f_{sQ}'}
         &
        \text{ for\,  $p>0$,}
        \\*[0.3cm]
       \displaystyle{
     s f_{sQ}-iQ f_{sQ}=f_{sQ}''-f_{sQ}'} &
        \text{ for\,  $p<0$}.
    \end{cases}
 \end{equation}
together with the following condition for the number conservation
\begin{eqnarray}
\left(f_{sQ}'+f_{sQ})\right|_{p=0^+}=\left(f_{sQ}'-f_{sQ})\right|_{p=0^-},\label{eq:numConG}
\end{eqnarray}
where $f_{sQ}$ is a compact notation for a function of two arguments, $s$ and $Q$, defined as
\begin{eqnarray}
f_{sQ}\equiv\int_0^\infty \rmd t\, \rme^{-st}\int \rmd z\, \rme^{-iQz} f(t,z,p).
\end{eqnarray}
After lengthy but straightforward mathematical manipulations, one finally obtains
\be
f_G(t,z,p; p_0)=\theta(p)f_G^+(t,z,p; p_0)+\theta(-p)f_G^-(t,z,p; p_0)\,,\ee
where (with $t \ge |z|$)
\bea\label{eq:fG+}
&&f_G^+(t,z,p; p_0)=\frac{ \rme^{-\frac{p_0-p}{2}-\frac{t}{4}}}{2 \sqrt{\pi t }}\left[\rme^{-\frac{(p-p_0)^2}{4 t}}-\rme^{-\frac{(p+p_0)^2}{4 t}}\right]\delta (t-z)\nn\\
&&+\frac{\rme^{-\frac{(p+p_0-z)^2}{4 t}-p}}{8 \sqrt{\pi } t^{5/2}} \left[t (t+2)-(p+p_0-z)^2\right] \text{erfc}\left(\frac{1}{2} \sqrt{t-\frac{z^2}{t}} \left(\frac{p+p_0}{t+z}-1\right)\right)\nn\\
&&+\frac{(t+z)  (p+p_0+t-z)}{4 \pi  t^2 \sqrt{t^2-z^2}}\,\rme^{-\frac{(p+p_0)^2}{2 (t+z)}+\frac{p_0-p}{2}-\frac{t}{4}}
,\eea
and 
\bea\label{eq:fG-}
&&f_G^-(t,z,p; p_0)=\frac{(t+z) (p_0+t-z) -p (t-z)}{4 \pi  t^2 \sqrt{t^2-z^2}} \,\rme^{-\frac{p^2}{2 (t-z)}+\frac{p+p_0}{2}-\frac{p_0^2}{2 (t+z)}-\frac{t}{4}}\nn\\
&&+\frac{\rme^{p-\frac{(p+p_0-z)^2}{4 t}} }{8 \sqrt{\pi } t^{5/2}}
\left[t (t+2)-(p+p_0-z)^2\right] \text{erfc}\left(\frac{1}{2} \sqrt{t-\frac{z^2}{t}} \left(\frac{p_0}{t+z}-\frac{p}{t-z}-1\right)\right)
.\eea
In these formul\ae, we have introduced
the complementary error function $\text{erfc}(x)\equiv 1 - \text{erf}(x)$ (cf. \eqn{eq:erfc}).
Also, we have used reduced variables $p\to p/T$, $t\to t/\trel$ etc.,  to simplify writing
(recall the discussion after \eqn{eq:eqstat}). The above expression for $f_G$ 
is normalized to unity w.r.t. these reduced variables:
$\int \rmd z\rmd p\,f_G(t,z,p; p_0)=1$. In order to obtain the properly normalized 
Green's function in physical units, one must divide the result  in 
Eqs.~\eqref{eq:fG+}--\eqref{eq:fG-} by the dimensionless product $T\trel$ and replace
all the reduced variables by their physical counterparts ($p\to p/T$, etc).


\begin{figure}[t] \centerline{\hspace*{-.3cm}
  \includegraphics[width=0.3\textwidth]{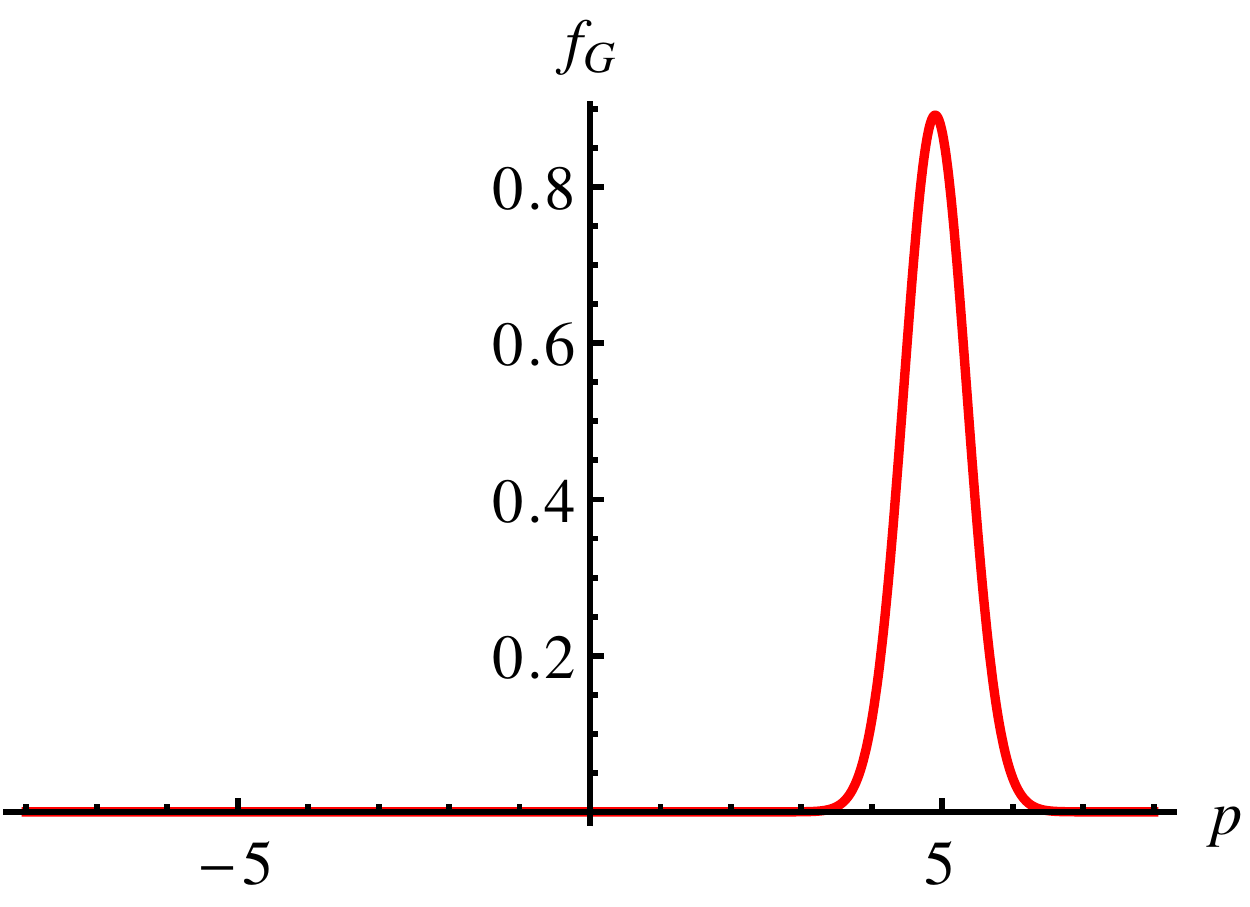}\quad
\includegraphics[width=0.3\textwidth]{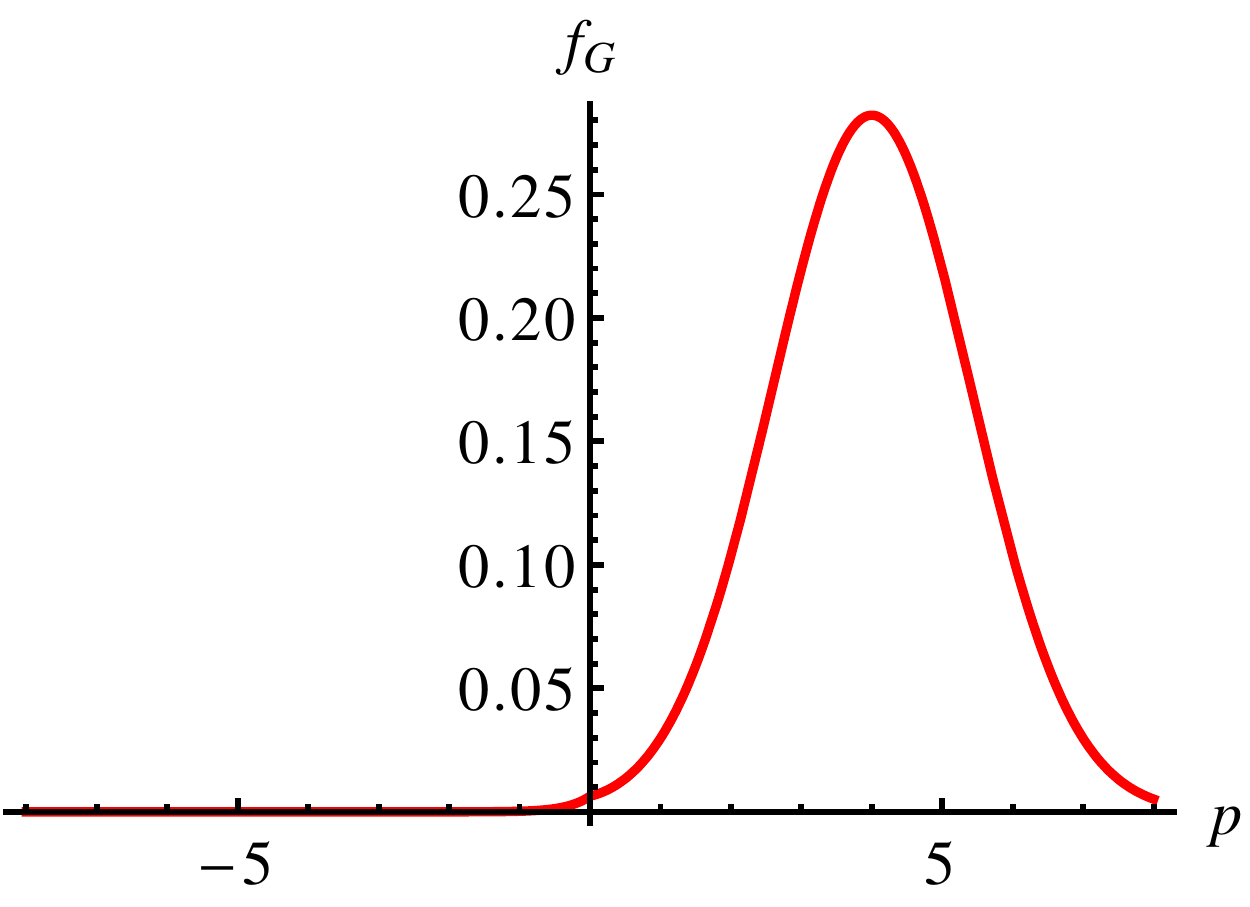}\quad
\includegraphics[width=0.3\textwidth]{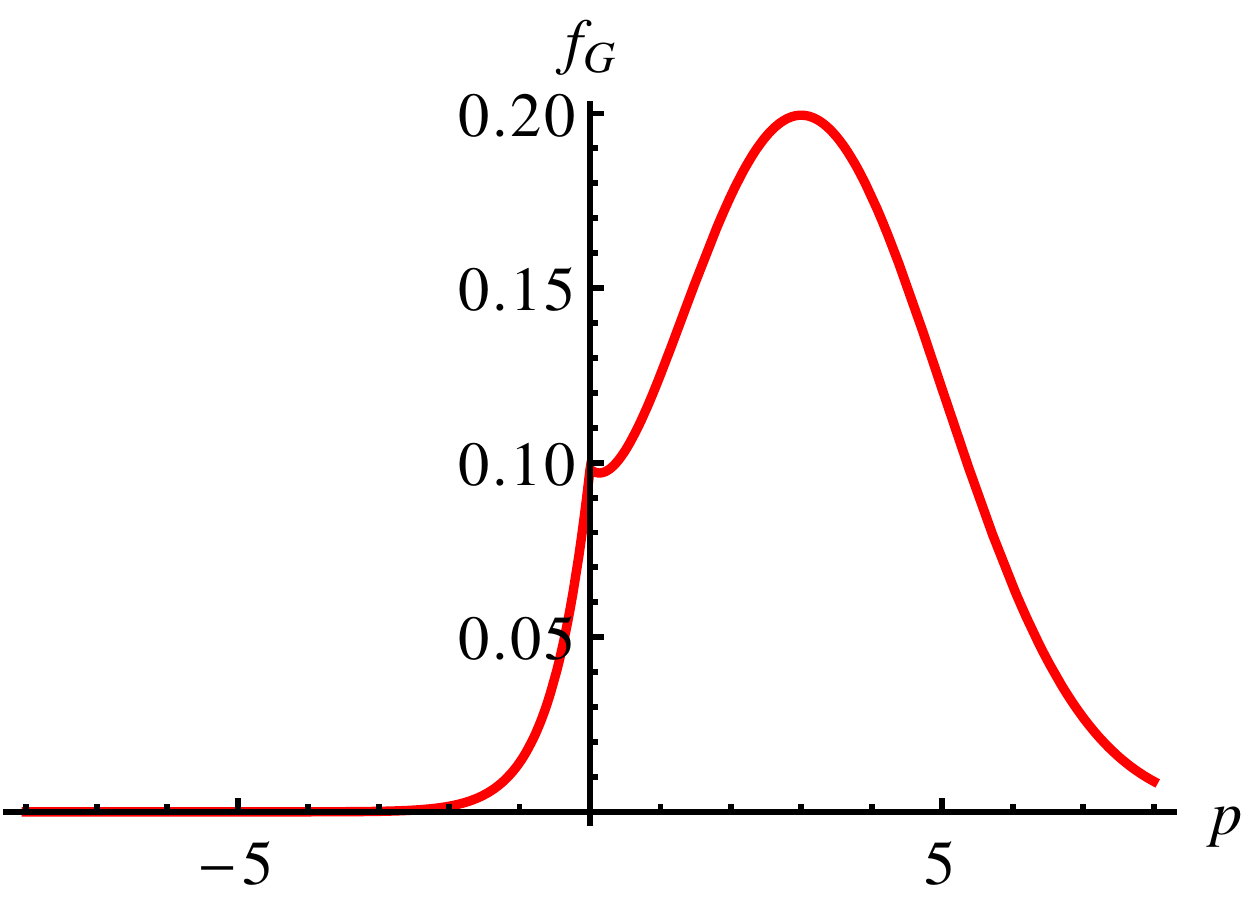}}
\vspace*{0.4cm}
 \centerline{\hspace*{-.3cm}
  \includegraphics[width=0.3\textwidth]{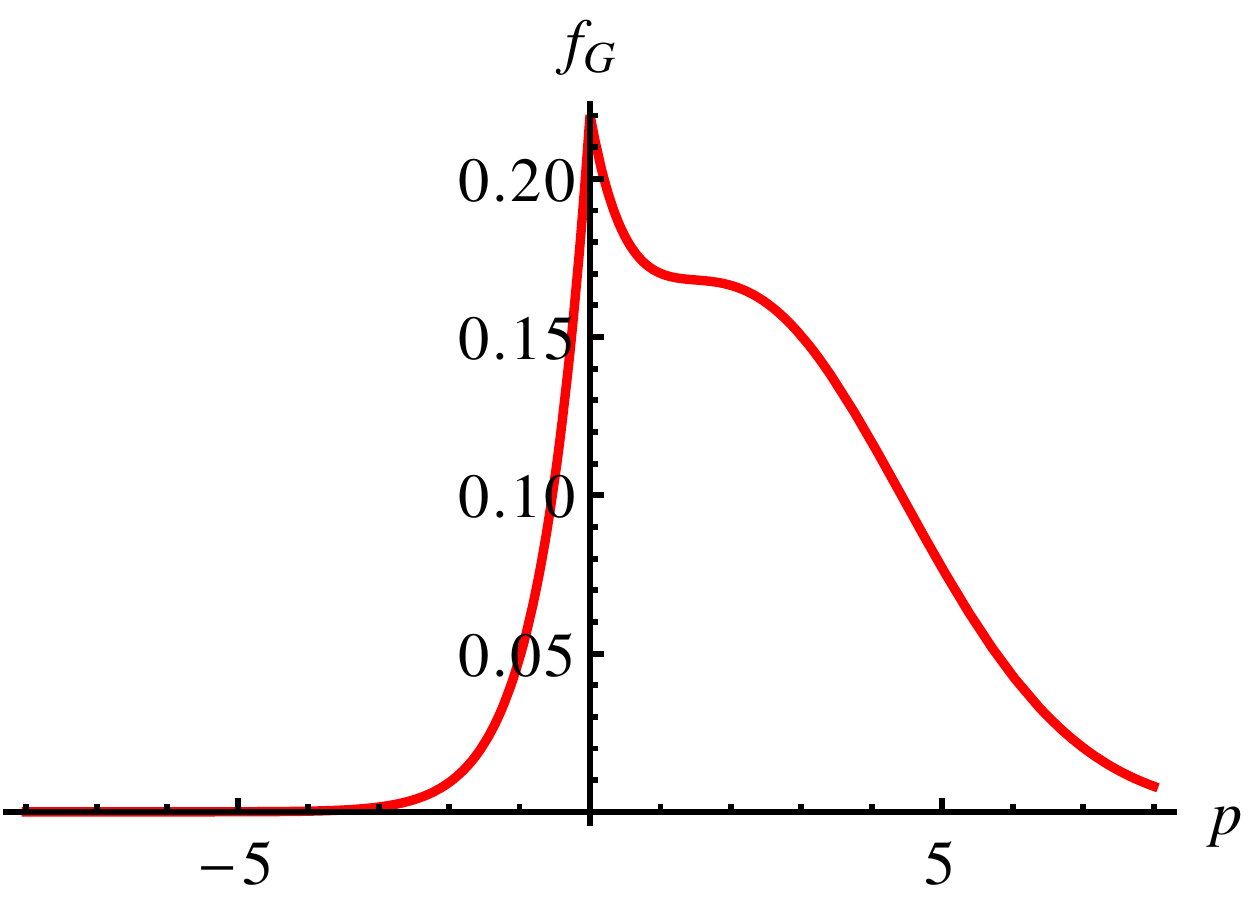}\quad
\includegraphics[width=0.3\textwidth]{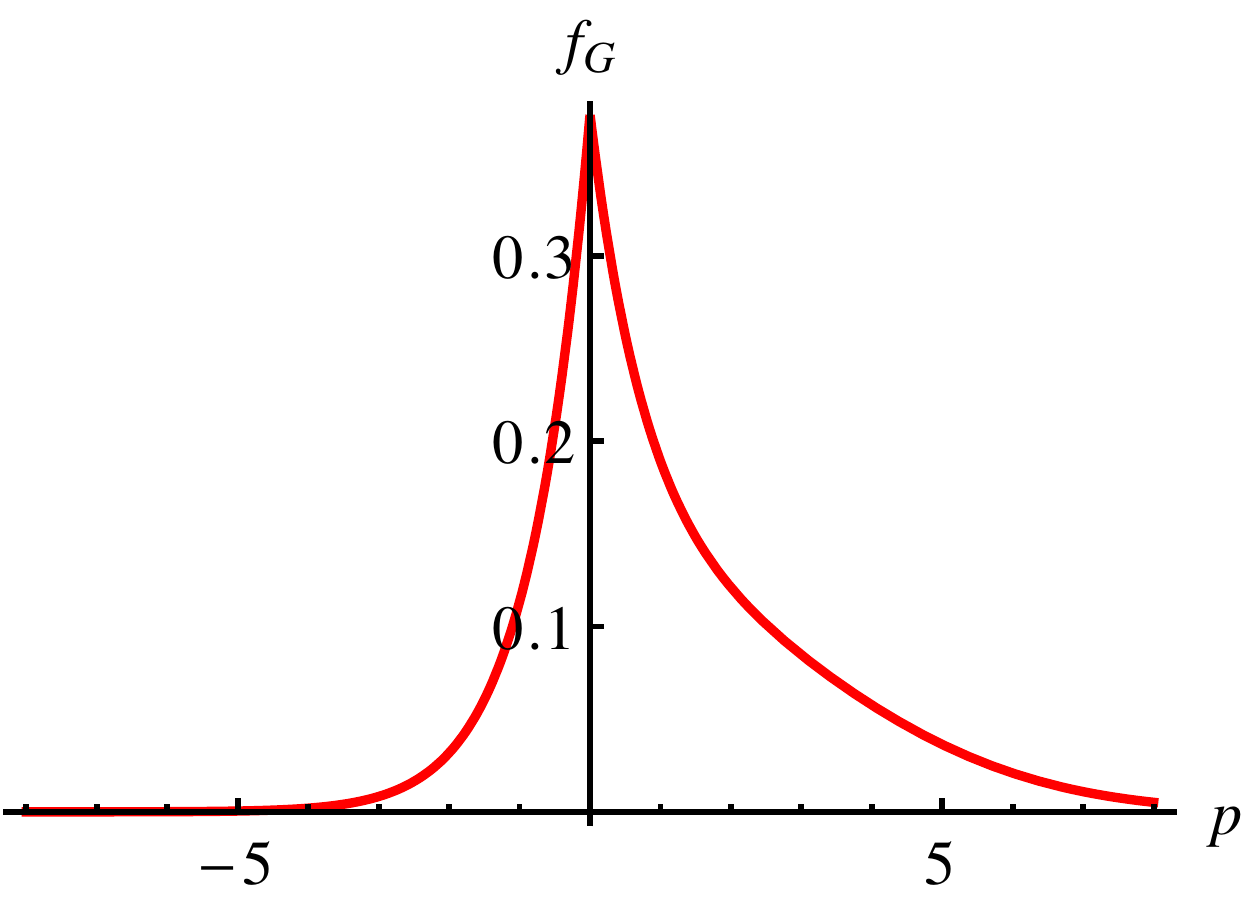}\quad
\includegraphics[width=0.3\textwidth]{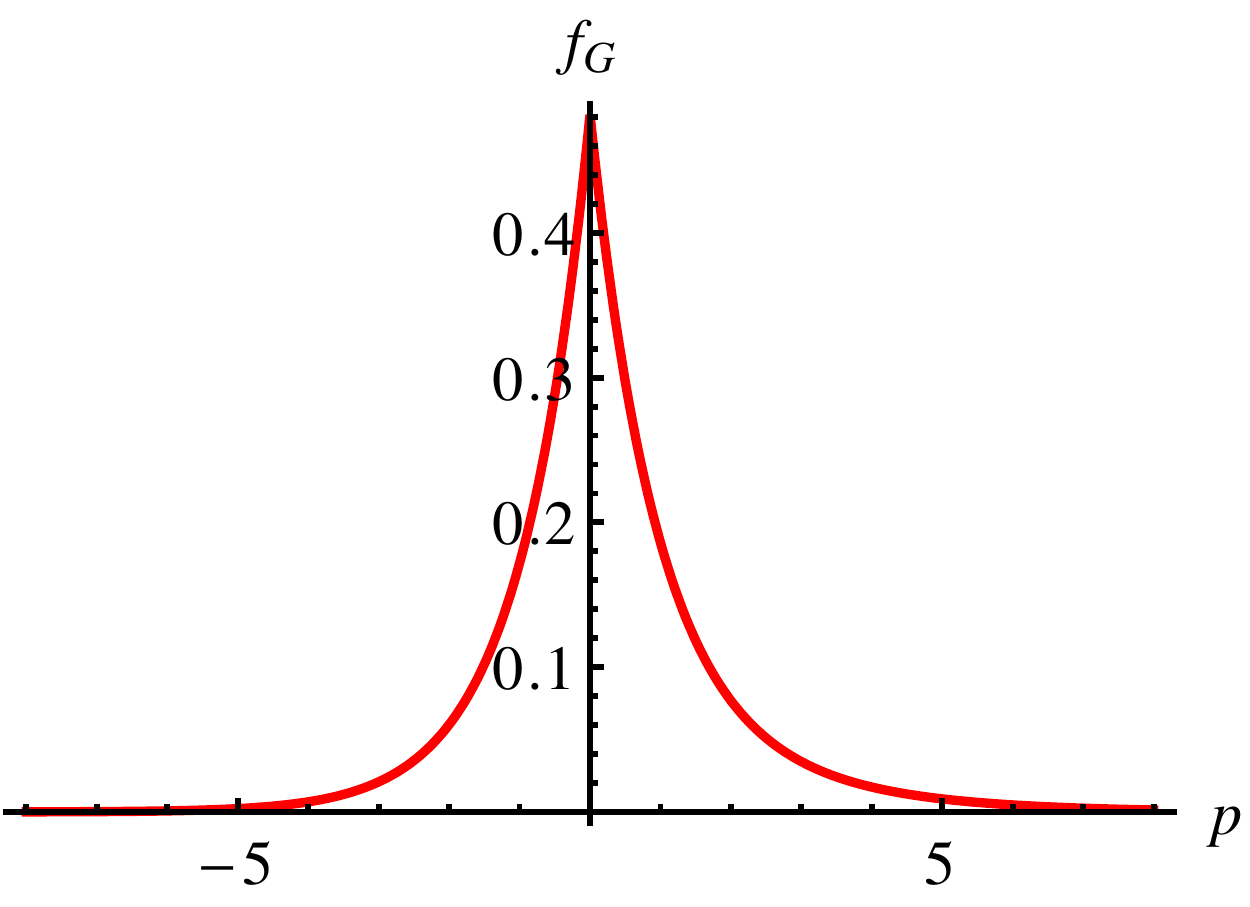}}
 \caption{\sl The `reduced' Green's function in \eqn{eq:fGhom} plotted as a function of
 $p$ for $p_0=5$ and 6 successive values of time: $t=0.1,\,1,\,2,\,3,\,5,\,10$.}
 \label{Gred}
\end{figure}

The above expression for $f_G$ looks quite involved. In order to unveil its physical content, it is useful
to first consider its simpler version obtained after integrating over $z$~:
 \begin{equation}\label{eq:fGhom}
    f_G(t,p;p_0)=
    \begin{cases}
        \displaystyle{\frac{1}{4}\,\rme^{-p}\,\text{erfc}\Big(\frac{p+p_0-t}{2\sqrt{t}}\Big)
        +\frac{1}{2\sqrt{\pi  t}}\,\rme^{-\frac{(p-p_0+t)^2}{4t}} }
         &
        \text{ for\,  $p>0$}
        \\*[0.4cm]
       \displaystyle{
        \frac{1}{4}\,\rme^{p}\,\text{erfc}\Big(\frac{-p+p_0-t}{2\sqrt{t}}\Big)
        +\frac{\rme^{p}}{2\sqrt{\pi  t}}\,\rme^{-\frac{(-p+p_0-t)^2}{4t}}
        }&
        \text{ for\,  $p<0$}.
    \end{cases}
 \end{equation}
 This function describes the relaxation of an initial perturbation
which is homogeneous in $z$ but localized in momentum: $f_G(0,p;p_0)=\delta(p-p_0)$.
For $p_0\gg T$ and sufficiently small times, such that $\f{t}{\trel}\ll \f{p_0}{T}$,
$f_G(t,p;p_0)$ is dominated by its Gaussian component at $p>0$, that is,
\bea\label{eq:Gauss}
f_G(t,p;p_0)\,\simeq\,\frac{1}{2\sqrt{\pi  t}}\,\rme^{-\frac{(p-p_0+t)^2}{4t}}\qquad
\mbox{when\ $t\ll p_0$}\,.\eea
This describes the damping of the original energy via drag and also the
broadening of the longitudinal momentum distribution due to diffusion, in agreement
with the discussion in Sect.~\ref{sec:therm}:
\bea
\langle p(t)\rangle \simeq p_0 - (T/\trel)t\,,\qquad 
\langle p^2\rangle - \langle p\rangle^2 \simeq 2\hat q t\,.
\eea
But already for such small values of time, there is a second component (represented by the two terms
proportional to the complementary error function) which starts growing around $p=0$.
This corresponds to particles which have essentially lost their original
energy and pile up around $p= 0$, via diffusion. 
For larger times $t\gtrsim (p_0/T)\trel$, this component becomes the dominant one
and rapidly approaches  the thermal distribution (notice
that $\text{erfc}(x)\to 2$ when $x\to -\infty$) :
\be\label{thermal}
f_G(t,p;p_0) \simeq \f{1}{2} \,\rme^{-|p|}\qquad
\mbox{when\ $t\gg p_0$}\,.
\ee

\begin{figure}[ht] \centerline{\hspace*{-.3cm}
  \includegraphics[width=0.45\textwidth]{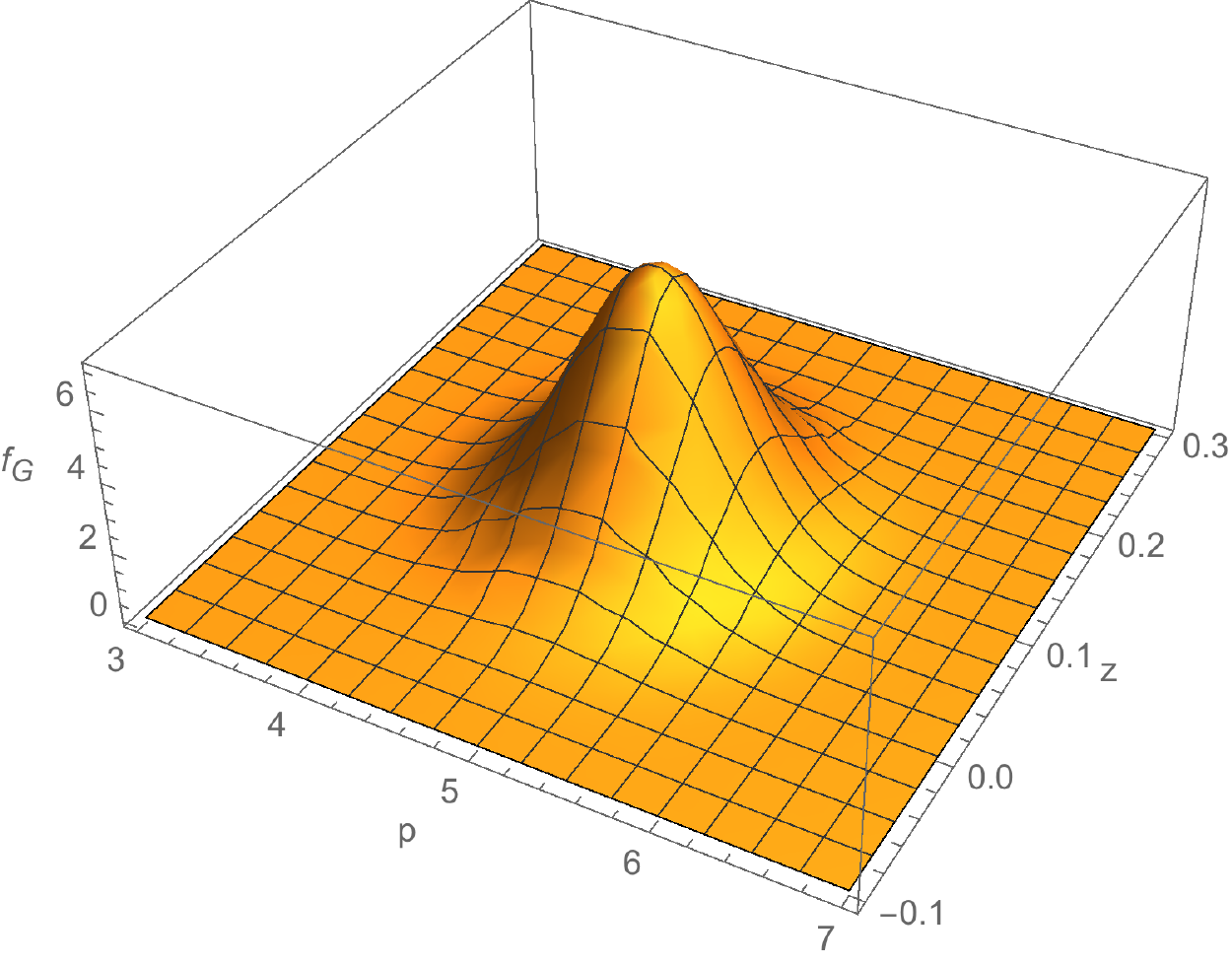}\quad
\includegraphics[width=0.45\textwidth]{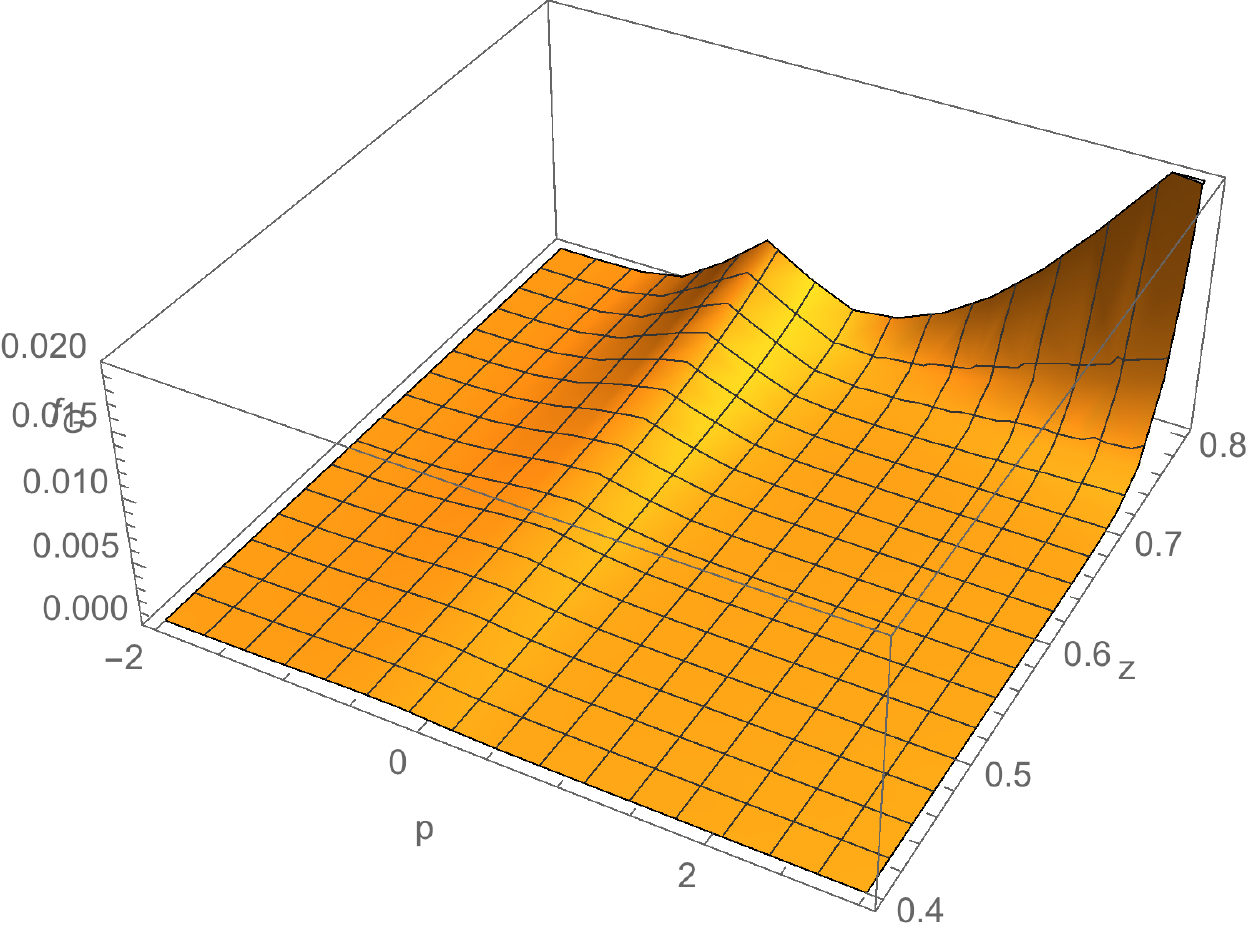}}
\vspace*{0.4cm}
 \centerline{\hspace*{-.3cm}
 \includegraphics[width=0.45\textwidth]{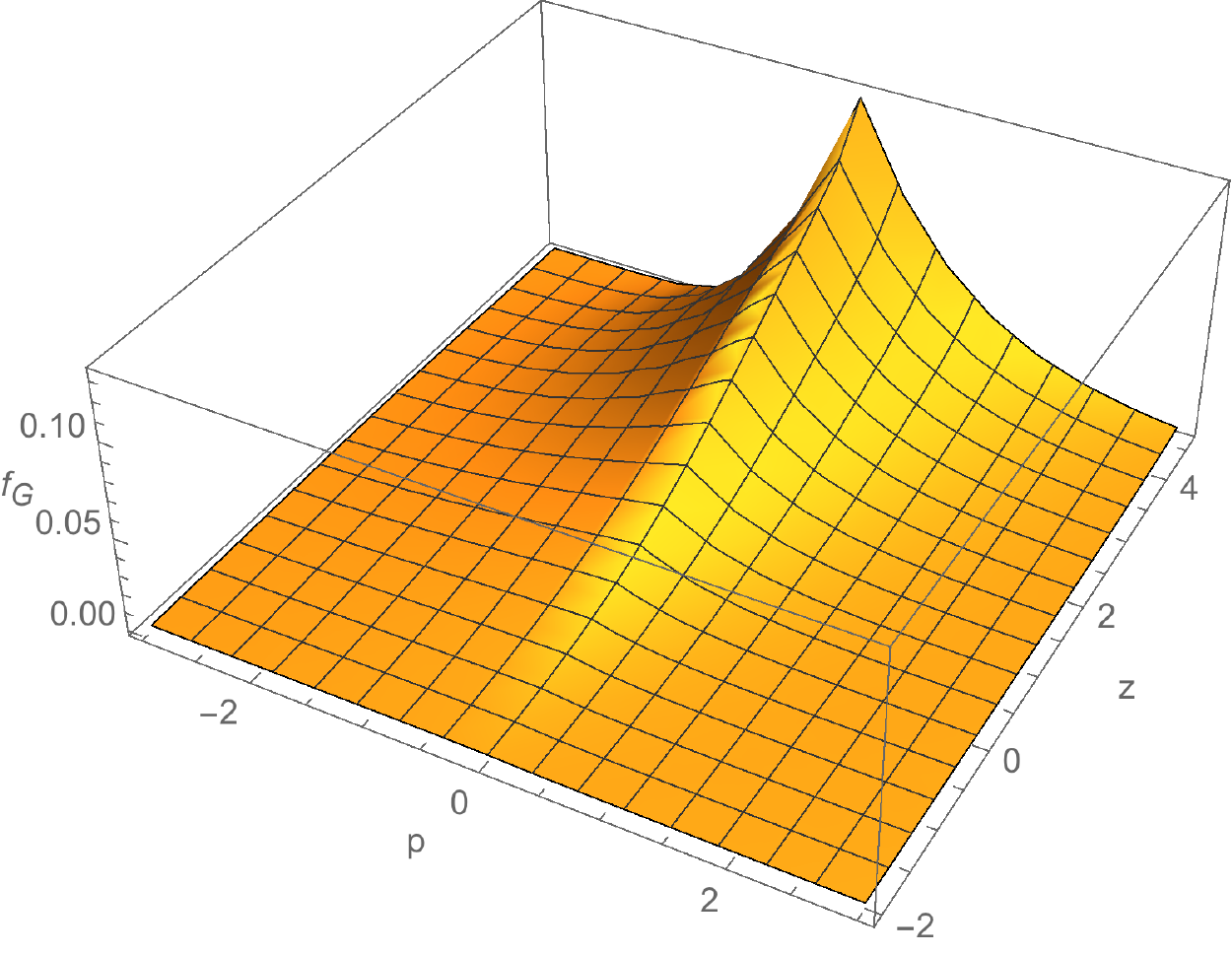}\quad
\includegraphics[width=0.45\textwidth]{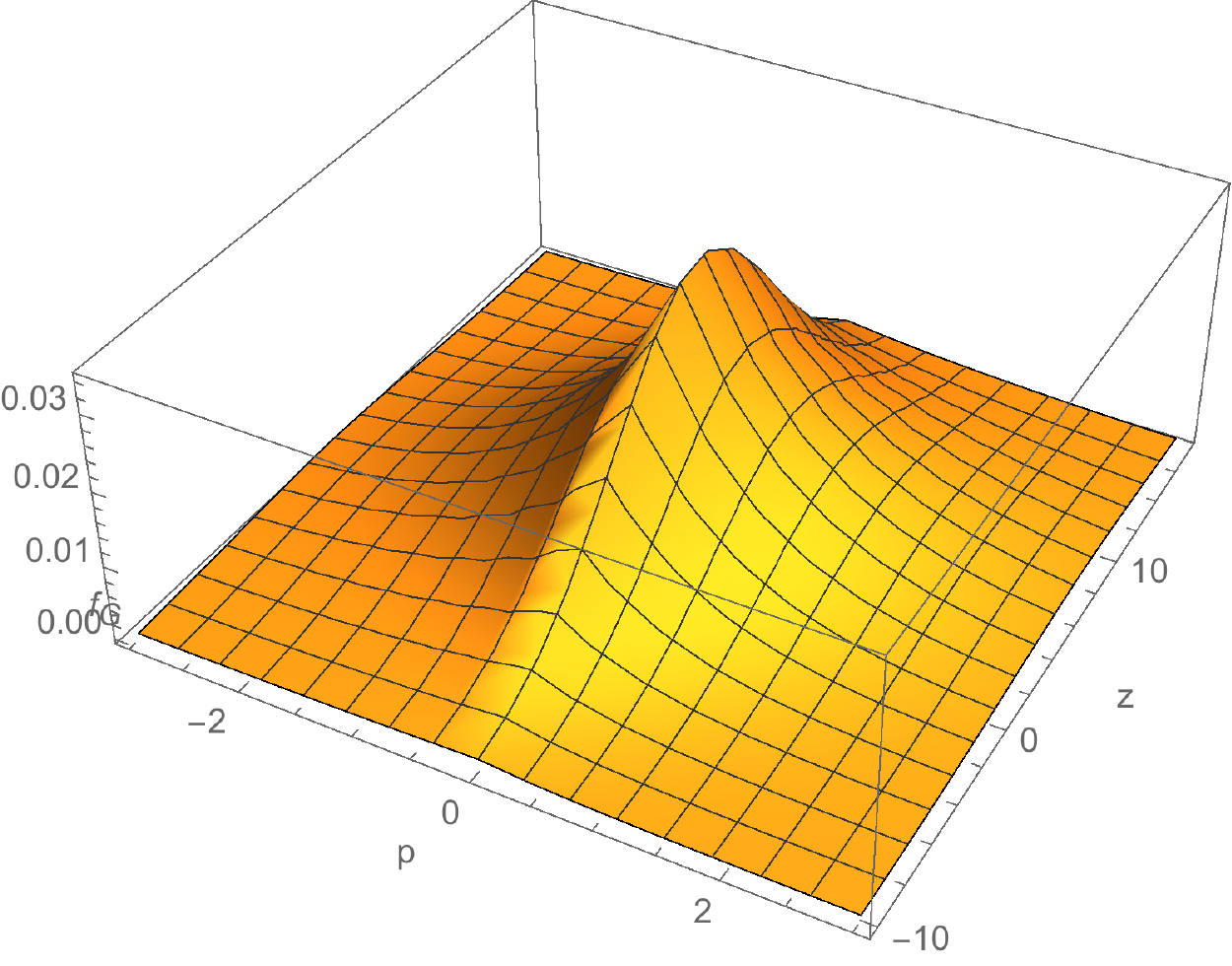}}
 \caption{\sl The Green's function in Eqs.~\eqref{eq:fG+}--\eqref{eq:fG-} plotted
as a function of $p$ and $z$ for $p_0=5$ and four values of $t$: upper line, left: $t=0.1$;
upper line, right: $t=1$; lower line, left: $t=5$; lower line, right: $t=20$.
 Note that the vertical scales and also the ranges in $p$ and $z$
 can significantly differ from one figure to another.
For $t\gtrsim 1$, one can see the emergence and growth of a thermal tail at  $|z|\ll t$,
which for $t\gg p_0$ is the Gaussian shown in \eqn{eq:dif}.}
 \label{Gpz}
\end{figure}

Turning now to the general case with $z$--dependence, 
Eqs.~\eqref{eq:fG+}--\eqref{eq:fG-}
show that, at early times, $t\ll (p_0/T)\trel$, there is a remnant of
the original perturbation --- the `jet front' localized at $z=t$, as  described by the first term 
in the r.h.s. of \eqn{eq:fG+} --- whose momentum
distribution is however degrading with time, in the same way as in \eqn{eq:Gauss}. 
With increasing time, this `jet front' is gradually 
washed out and a new distribution develops around $p=0$, which is 
slowly varying in $z$ (for $|z|\ll t$ at least). For sufficiently large times, $t\gg (p_0/T)\trel$,
and sufficiently far behind the front, $z \lesssim t-\trel$, this new distribution
is thermal in $p$  :
 \be\label{eq:dif}
f_G(t,z,p; p_0)\simeq \frac{\rme^{-|p|}}{2}\,
 \frac{\rme^{-\f{(z-p_0)^2}{4 t}}}{2 \sqrt{\pi  t}}
 \quad\mbox{when\ $t\gg p_0\gg 1$ and $|z|\ll t$}
\,.
\ee
This is recognized as the product between the Maxwell-Boltzmann distribution in momentum, 
cf. \eqn{thermal}, and the one-dimensional heat kernel describing diffusion in $z$.
\eqn{eq:dif} shows that the large-time distribution is centered around $z=(p_0/T)\trel$ (the maximal
distance travelled by the original perturbation until it has lost all its energy, due to drag)
and that its longitudinal extent grows with time, according to
 $\Delta z(t)\simeq \sqrt{4\trel t}$. Thus, remarkably, the momentum--space diffusion
 originally encoded in the Fokker--Planck equation has also generated {\em spatial}
 diffusion. One may chose as a criterion for having a quasi-homogeneous distribution
 the condition that the longitudinal extent $\Delta z(t)$ be larger than the location $(p_0/T)\trel$ of
 the center. This happens for
 \be
 t\,\gtrsim\,\f{1}{4}\left(\f{p_0}{T}\right)^2\trel\,,\ee
a time scale considerably larger than that required by the thermalization of the momentum
distribution. Note finally that by integrating \eqn{eq:dif} over $z$ and over $p$ we recover,
 to the accuracy of interest, the normalization of the initial perturbation,
 i.e. $\int \rmd z\rmd p\,f_G(t,z,p)=1$, which confirms that the whole perturbation has 
 thermalized. If on the other hand one computes the total energy contained in the thermalized
 distribution \eqref{eq:dif} one finds, clearly, $\int \rmd z\rmd p\,|p|\,f_G(t,z,p)\simeq T$ for
 $t\gg (p_0/T)\trel$. That is, out of the total energy $p_0 > T$ of the original perturbation,
 a fraction $T/p_0$ is carried at large times by the thermalized  distribution, whereas the 
 remaining fraction $(p_0-T)/p_0$ has been transmitted to the medium, via drag.  This conclusion
 on the energy loss is similar
 to our previous findings for the case of a stationary source in Sect.~\ref{sec:steady}.

\subsubsection{A physical source generated by the branching process}
 \label{sec:source}
 
In this subsection we apply the Green's function method to the main physical problem of interest,
namely a source generated by a branching process. More precisely, we 
consider an {\em ideal} branching process, for which the splitting 
dynamics at $p>p_*$ is not at all influenced by elastic collisions: the medium solely 
acts as a `perfect sink' which absorbs the energy of the gluon cascade at the `infrared' scale $p_*\sim T$.
(A more general situation will be studied in Sect.~\ref{sec:numerics}.)
Under these circumstances, the source has the structure shown in \eqn{eq:Source} with
$\Gamma(t,p_*)= \F(E,p_*,t)/p_*$. Here,  $\F(E,p_*,t)$ is the energy flux at $ p_*$ generated at time $t$ 
by a cascade that was initiated at $t=0$ by a leading particle with initial energy $E\gg T$
(recall the discussion in Sect.~\ref{sec:branch}). This source truly represents a bunch of 
relatively soft particles, which carry all the same energy $p_*$, move together at the 
speed of light, and whose number is evolving in time due to the branching dynamics. 

Within the context of the ideal cascade, one was able to obtain an exact analytic result for this
function $\F(E,p_*,t)$  \cite{Blaizot:2013hx}. Strictly speaking,
the analysis in \cite{Blaizot:2013hx} required two additional assumptions, which are
not essential from the viewpoint of physics, but simplify the mathematical manipulations:

\texttt{(a)} the energy $E$ of the LP is not {\em too} large,
 namely it obeys\footnote{The generalization of the subsequent results to
more energetic jets with $E\gg \omega_c(L)$ can be found in Ref.  \cite{Fister:2014zxa}.}  
$E\le \omega_c(L)\equiv \hat q L^2$, where the upper limit $\omega_c(L)=\omega_{\rm br}(L)/\abar^2$ 
is parametrically larger than $\omega_{\rm br}(L)$ at weak coupling;

\texttt{(b)} the kernel
\eqref{Kbr} within the BDMPSZ splitting rate is replaced by its simplified version
${\cal K}_0(x)\equiv 1/{[x(1-x)]^{\frac{3}{2}}}$ , which preserves the correct behavior
near the singular endpoints at $x=0$ and $x=1$.

Under these assumptions, the energy flux is obtained as \cite{Blaizot:2013hx}
\begin{align}\label{eq:flow}
  \F(E,p_*,t)\,=\,2\pi E \frac{t}{\tbr^2(E)}\,
    {\rm e}^{-\pi t^2/\tbr^2(E)}\,= \,
  \frac{\rmd}{\rmd t}\, {\Delta E}_{\rm flow}\,,
  \end{align}
where $\tbr(E)$ is the branching time introduced in \eqn{tbr} (the typical time
after which a parton with energy $E$ undergoes the first democratic branching) and 
\be\label{Econd}
{\Delta E}_{\rm flow}(E,p_*,t)\,=\,E\,\big[ 1-
    {\rm e}^{-\pi t^2/\tbr^2(E)}\big]\,=\,E\,\big[ 1-
    {\rm e}^{-\pi \omega_{\rm br}(t)/E}\big]\ee
with $\omega_{\rm br}(t)=\abar^2\hat q t^2$, is the energy which accumulates into the soft modes with $p\le p_*$
after a time $t$. 
The above results strictly apply for $p_*\ll \omega_{\rm br}(t)$, or, equivalently $t\gg \tbr(p_*)\sim \trel$,
and within that regime they are independent of $p_*$. In fact, in the absence of the sink at $p_*$,
this whole energy would accumulate in a  condensate at $p=0$ \cite{Blaizot:2013hx}.

 \eqn{Econd} confirms that the time scale $\tbr(E)$ plays the role of the lifetime of the
 leading particle w.r.t. democratic branchings.
At small times $t\ll \tbr(E)$, one can expand the exponential there to lowest order and thus find
\be\label{DEflowpi}
{\Delta E}_{\rm flow}(t)\,\simeq\,  \pi \omega_{\rm br}(t)\quad\mbox{when \ $t\ll \tbr(E)$}\,.\ee
Alternatively, this estimate is correct for a given time $t$ provided the energy $E$ of the
LP is sufficiently high, $E\gg \omega_{\rm br}(t)$. This result is in agreement
with \eqref{DEflow} up the replacement $\upsilon\to 2\pi$, 
due to the use of the approximate kernel ${\cal K}_0(x)$.  It shows that, at small times,
the energy loss via flow is independent of $E$ (essentially, because the LP does not `feel'
the change in its own energy due to flow) and that it grows with time like $t^2$.
As explained in Sect.~\ref{sec:phys}, this
`small time' (or `high energy') regime is the most interesting one for jets at the LHC,
where one indeed has $E\gg \omega_{\rm br}(L)$, with $L$ the size of the medium.

For larger times, such that $t\sim \tbr(E)$ or, equivalently, $E\sim \omega_{\rm br}(t)$,
the LP disappears via democratic branching and its whole initial energy is carried away
by the flow. This is indeed consistent with \eqn{DEflowpi} which shows that
${\Delta E}_{\rm flow}(t)\simeq E$ when $t\gtrsim \tbr(E)$. From a physical viewpoint,
this `large time' (or `low energy') regime better corresponds to the primary gluons 
radiated by the LP, which evolve into  `mini-jets'.

It is also interesting to notice the time dependence of the energy flux in \eqn{eq:flow}:
this rises linearly with $t$ at small times, then reaches a maximal value around $t=\tbr(E)$, and rapidly
vanishes at larger times. This means that the production rate for soft gluons is largest
towards the late stages of the branching process, i.e. for $t\sim\tbr(E)$, just before the LP dies away.

\begin{figure}[t] \centerline{\hspace*{-.3cm}
  \includegraphics[width=0.45\textwidth]{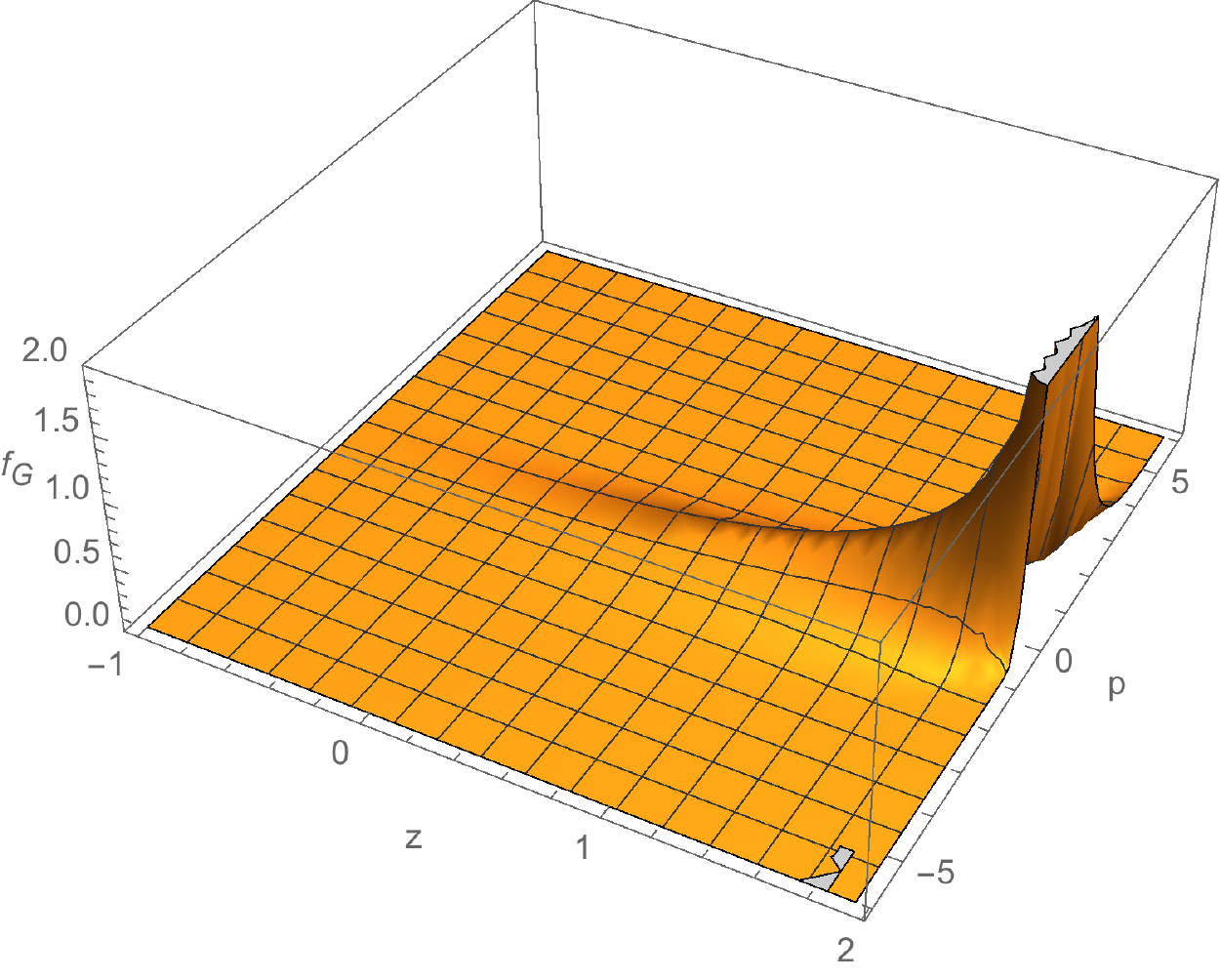}\quad
\includegraphics[width=0.45\textwidth]{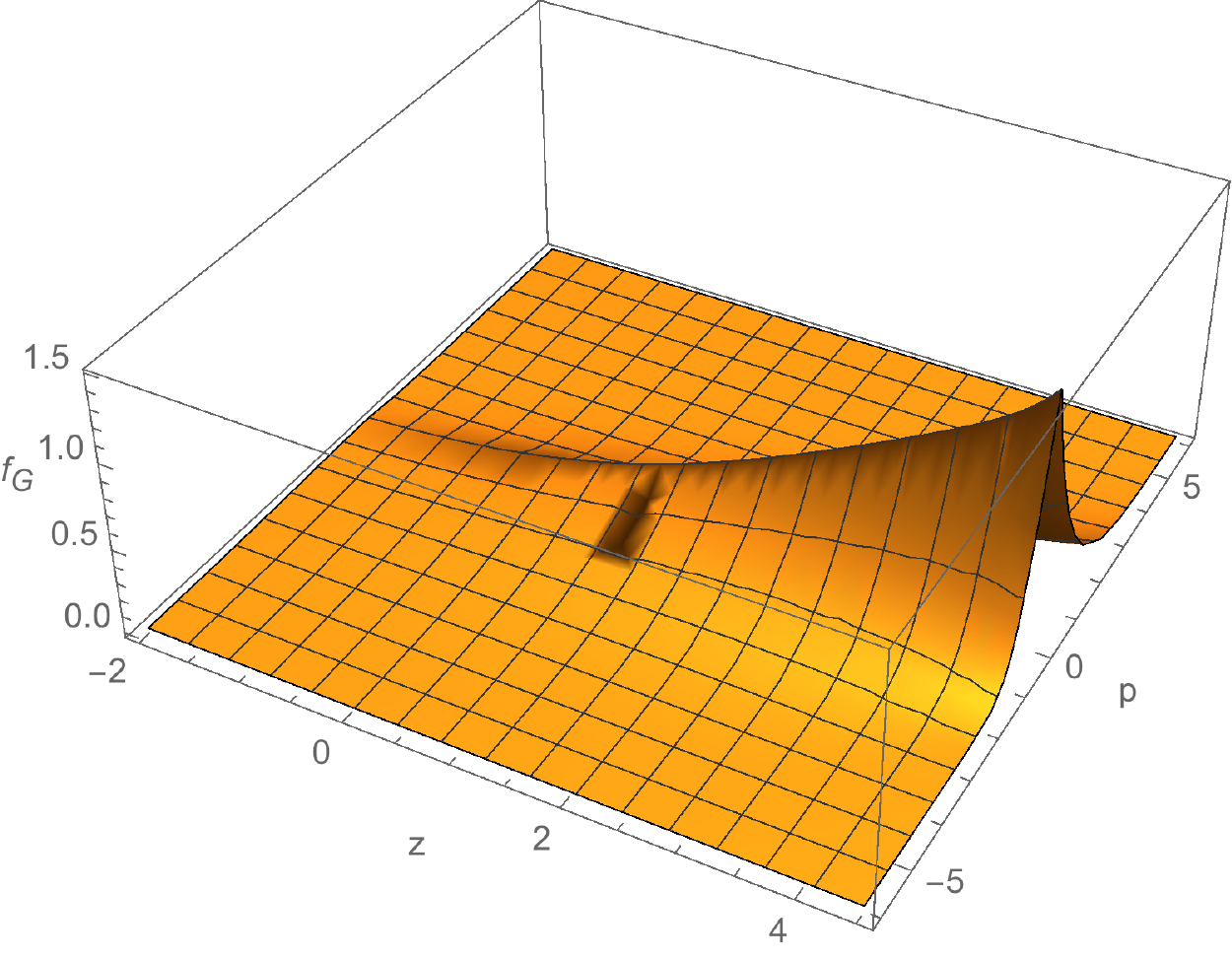}}
\vspace*{0.4cm}
 \centerline{\hspace*{-.3cm}
 \includegraphics[width=0.45\textwidth]{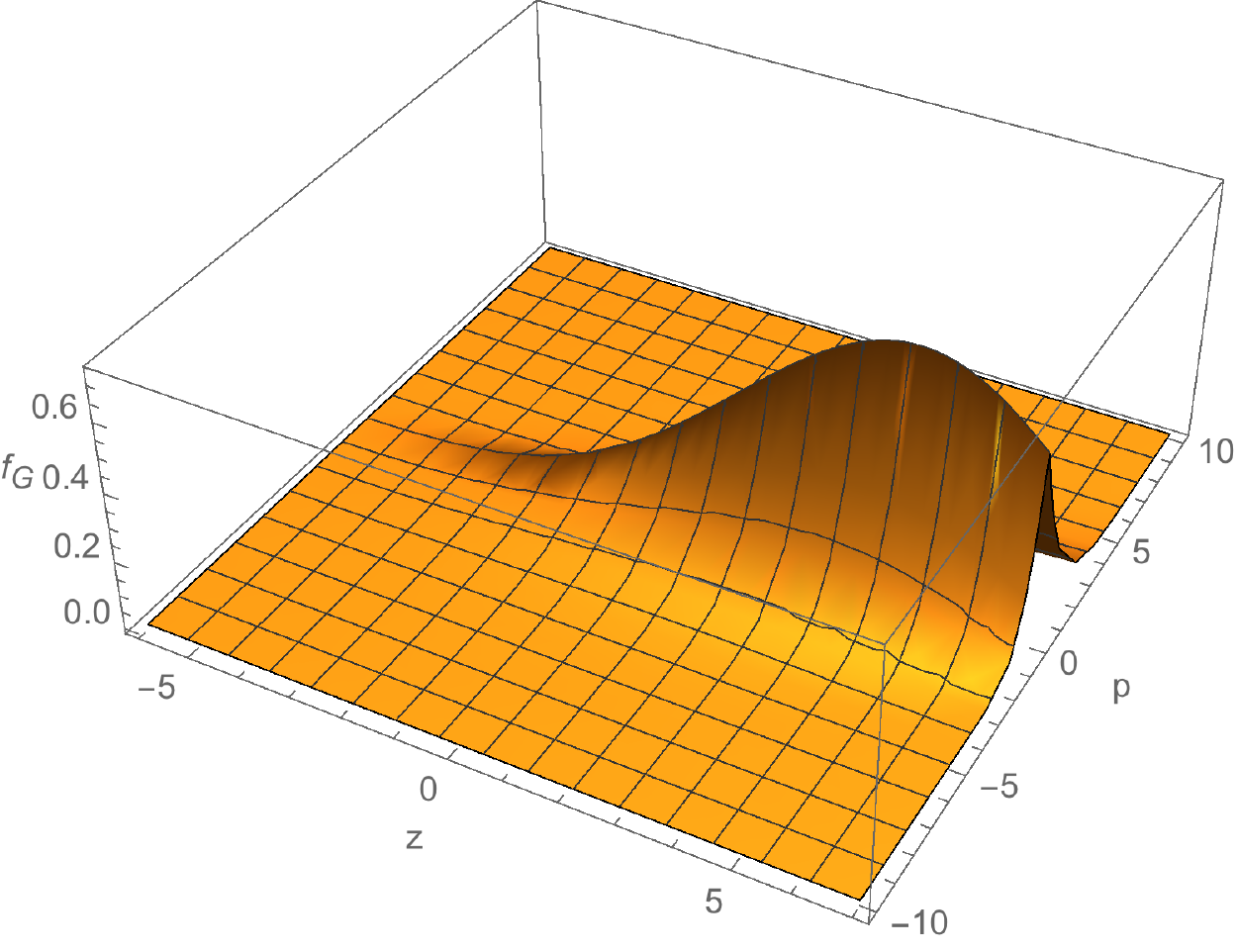}\quad
\includegraphics[width=0.45\textwidth]{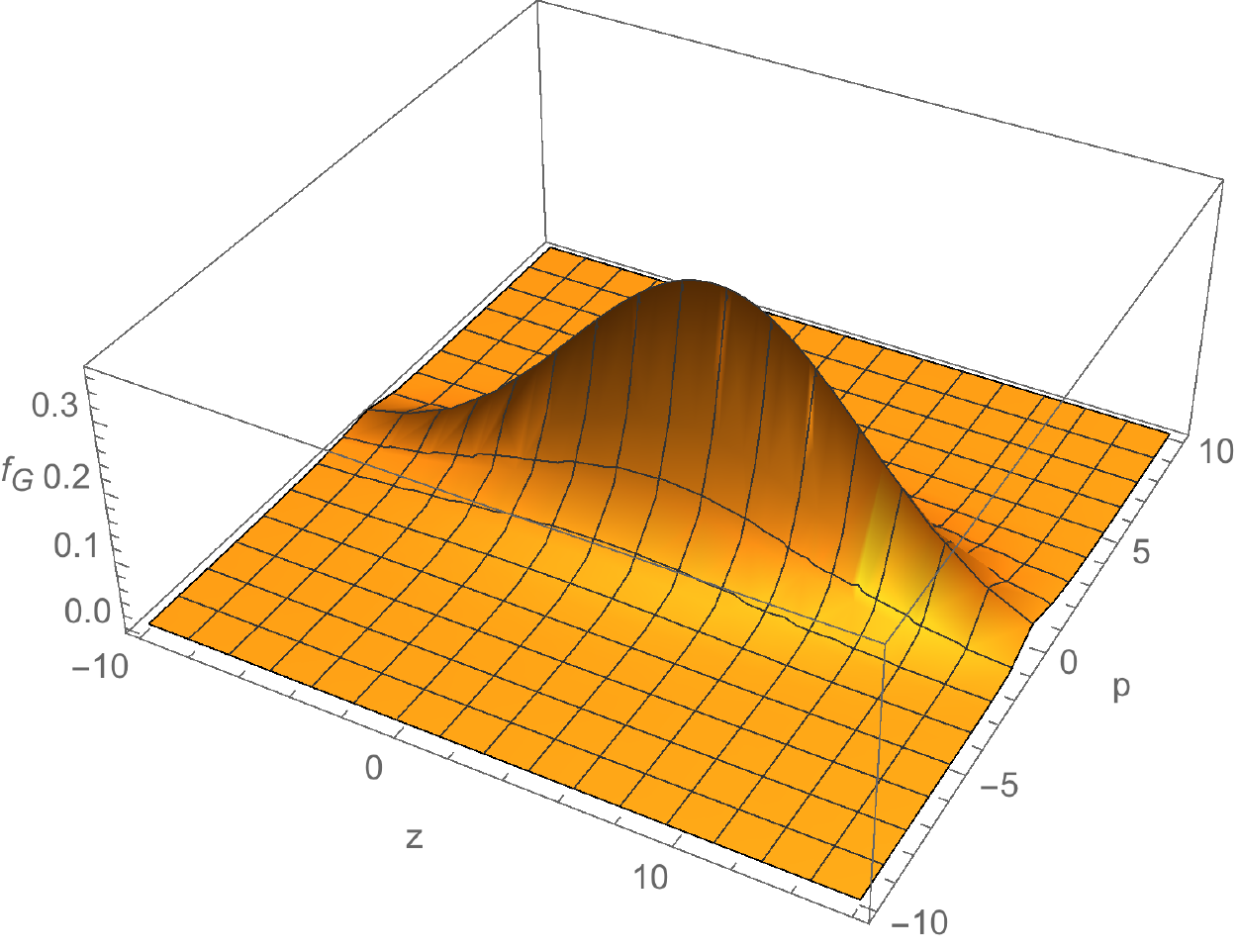}}
 \caption{\sl The distribution \eqref{eq:Fsol} produced by the source in
 Eq.~\eqref{eq:flow} with $\tbr(E)=6$ and $ p_*=1$ is plotted
as a function of $p$ and $z$ for four values of $t$: upper line, left: $t=2$;
upper line, right: $t=5$; lower line, left: $t=8$; lower line, right: $t=20$.
For relatively small times $t < \tbr(E)$, the source is still active and the distribution 
is thermal only in the tail at $z<t-\trel$. For larger times $t > \tbr(E) + p_*$, 
the source has essentially decayed and the distribution is thermal at any $z$.
For $t=20$ there is no significant difference between the exact result displayed above and the 
respective prediction of the diffusion approximation \eqref{eq:Fsoldiff}.
}
 \label{Spz}
\end{figure}

We now return to the solution to the Fokker--Planck equation for the physical source at hand.
Using \eqn{eq:Source}, the integrals over $z_0$ and $ p_*$ in \eqn{eq:gensol}
can be immediately performed to yield 
\bea\label{eq:Fsol}
f(t,z,p)\,= \,\frac{1}{ p_*} \int_0^t \rmd t' f_G(t-t',z-t',p; p_*) \,\F(t')\,.
\eea
It seems difficult to analytically perform the remaining integral over $t'$,
but this can be numerically computed, with the
results shown in Fig.~\ref{Spz} (in terms of reduced variables\footnote{
\label{f10} When rewriting  
\eqn{eq:FP} in terms of reduced variables, we find that the dimensionless version of the 
source in \eqn{eq:Source} reads (we restore the hat on reduced quantities, for more clarity):
$\hat{\mathcal{S}}= \theta(\hat t) \delta(\hat t-\hat z)\delta(\hat p- \hat p_*)({\hat{\F}}/{\hat  p_*})$,
where $\hat{\F}\equiv \F/T^2 = 8\pi \abar^2\hat t\exp\{-\pi\hat t^2/{\hat t}_{\rm br}^2(E)\}$ 
depends upon the energy $E$ only via the reduced branching time 
${\hat t}_{\rm br}(E)\equiv t_{\rm br}(E)/\trel$.}). These results can be understood as follows:
For relatively small times, $t\ll \tbr(E)$, the source  (the leading particle) is still present and
the gluon distribution is quite similar to that produced by a stationary source, as shown in
Fig.~\ref{fig:sourced}: it exhibits a front at $z=t$ and $0<p< p_*$ (but with the edges 
smeared out by diffusion), which represents the gluons that have been recently emitted, 
and with a tail at $z<t$ and peaked in momentum at $p=0$, which describes the
gluons which have experienced the effects of collisions. For later times $t\gtrsim\tbr(E)$, the source
has disappeared via democratic branching and the gluon distribution looks quite similar to that
produced by a localized source at late times, cf. Fig.~\ref{Gpz}: a thermal distribution in momentum
which extends in $z$ via diffusion and which peaks around $z=\tbr(E)$ (the maximal displacement of 
the source before it dies away).

In fact, for sufficiently large time, $t- \tbr(E)\gg  (p_*/T)\trel\sim \trel$, 
one can use the diffusion approximation for the Green's function,
\eqn{eq:dif}, to deduce (in reduced variables, cf. footnote \ref{f10})
\bea\label{eq:Fsoldiff}
f(t,z,p)\,\simeq \, 4\pi\abar^2\, \frac{\rme^{-|p|}}{p_*} 
 \int_0^t {\rmd t' \,t'} \,\frac{\rme^{-\f{(z-t')^2}{4 (t-t')}}}{\sqrt{4\pi  (t-t')}} \,
\,{\rm e}^{-\pi t'^2/\tbr^2(E)}\,.
\eea
This represents the medium perturbation that would be left over  by a relatively soft mini-jet
($E \ll \omega_{\rm br}(t)$), after it thermalizes. 
By integrating this late-time distribution over $z$ and $p$, it is easy to check that it
encompasses all the particles generated by the source, $\int \rmd z\rmd p f(t,z,p)
= {E}/{p_*}$, as it should, and that is contains a fraction $T/p_*$ of the total energy:
 \be
\Delta E_{\rm ther}\,= \int \rmd z\rmd p\,|p|\,f(t,z,p)\Big |_{\rm large \,\,time}
\,\simeq\,
\frac{T}{p_*}\,E\,\quad\mbox{when}\ \ t\,\gg\,\tbr(E)\,.\ee
We thus conclude that by choosing $p_*=T$ one can ensure that the whole initial energy of
the source is eventually recovered in the thermalized gluon
distribution: the energy loss via viscous drag is negligible
since the gluons are directly injected at the thermal scale.

\section{Numerical studies of the kinetic equation}
\label{sec:numerics}

The general discussion and the parametric estimates for the energy loss presented in Sect.~\ref{sec:phys},
as well as the explicit calculations using the Green's function method in Sect.~\ref{sec:source}, were based on 
an important physical assumption: the fact that the gluon cascade generated via multiple branchings
is not modified by the elastic collisions responsible for thermalization and hence it can be 
modeled as an ideal branching process, along the lines of Refs. \cite{Blaizot:2013hx,Fister:2014zxa} 
--- that is, a turbulent cascade, for which the medium acts as a perfect sink at the lower end
($p\sim T$) of the cascade. The validity of this assumption is far from being obvious,
as shown  by the following argument: via elastic collisions, the soft gluons are redistributed in 
phase space, in such way to match a thermal distribution, and for $p\sim T$ 
the latter is quite different from the scaling spectrum produced by the turbulent cascade. 

In this section, we shall give up the `perfect sink' assumption and present a detailed numerical study 
based on the kinetic equation  \eqref{eq:equationL}. This equation too is a rather simplified
version of the actual dynamics, as explained in Sect.~\ref{sec:kin}, but as compared to 
the source approximation in Sect.~\ref{sec:analytic} it has the merit to include 
an explicit infrared cutoff $p_*\sim T$ in the branching process and also 
the interplay between branching and thermalization at $p > p_*$.

\subsection{Setting-up the problem}

As in the previous section, it is convenient in practice to measure all the momenta in units of $T$ 
and all the space-time scales in units of $\trel$, that is, to use the reduced variables introduced
in \eqn{eq:eqstat}. In terms of these variables, 
the kinetic equation \eqref{eq:equationL} reads (with $v\equiv {p}/{|p|}$)
\begin{align}\label{eq:Lred}
\left({\partial_t}+v{\partial_z}\right)f(t,z,p)  & = \partial_p(\partial_p+v)f(t,z,p)  \nn\\*[0.3cm]
&
+\f{\trel}{\tbr(T)p^{\frac{1}{2}}}\int\limits_{r} \rmd x\,  {\cal K}(x)\left[\f{1}{\sqrt{x}}\,
f\left(t,z,\f{p}{x}\right) - \frac{1}{2} f(t,z,p) \right]\,,
\end{align}
where we recall that the subscript $r$ on the integral over $x$ indicates the condition
that both daughter gluons in a splitting process be harder than the  `infrared'  scale $p_*\sim T$
 (the lower end of the cascade).
The precise kinematical conditions are as follows: for the gain term, $p>p_*$ and $p(1-x)/x>p_*$, whereas
for the loss term, $x p> p_*$ and $(1-x) p>p_*$. For definiteness,  we chose this 
scale $p_*$ to be exactly equal to $T$ (i.e. $p_*=1$ in \eqn{eq:Lred}). 

The ratio $\trel/\tbr(T)$ which appears in front of the branching term
in \eqn{eq:Lred} is parametrically of order one for the
 weakly-coupled quark-gluon plasma. For what follows, it is convenient to 
choose this ratio to be {\em exactly} equal to one. This choice is also reasonable from a physics standpoint:
it corresponds to the condition $\hat q = 16\abar^2 T^3$ (or, equivalently, $4\abar^2 T\trel =1$),
 which is satisfied by the following values for the physical parameters:
\be\label{parameters}
\abar\,=\,0.3\,,\quad T\,=\,0.5~{\rm GeV}\,,\quad \hat q\,=\,1~{\rm GeV}^2/{\rm fm}\,\simeq\,0.2
~{\rm GeV}^3\,,\quad \trel\,=\,1~{\rm fm}\,,\ee
which are indeed consistent with the current phenomenology.

The equation thus obtained will be solved numerically, with the initial condition
\be\label{eq:init}
f(t=0,z,p)=\delta(p-E)\delta(z)\,\to\, \f{10}{\pi}\, \rme^{-10(p-E)^2-10z^2}\,,\ee
where 
the product of $\delta$--functions is regulated as shown in the r.h.s. 
The picture that we expect in the light of the general
discussion in Sect.~\ref{sec:phys} is as follows:

 \texttt{(i)} For sufficiently small times $t\ll \tbr(E)$, the leading particle should survive and carry
most of the total energy. The energy lost towards the medium 
should be comparatively small and follow the law shown in \eqn{DEflow}~;
that is, it should be of order $\omega_{\rm br}(t)$ and thus grow with $t$ as $t^2$.

 \texttt{(ii)} For larger times $t\gtrsim \tbr(E)$, the LP should disappear 
via democratic branching and the energy loss should be of the order of the total energy $E$.


For what follows, one should keep in mind that some of the assumptions underlying
this picture might not be well satisfied when solving \eqn{eq:Lred} in practice. For instance,
in Sect.~\ref{sec:phys} we have assumed the medium to act as
a perfect sink for the energy carried away by the branching process, which in turn 
allowed for a well-developed phenomenon of turbulence.  
This is an important hypothesis, implicitly assumed in the previous literature, 
for which our subsequent study will provide an explicit test. 

\subsection{The gluon spectrum}
\label{sec:spectrum}

Before we describe the full picture of the gluon cascade in the longitudinal phase-space $(z,p)$, let us
present the results for the distribution integrated over $z$, that is, the gluon spectrum 
\be
f(t,p)\,\equiv\,\f{\rmd N_g}{\rmd p}\,=\int \rmd z\,f(t,z,p)\,.\ee
Clearly, this function can be obtained by solving
directly the homogeneous (in the sense of independent of $z$) version of \eqn{eq:Lred}, with
initial condition $f(0,p)=\delta(p-E)$. This is a relatively simple numerical problem, which in particular
allows for an extensive study of the role of the lower cutoff at $p=p_*$ on the branching process.
To that aim, we shall consider three cases: (a) an `ideal' branching process, which involves 
no infrared cutoff and develops a clear phenomenon of turbulence 
(the corresponding equation is obtained by keeping only the branching
term in the r.h.s. of \eqn{eq:Lred} and letting $p_*=0$); (b) a branching process with a sharp infrared cutoff
at $p_*\ll E$, as described by \eqn{eq:Lred} without the Fokker-Planck terms, 
and (c) the complete dynamics (branchings with infrared cutoff $p_*$ and elastic collisions),
as described by the homogeneous version of \eqn{eq:Lred}.

\begin{figure}
\includegraphics[width=0.4\textwidth]{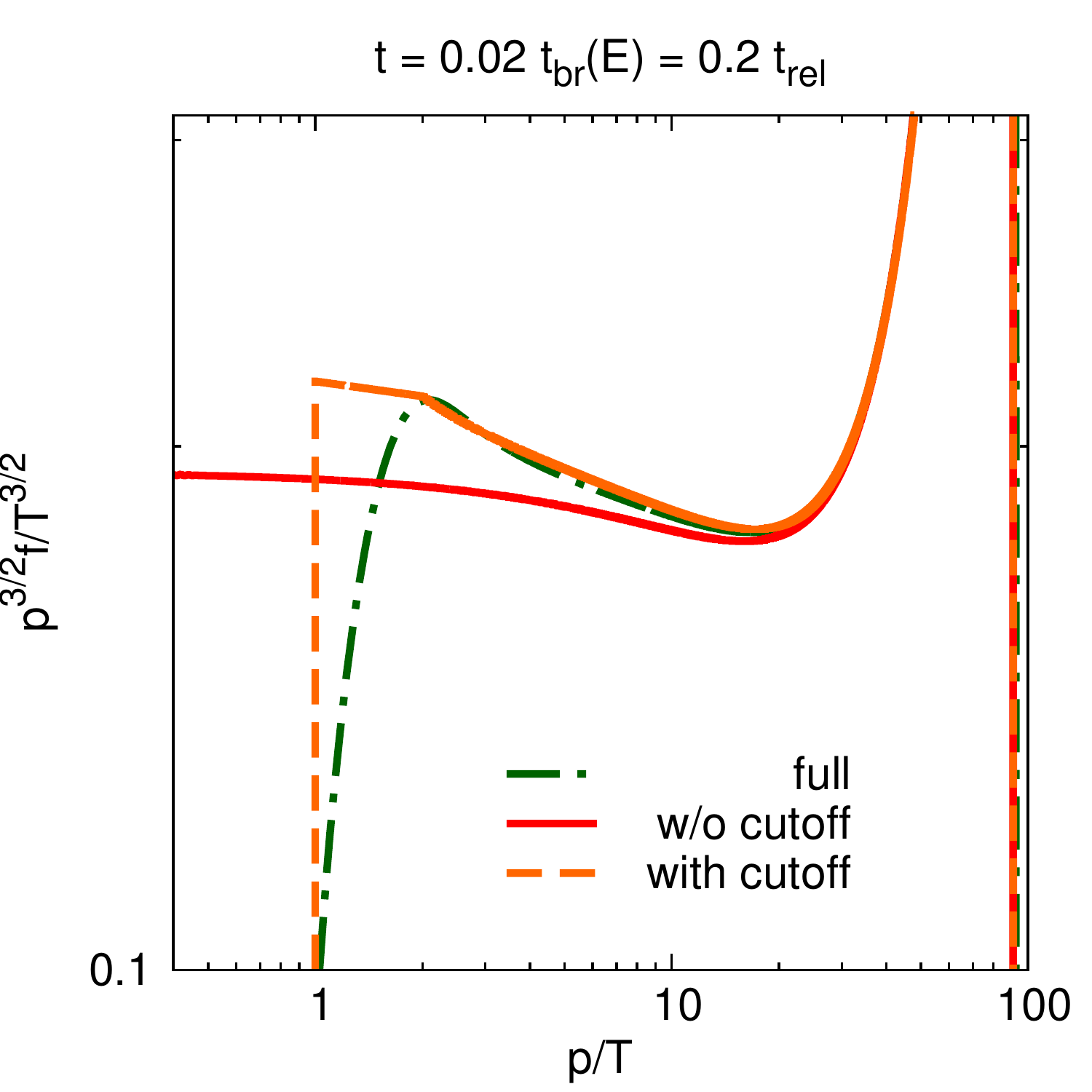}
\hspace{0.1\textwidth}
\includegraphics[width=0.4\textwidth]{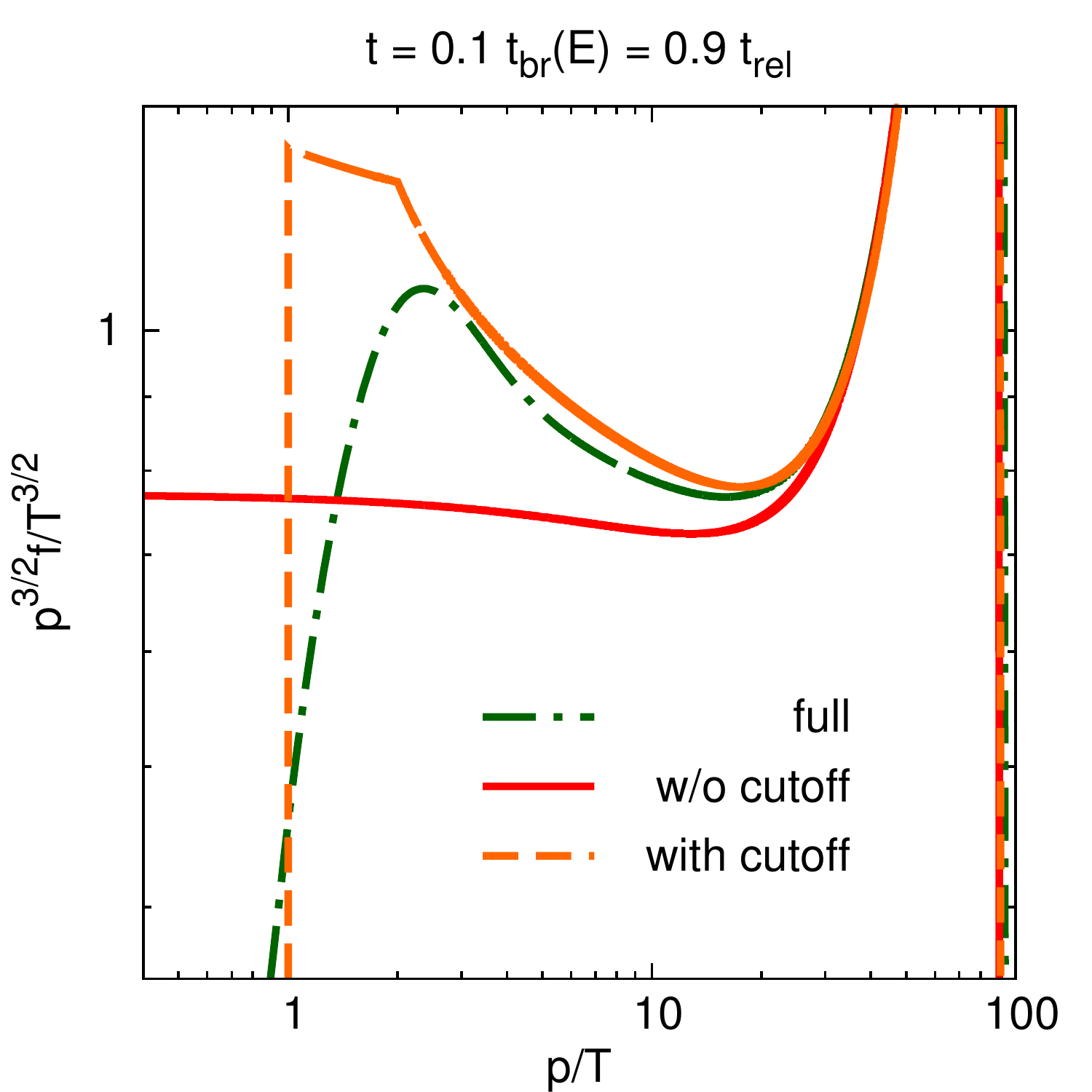}\vspace{0.01\textheight}
\\
\includegraphics[width=0.4\textwidth]{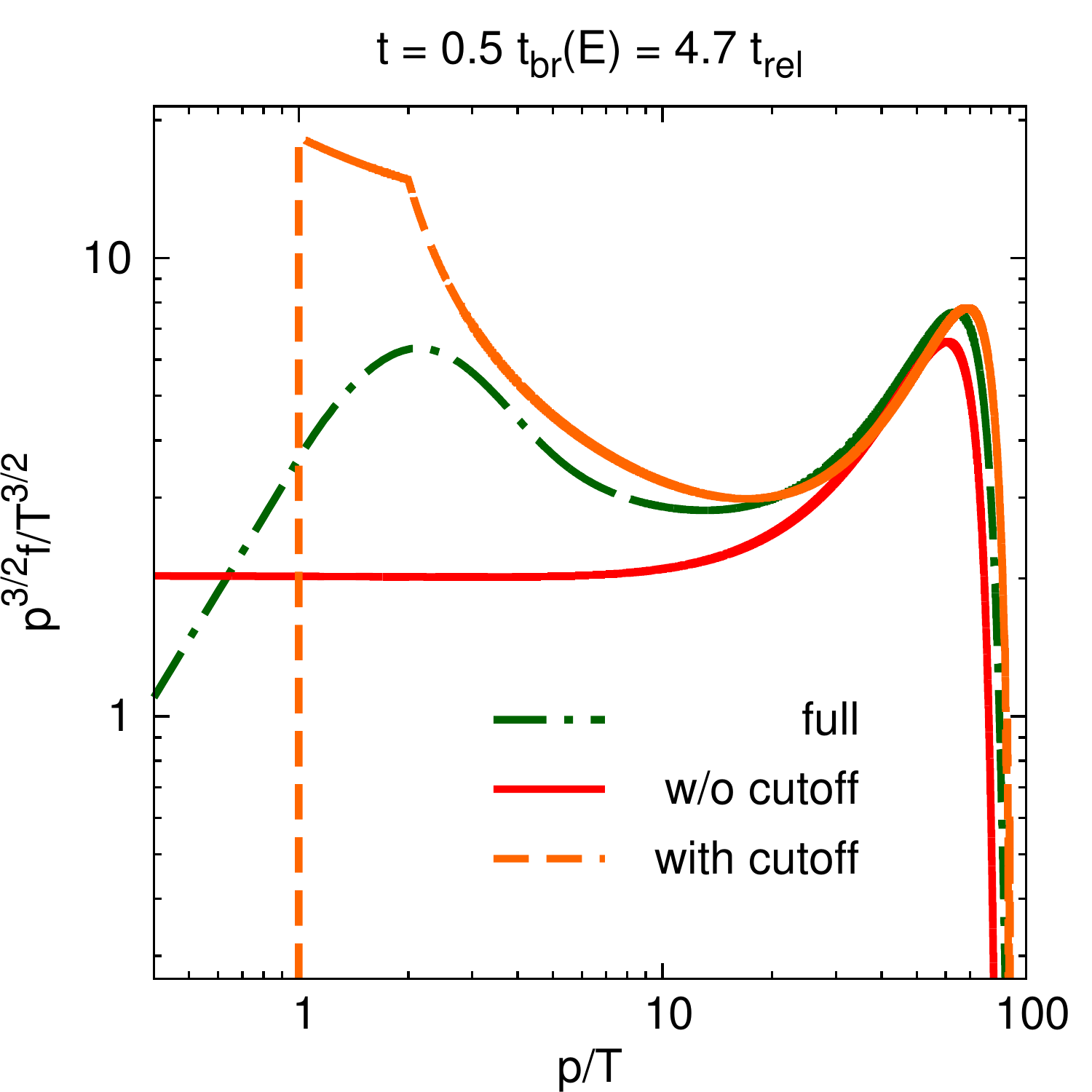}
\hspace{0.1\textwidth}
\includegraphics[width=0.4\textwidth]{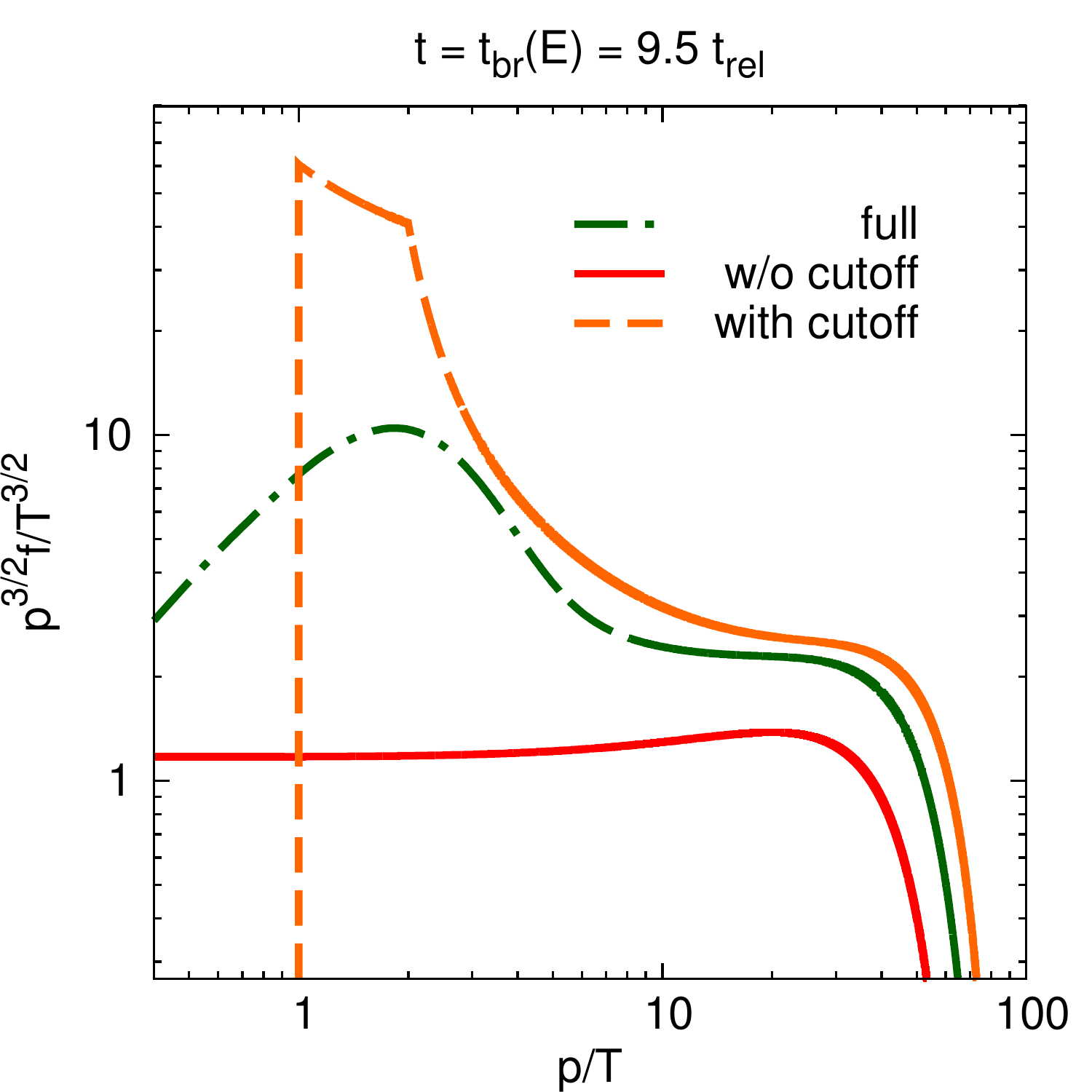}
\caption{The gluon spectrum $f(t,p)$ for a spatially homogeneous distribution, for an initial energy $E=90T$
and 4 values of time. The three curves correspond to (a) an ideal branching process
(the red, continuous, line), (b) a branching process with infrared cutoff $p_*=T$ (orange, dotted line), and (c)
the full process with branchings and elastic collisions (green, dashed-dotted line).
On the vertical axis, the spectrum is multiplied by $(p/T)^{3/2}$
to render manifest the scaling behavior for the ideal branching process.}\label{fig:spatialHomo}
\end{figure}

As already mentioned, the ideal branching process has been extensively studied in the literature and,
in particular, exact analytic solutions have been obtained  \cite{Blaizot:2013hx} for the case of a simplified kernel 
${\cal K}(x)\to {\cal K}_0(x)\equiv 1/{[x(1-x)]^{\frac{3}{2}}}$. In our numerical
study, we use the full kernel in \eqn{Kbr}, but the solution is qualitatively similar to that presented
in Ref.~\cite{Blaizot:2013hx}. Namely, for sufficiently small momenta $p\ll E$, the spectrum exhibits the
scaling law $f(t,p)\propto 1/p^{3/2}$, which is a fixed point of the branching kernel and the signature of 
wave turbulence.  For not too large times
$t\ll \tbr(E)$, the leading particle is visible in the spectrum, as a pronounced peak just below $p=E$. The
width of this peak increases with time (due to radiation) and eventual becomes of order one --- meaning
that the LP undergoes its first democratic branching --- when $t\sim \tbr(E)$. For even larger times, the scaling 
law $\sim  1/p^{3/2}$ is still visible at small $p$, but the spectrum is suppressed as a whole, since the
energy flows via multiple branching and accumulates at $p=0$. When $t\gg \tbr(E)$, 
the whole energy $E$ ends up in this `condensate'.  This behavior
 is clearly visible in the numerical results displayed in Fig. \ref{fig:spatialHomo}.

After introducing the infrared cutoff $p_*$, the gluons with $p\le 2p_*$ cannot
split anymore, as there is no phase-space available to the daughter gluons. Hence, instead of falling
at $p=0$, gluons start accumulating in the bins at $p\gtrsim p_*$. As a result, 
the spectrum above  $p_*$ deviates from the scaling spectrum:
it shows an excess (`pile-up'), which is particularly marked at $p_*<p<2p_*$, where it looks like a
bump. Both the size of this excess and its extent in $p$ above  $p_*$ are increasing with time,
as clearly visible in Fig. \ref{fig:spatialHomo}. This can be understood as follows: a gluon with, say, 
$p=3p_*$ has more chances to be created via the decay of parent gluons with $p\gg p_*$
(for which the kinematical constraint is relatively unimportant) than to disappear via a
decay (since a significant  fraction of the phase-space for its decay, that
at $p\le p_*$, is not accessible anymore). 

On physical grounds, it is quite clear that this pile-up cannot be entirely physical: gluons with $p\sim T$ can
efficiently lose energy towards the medium via elastic collisions and hence they should fall into the bins at
lower energies $|p|< T$. We thus expect the pile-up to be considerably reduced and possibly washed out
after also including the elastic collisions, as represented by the Fokker-Planck terms in the r.h.s. of \eqn{eq:Lred}.
This expectation is confirmed by the numerical solution to the homogeneous version of \eqn{eq:Lred}
(whose results are shown too in Fig. \ref{fig:spatialHomo}), but only partially: the pile-up in the spectrum is indeed reduced
by the elastic collisions, but the deviation with respect to the scaling spectrum remains quite large --- 
so large, that there seems to be no scaling window in practice. If true, the last conclusion would also imply
that the physical results are very sensitive to the details of the mechanism which stops the
branching process and which in our analysis has been only crudely mimicked by the 
infrared cutoff $p_*$. Fortunately though, these last conclusions are not fully right and the numerical results
exhibited in Fig. \ref{fig:spatialHomo} are in this respect quite misleading. The gluons which appear to
accumulate on top of the scaling spectrum in this figure are actually located at {\em different} values of $z$.
These are relatively soft gluons, which undergo strong diffusion as a consequence of collisions and thus
separate from each other and also from the more energetic constituents of the jet (which keep 
propagating along the light-cone at $z=t$). Hence the `pile-up' visible in the curves denoted as
`full' in Fig. \ref{fig:spatialHomo} is merely an artifact of integrating the gluon distribution over the
longitudinal coordinate $z$: it comes from the superposition of a nearly ideal branching spectrum in the
front of the jet at $z\simeq t$ 
and of nearly thermal spectra in the tail of the jet at $z\ll t$. This will be demonstrated by the subsequent
analysis, where the $z$--distribution is kept explicit.

\subsection{Jet evolution in longitudinal phase-space}

In this subsection we present 
numerical solutions to the complete equation \eqref{eq:Lred}
with initial conditions of the type shown in \eqn{eq:init}. We consider two 
values for the initial energy, $E=25\,T$ and $E=90\,T$, which correspond to rather distinct physical situations.
The first value $E=25\,T$ ($=12.5$~GeV according to \eqn{parameters}) is quite low 
and is representative for a {\em mini-jet} radiated by 
a leading particle with a much higher energy $E_0\ge 100$~GeV. 
This is one of the typical mini-jets which control the energy loss  in the case where the LP
crosses the medium along a distance $L\simeq\tbr(E) =5\,\trel =5$~fm. 
The second value $E=90\,T=45$~GeV is closer to the 
energy of an actual jet at the LHC and in particular is large enough to ensure that the respective
leading particle does not disappear into the medium: indeed, the respective branching
time $\tbr(E)\simeq 9.5$~fm is larger than the typical distance $L\lesssim 8$~fm that 
the LP might travel across the medium in the experimental situation at the LHC.

\begin{figure}[t]
\begin{center}
\includegraphics[width=0.4\textwidth]{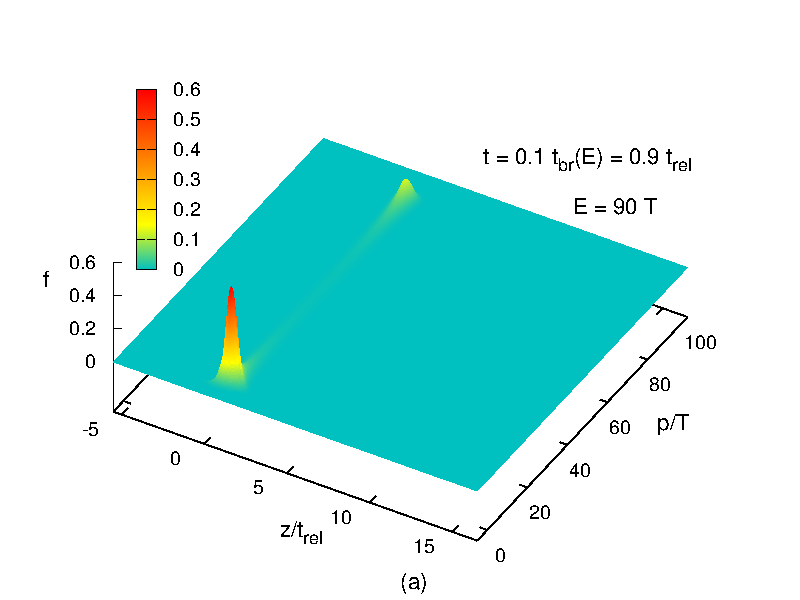}\hspace{0.01\textwidth}\includegraphics[width=0.4\textwidth]{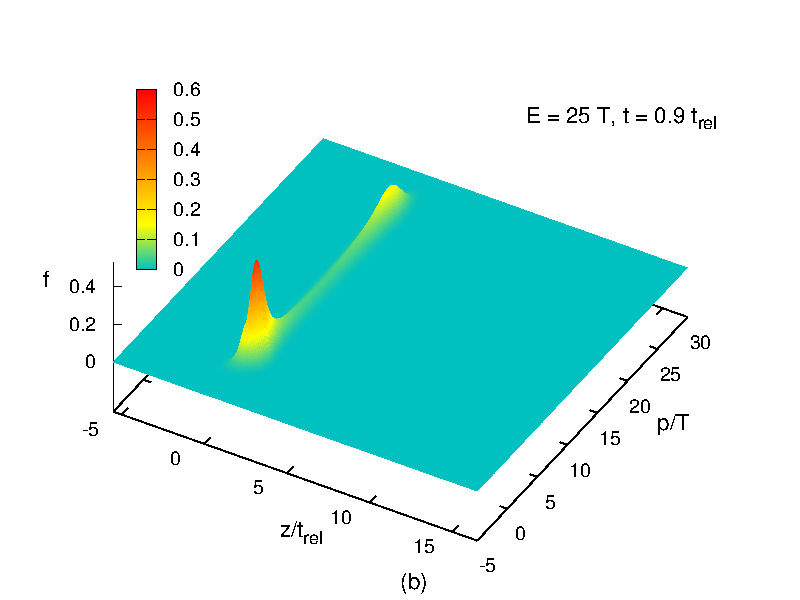}\\
\includegraphics[width=0.4\textwidth]{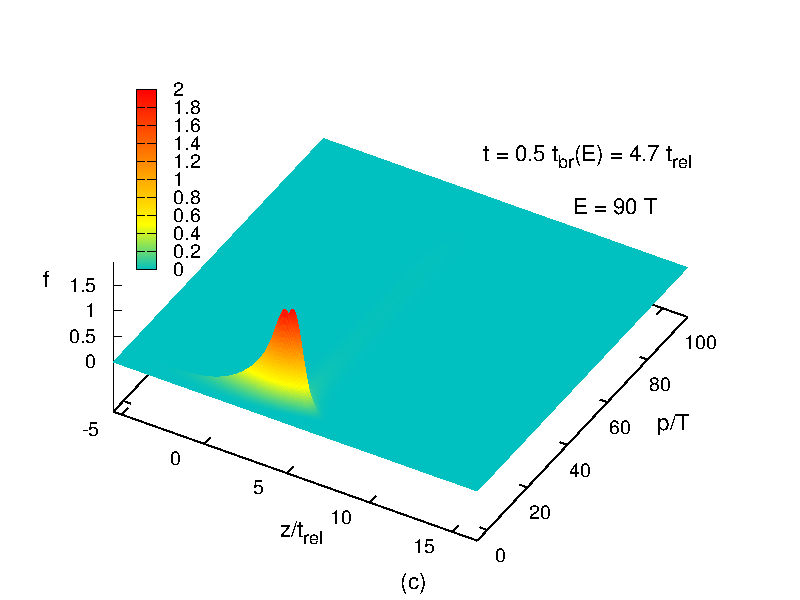}\hspace{0.01\textwidth}\includegraphics[width=0.4\textwidth]{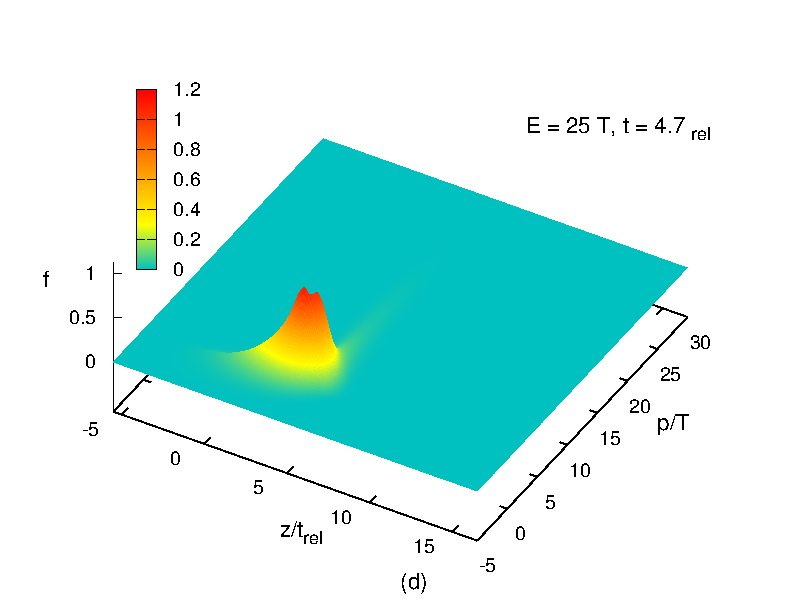}\\
\includegraphics[width=0.4\textwidth]{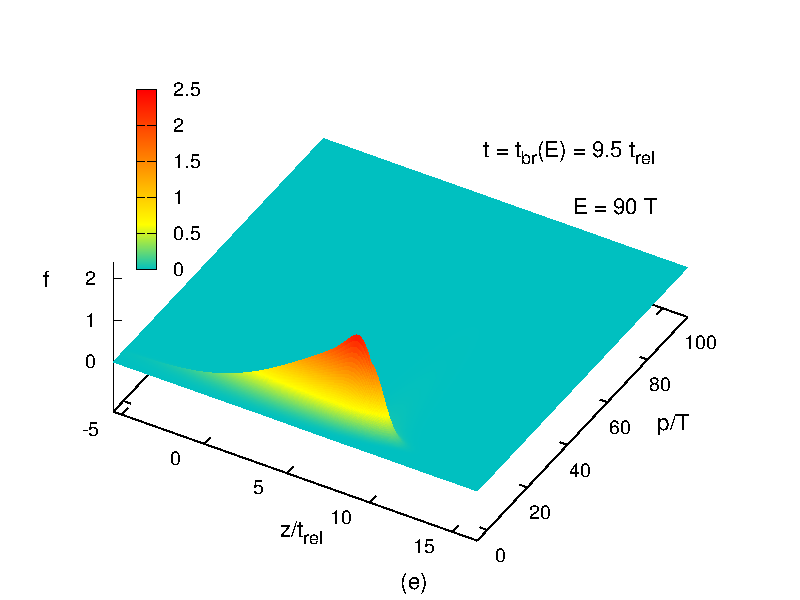}\hspace{0.01\textwidth}\includegraphics[width=0.4\textwidth]{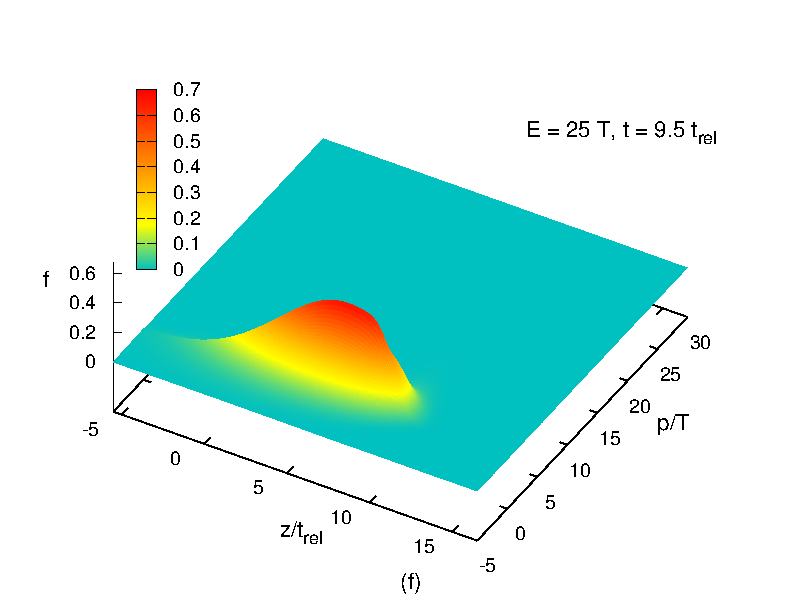}\\
\includegraphics[width=0.4\textwidth]{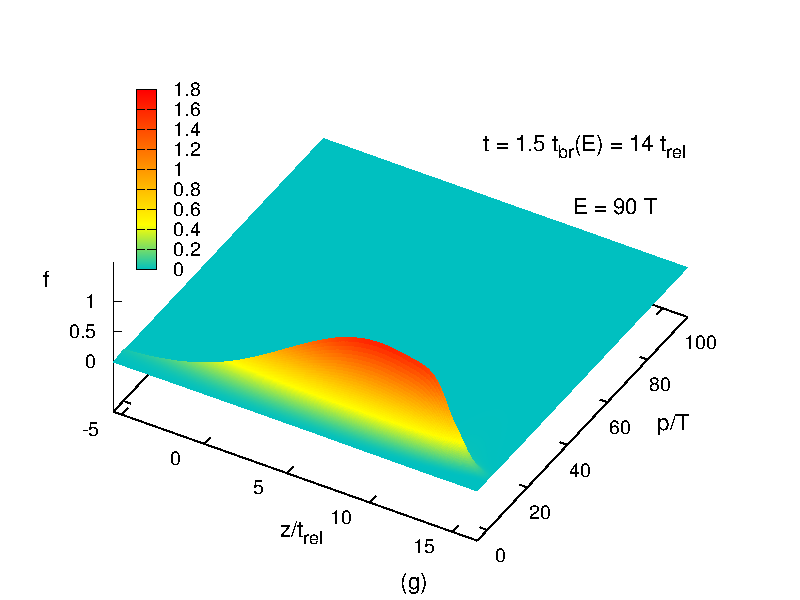}\hspace{0.01\textwidth}\includegraphics[width=0.4\textwidth]{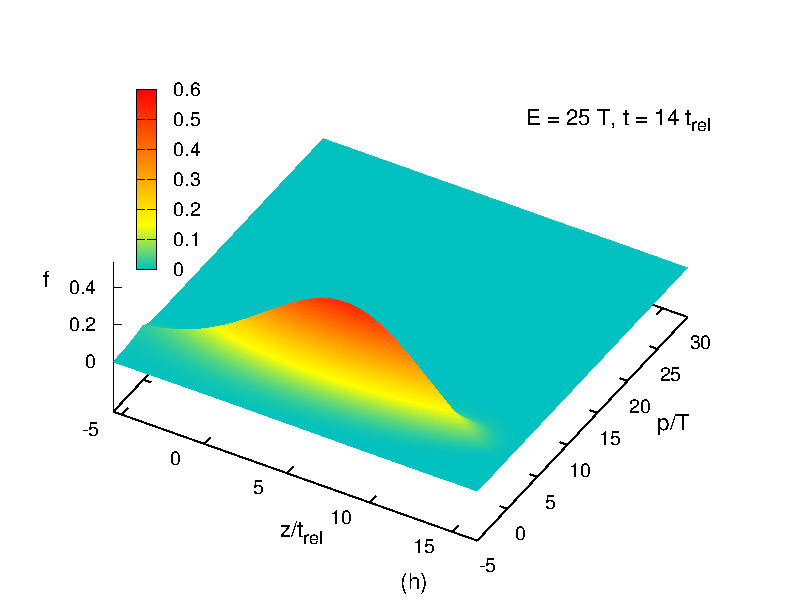}
\caption{The time evolution of the phase-space distribution $f(t,z,p)$ produced by a jet
with initial energy $E=90\,T$ (left) and respectively $E=25\,T$ (right), plotted for exactly the same 
values of time.
}\label{fig:fEvol}
\end{center}
\end{figure}

\begin{figure}[t]
\begin{center}
\includegraphics[width=0.4\textwidth]{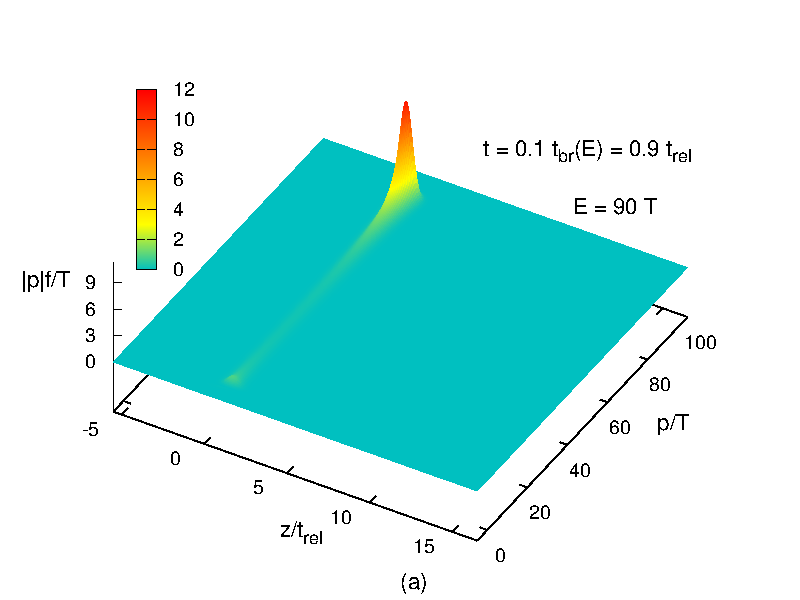}\hspace{0.01\textwidth}\includegraphics[width=0.4\textwidth]{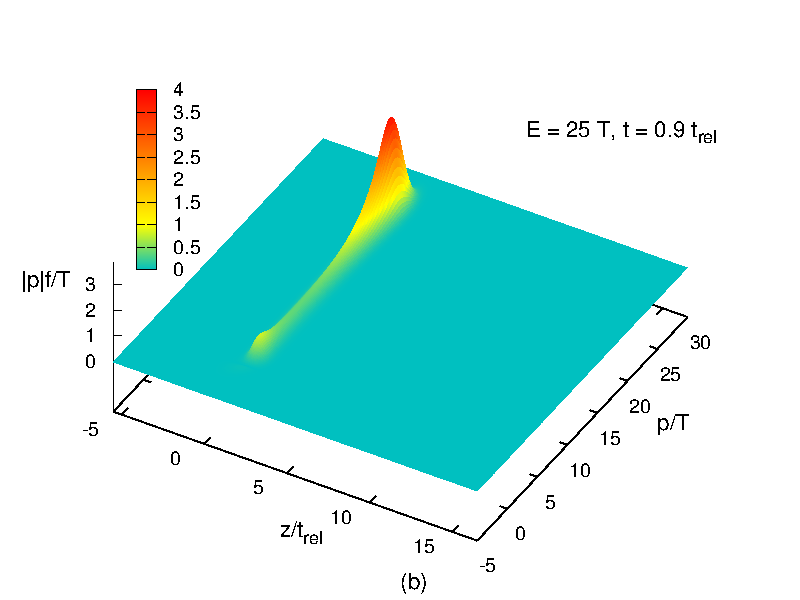}\\
\includegraphics[width=0.4\textwidth]{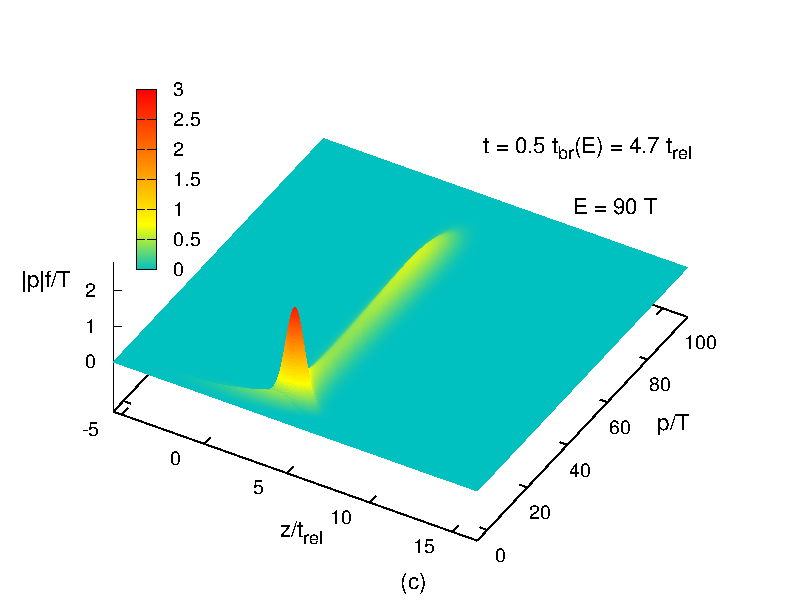}\hspace{0.01\textwidth}\includegraphics[width=0.4\textwidth]{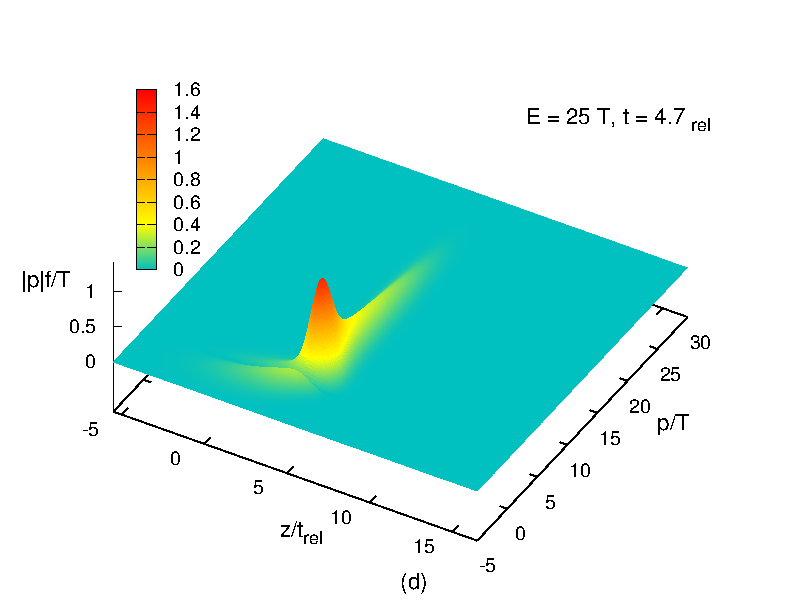}\\
\includegraphics[width=0.4\textwidth]{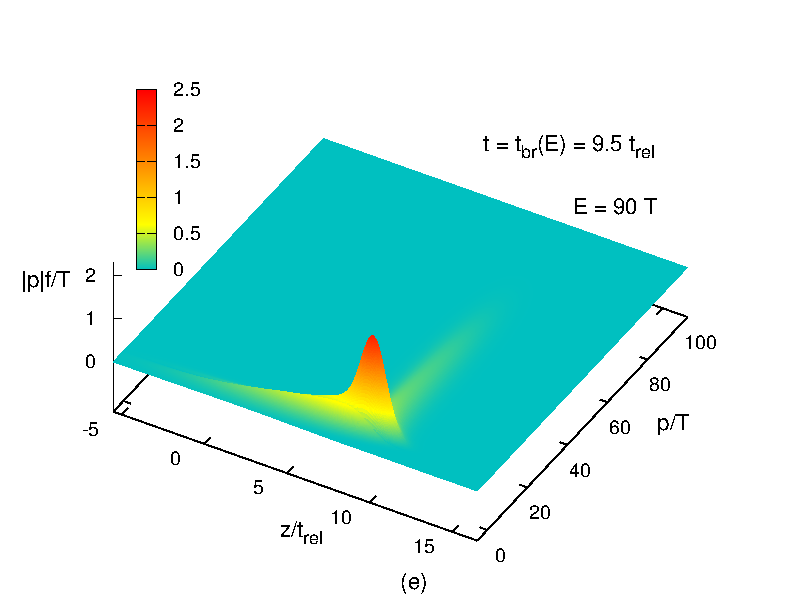}\hspace{0.01\textwidth}\includegraphics[width=0.4\textwidth]{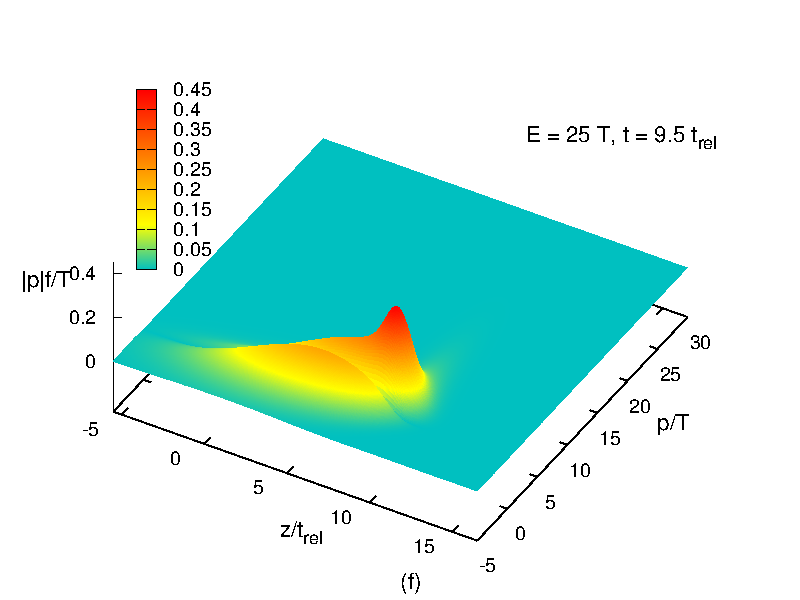}\\
\includegraphics[width=0.4\textwidth]{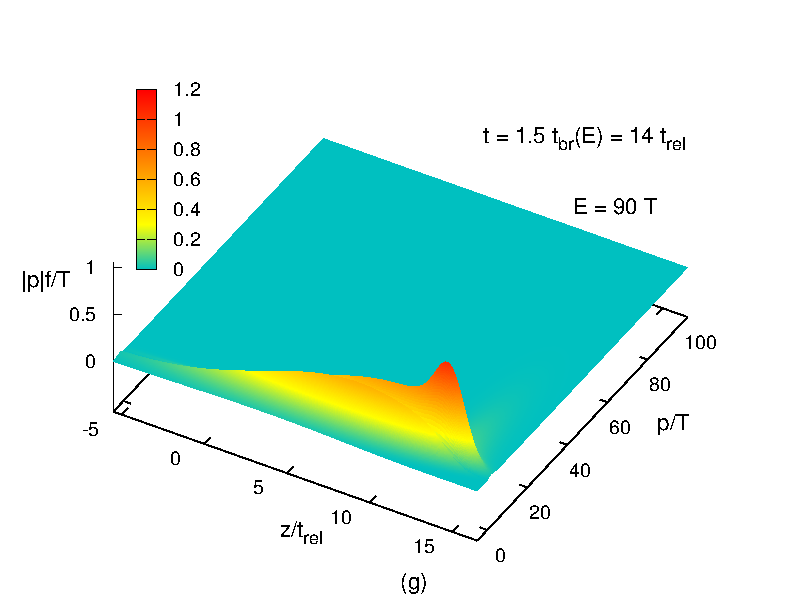}\hspace{0.01\textwidth}\includegraphics[width=0.4\textwidth]{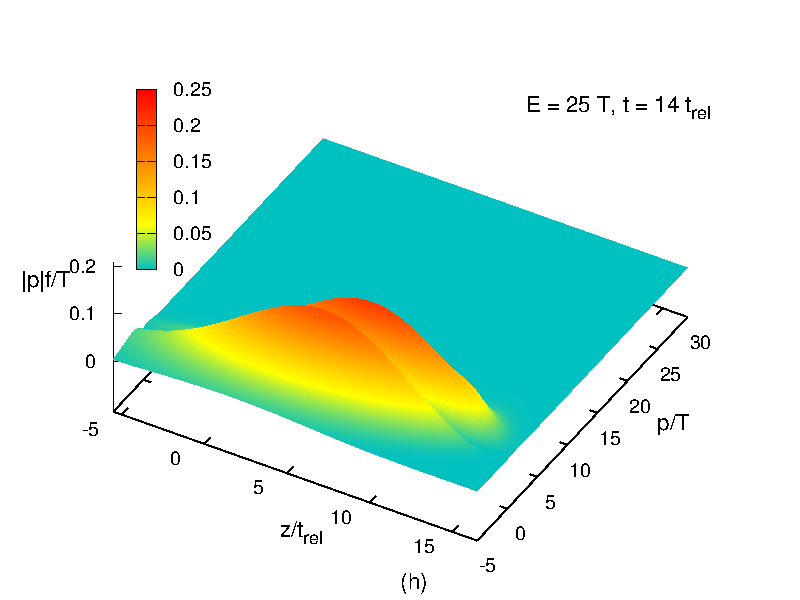}
\caption{The time evolution of the phase-space energy density $|p|f/T$, for the same conditions
as in Fig. \ref{fig:fEvol}.}\label{fig:eEvol}
\end{center}
\end{figure}

\subsubsection{The gluon distribution and the energy density}

The general features of the evolution of the gluon cascade produced by a high-energy jet can be appreciated
by inspection of Figs.~\ref{fig:fEvol} and \ref{fig:eEvol}, which show the phase-space distribution of the gluon number 
$f(t,z,p)$ and, respectively, the gluon energy $|p| f(t,z,p)$, for the two energies of the LP,
$E=90\,T$ and $25\,T$, and four values of time: $t/\trel=0.95,\,4.7,\,9.5,$ and $14$. These particular values 
for $t$ have been chosen since, with our present conventions (i.e. $\tbr(E)/\trel = \sqrt{E/T}$), they
correspond to rather special values for the case of a jet with $E=90\,T$; namely, they amount to
$t/\tbr(90)\simeq 0.1,\,0.5,\,1,$ and 1.5, respectively. Since the natural time scale for the jet evolution is
$\tbr(E)$, let us also list the corresponding values for the softer jet with $E=25\,T$~: one roughly has
$t/\tbr(25)\simeq 0.2,\,1,\,2,$ and 3, respectively. To gain more intuition about these time scales in physical 
units, it is useful to recall that $\trel=1$~fm for the medium parameters in \eqn{parameters}.

\begin{figure}[h]
\begin{center}
\includegraphics[width=0.3\textwidth]{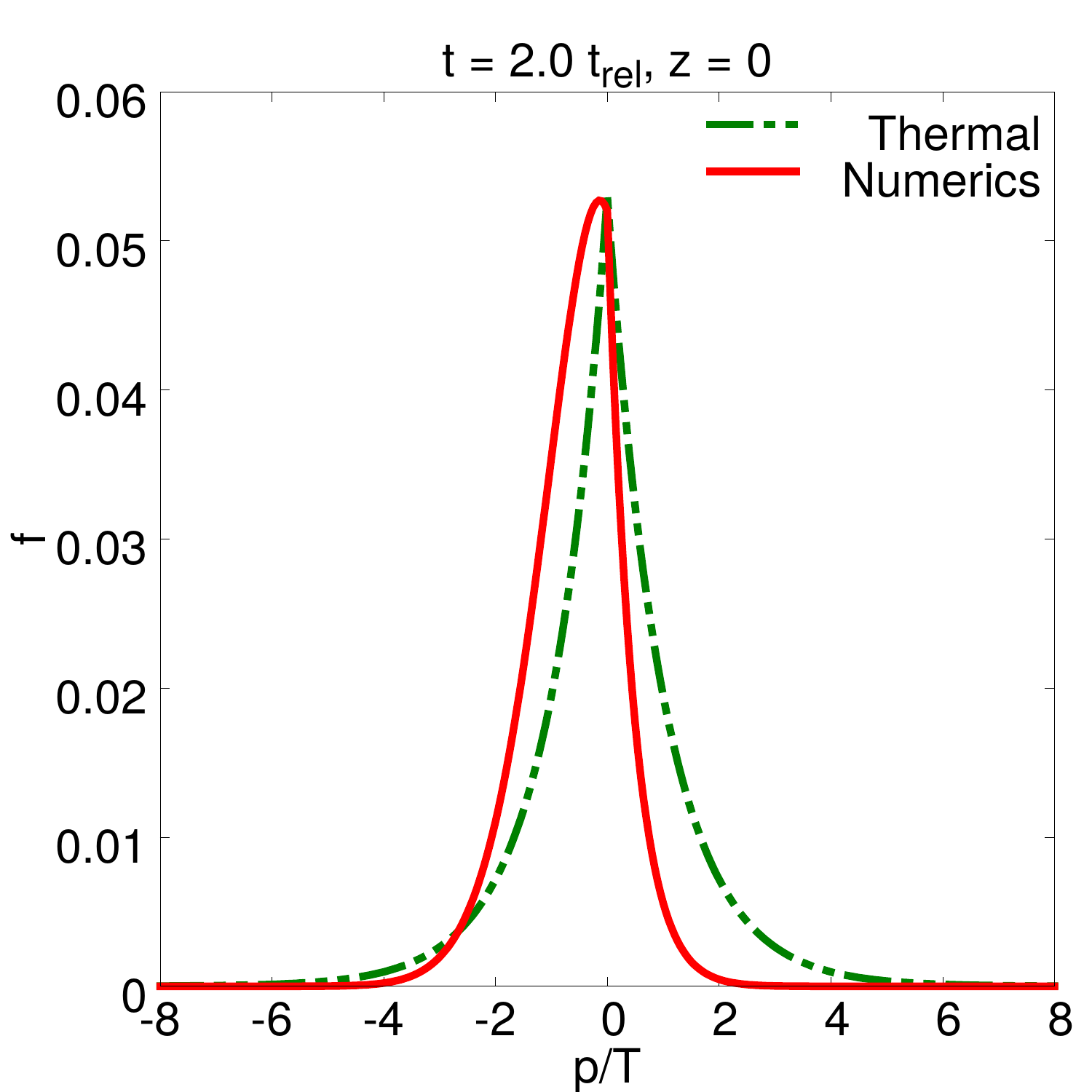}
\includegraphics[width=0.3\textwidth]{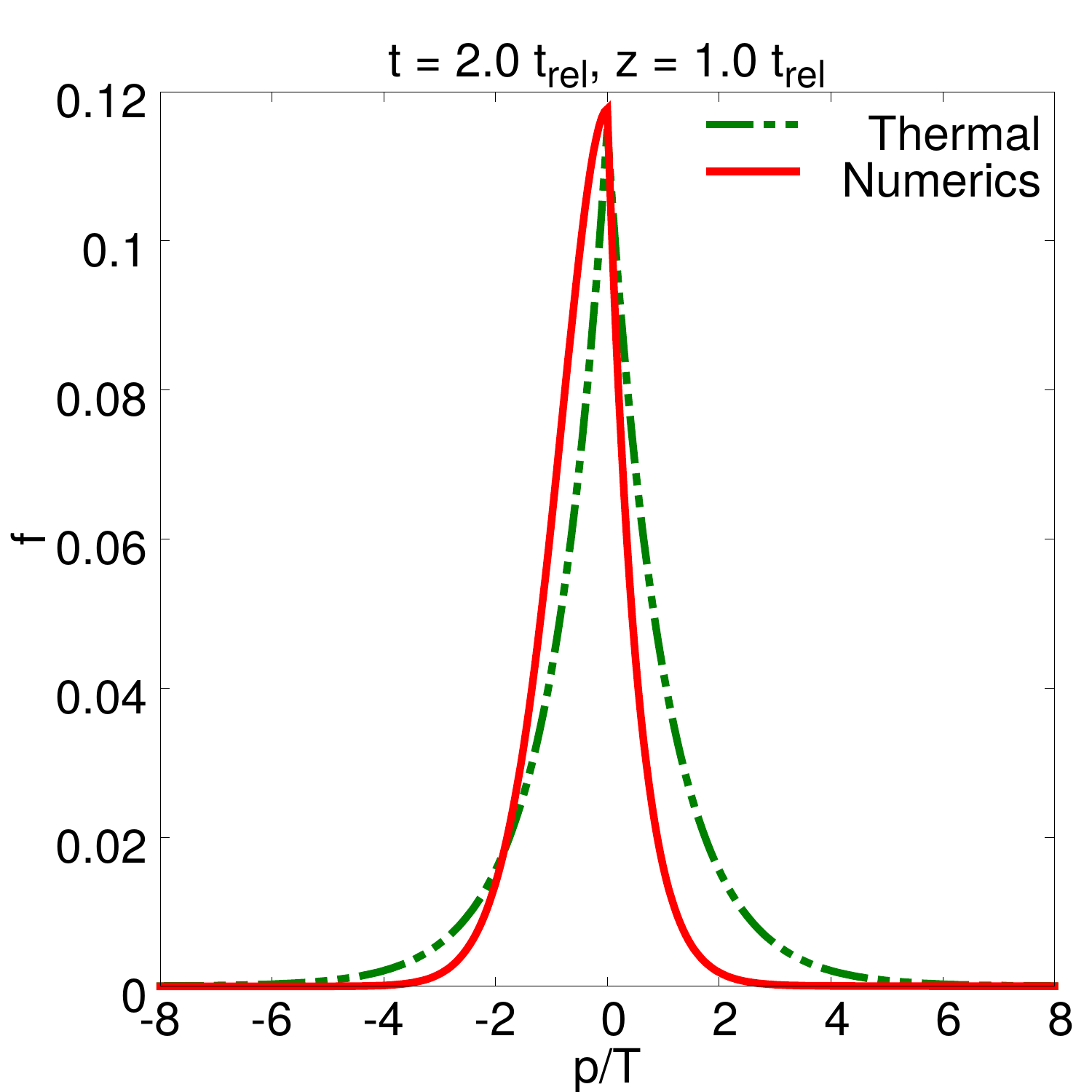}
\includegraphics[width=0.3\textwidth]{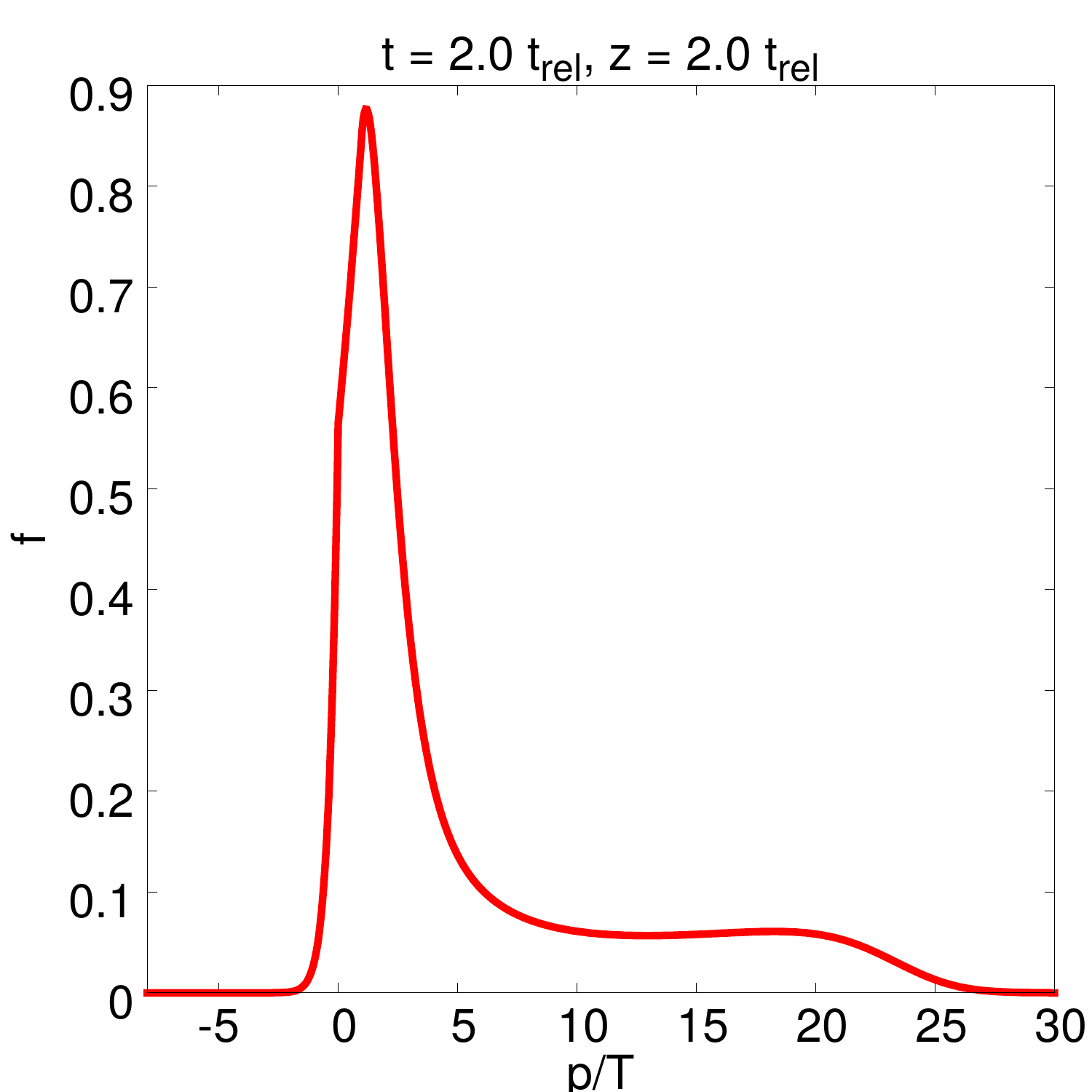}
\caption{The gluon distribution $f(t,z,p)$ for $E=25\,T$ is represented as a function of $p$ at time
$t = 2t_{\rm rel}$ and for 3 different values of $z$~: $z=0$, $z=\trel$, and $z=2\trel$. For $z<t-t_{\rm rel}$, 
the momentum distribution is nearly thermal,  
but with an overall strength which depends upon $z$~: $f(z,p)\simeq f_0(z)\rme^{-|p|/T}$.
For $z=t$ on the other hand, the distribution is far from thermal equilibrium, albeit it is still strongly peaked
near $p=T$. Similar conclusions hold for the distribution at later times and for a jet with $E=90\,T$.}\label{fig:fatt2}
\end{center}
\end{figure}

When discussing Figs.~\ref{fig:fEvol} and \ref{fig:eEvol}, it is natural to group together those
plots which correspond to different values of the energy $E$, but similar values of $t/\tbr(E)$, because
they refer to similar stages in the evolution of the jet via branching. But even for identical values of
$t/\tbr(E)$, one should still expect some differences between the two cases, $E=90\,T$ and 
$E=25\,T$, because the physics of thermalization introduces an additional energy 
scale in the problem --- the infrared cutoff $p_*$.

Consider first Figs. \ref{fig:fEvol} (a)-(c) and Figs. \ref{fig:eEvol} (a)-(c), which illustrate 
the evolution at early stages, $t < \tbr(E)$. Figs. \ref{fig:fEvol} (a)-(c) show that, already
for such early times, most of the particles are relatively soft ($p\sim p_*=T$), 
meaning that they are products of radiation. 
In particular, those particles which at a given time $t$ have an energy $|p|$ 
smaller than $\omega_{\rm br}(t)=\bar\alpha^2 \hat{q} t^2$, are generally
produced via {\em multiple branchings}, that is, they belong to gluon cascades generated
via democratic branchings by primary gluons with $p\sim \omega_{\rm br}(t)$. 
But so long as
$t\ll \tbr(E)$, most of the energy is still carried by the LP, as manifest in Figs. \ref{fig:eEvol} (a)-(b): 
the energy distribution is peaked at a value which is smaller than, but comparable to, the original energy
$E$.
When $\trel\lesssim t \ll \tbr(E)$, the energy loss by the LP and its longitudinal broadening are
both controlled by soft branchings and hence they grow with time like $t^2$.

\begin{figure}[h]
\begin{center}
\includegraphics[width=0.4\textwidth]{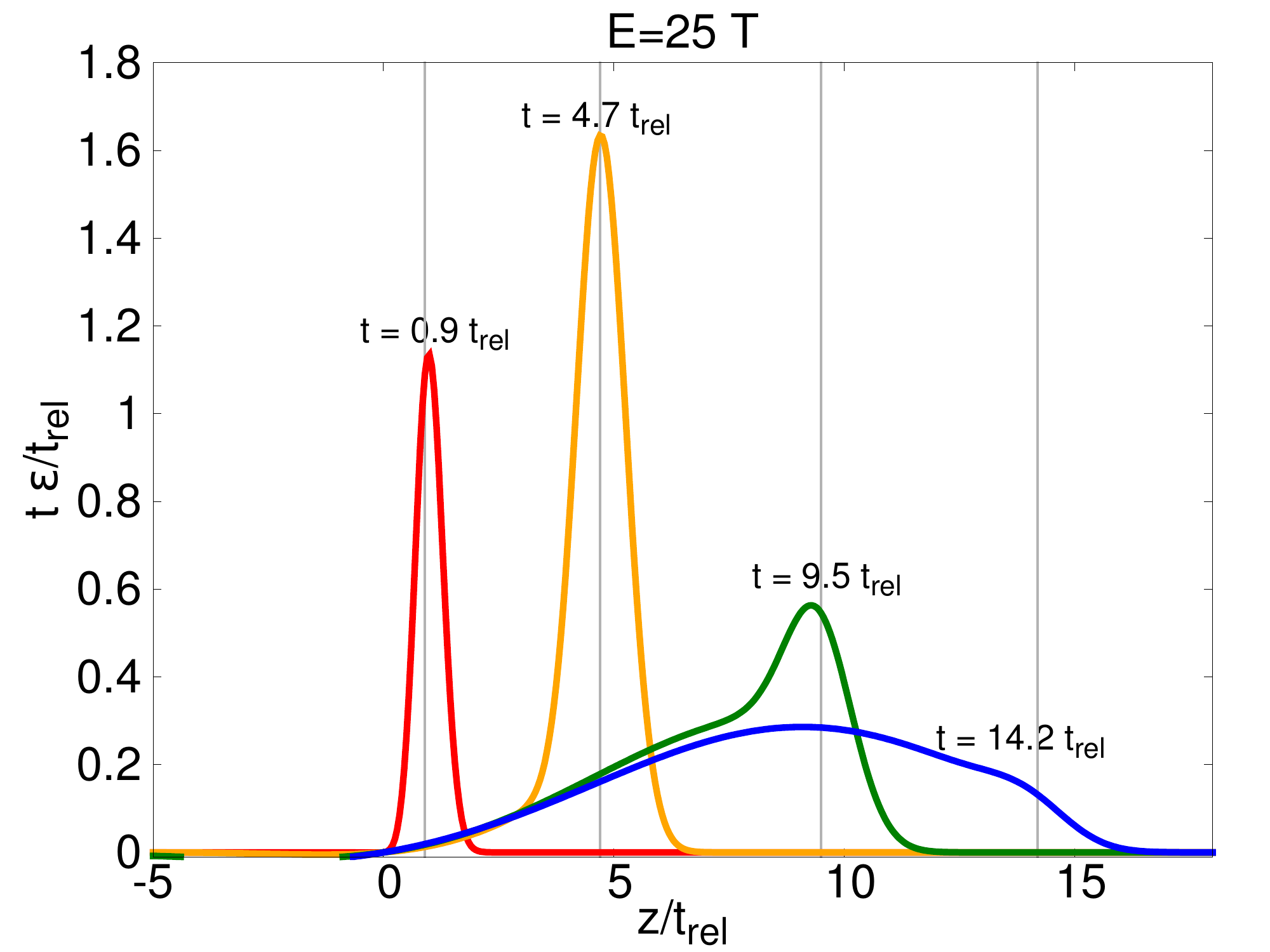}\hspace{0.1\textwidth}\includegraphics[width=0.4\textwidth]{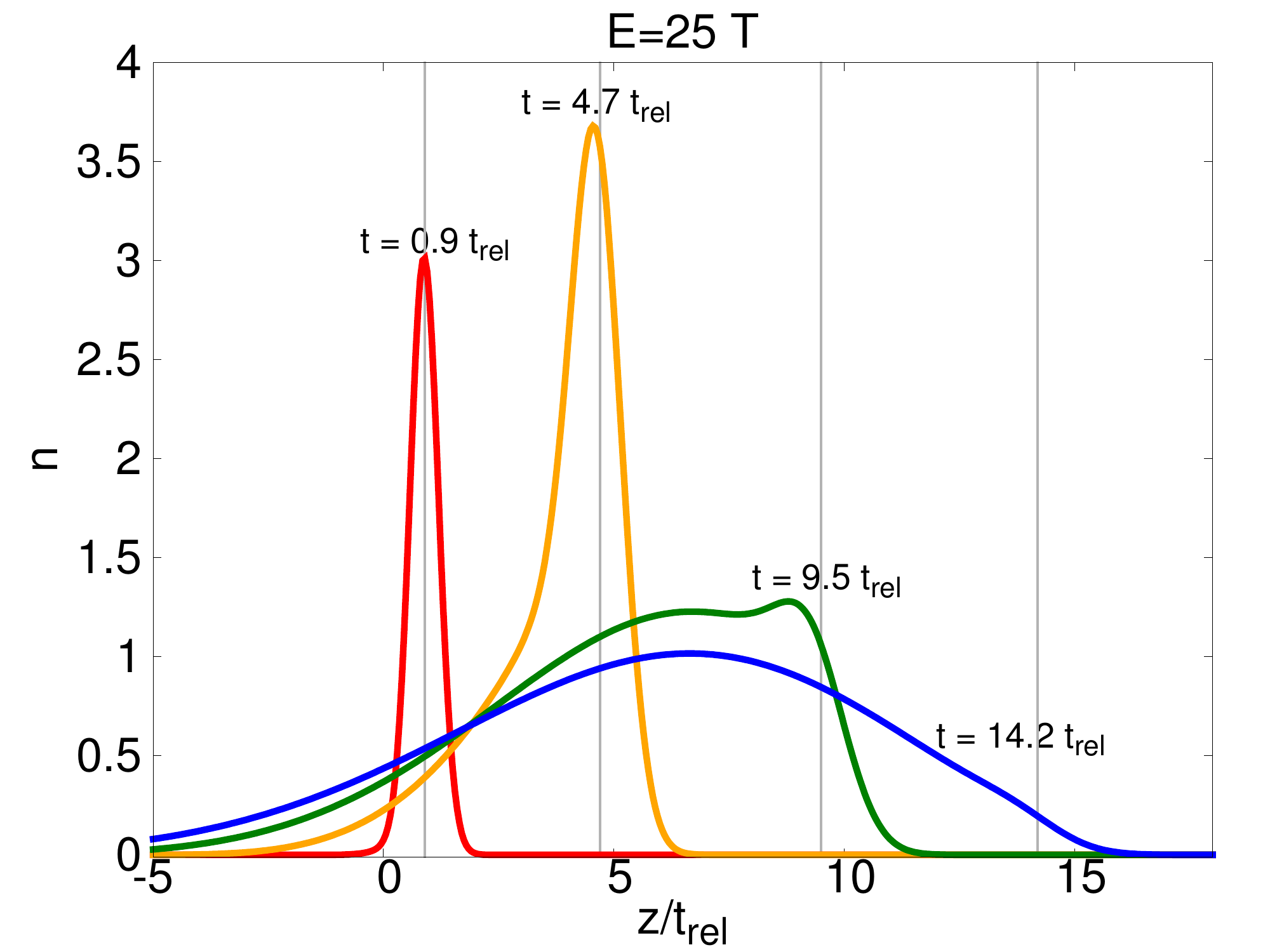}\hspace{0.05\textwidth}
\vspace{0.01\textheight}\\
\includegraphics[width=0.4\textwidth]{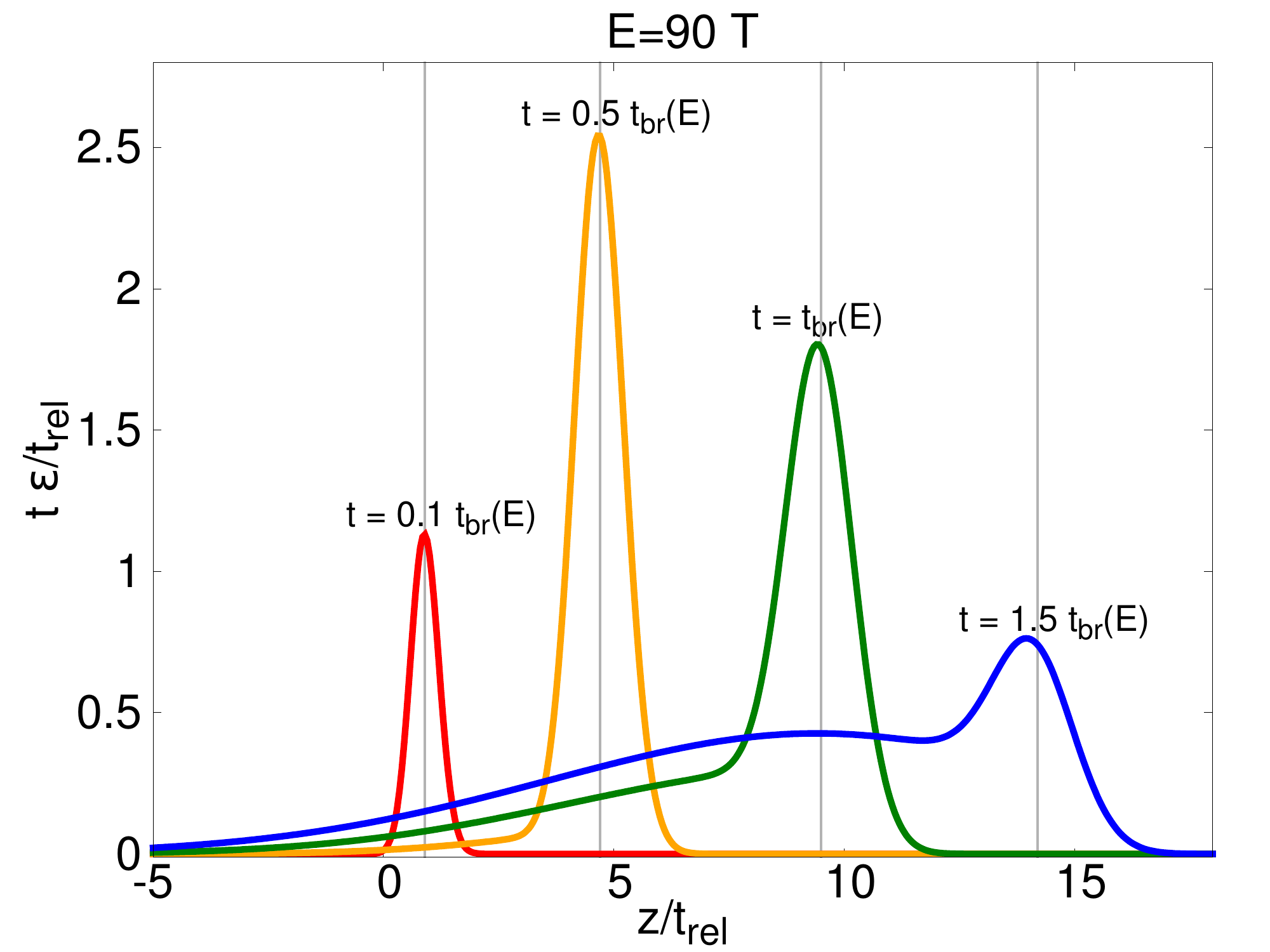}\hspace{0.1\textwidth}\includegraphics[width=0.4\textwidth]{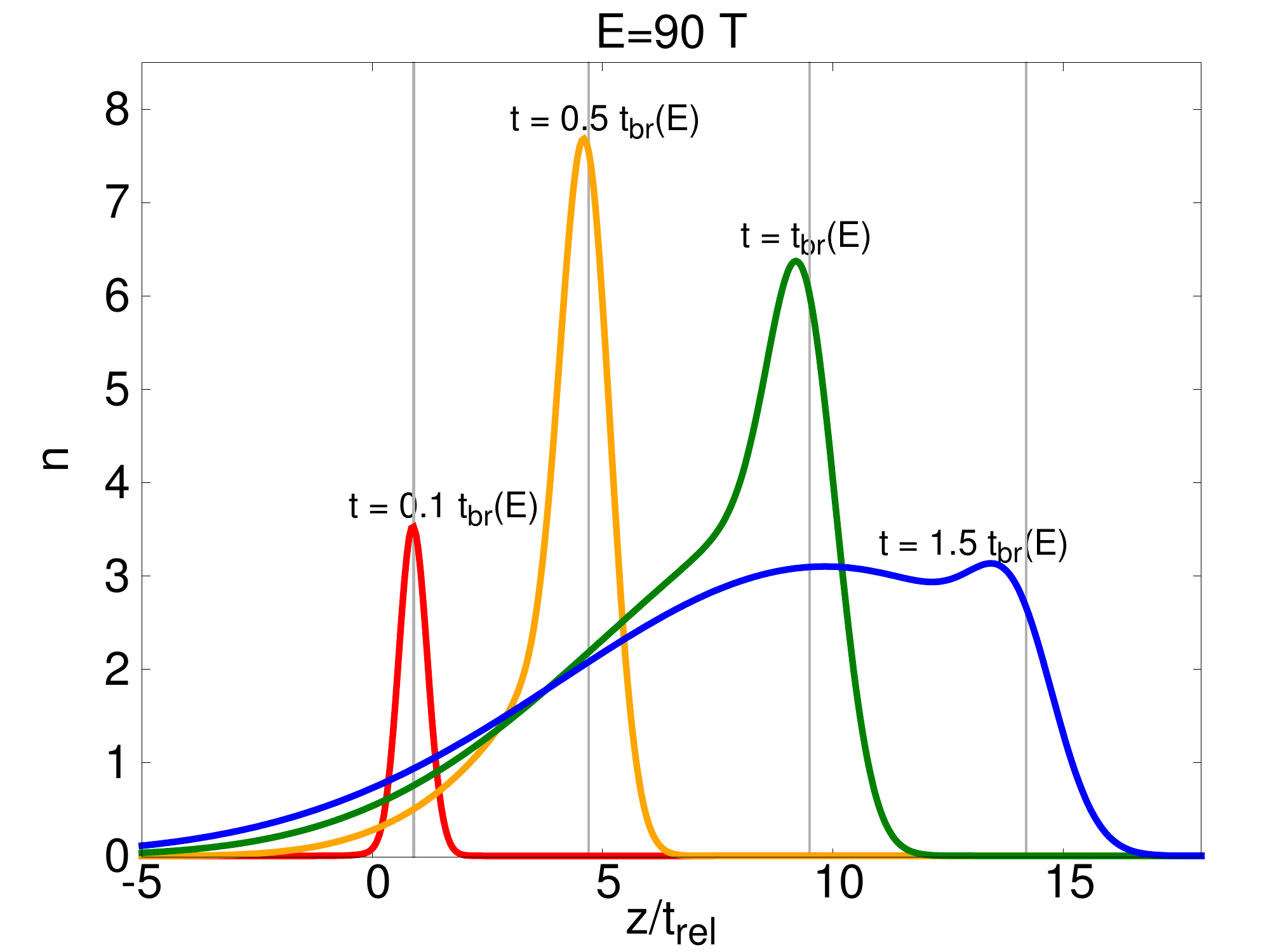}\hspace{0.05\textwidth}
\caption{The energy density and the gluon number density are shown as functions of $z$ at different times.
The grey vertical lines indicate the location of the light-cone, that is, $z=t$, for each value of
$t$.}\label{fig:fEz}
\end{center}
\end{figure}

Fig. \ref{fig:eEvol} (c) shows another interesting feature:  for $t\simeq 0.5\, \tbr(E)$, one sees 
a second peak emerging in the energy distribution at $p\sim T$. This demonstrates the
strong accumulation of gluons towards the lower end of the spectrum which
in turn reflects a limitation of the medium capacity to act as a `perfect sink'. 
We shall later return to a more detailed study of this phenomenon (see notably Figs.~\ref{fig:fEz}
and \ref{fig:deviScal} and the associated discussions).

As also visible in Figs. \ref{fig:fEvol} (a)-(c), the approach to thermalization in the tail
of the distribution at $z < t$ is noticeable already at such early times $t < \tbr(E)$. 
There is indeed a substantial number of gluons 
which remain behind the LP (i.e., which do not travel at the speed of light) and whose momentum 
distribution is nearly thermal. This is more clearly illustrated by the plots in Fig.~\ref{fig:fatt2}, 
corresponding to $E=25\,T$, which show the 
momentum distribution at $t=2\,\trel= 0.4\,\tbr(25)$ and for different values of $z$:
the shape of this distribution is close to the exponential $\rme^{-|p|/T}$ at any $z\lesssim t-\trel$.

Consider now later times $t\gtrsim \tbr(E)$, where one expects the LP to disappear
via democratic branching. Figs. \ref{fig:eEvol} (d) and (e) confirm that, when $t\sim \tbr(E)$,
there is no visible trace of the LP, albeit a few semi-hard particles, with $T \ll p\ll E$, still exist.
For even larger times, these semi-hard particles will themselves disappear via democratic branchings,
so there will be an increasing fraction of the total energy which is carried by the soft gluons 
with $p\lesssim T$. This trend is indeed visible in Figs. \ref{fig:eEvol} (f)-(h).  However,
one should not conclude that all this energy has already thermalized:  the soft gluons which propagate
together with their (semi-)hard sources along the light-cone $z=t$ cannot be thermal. This is
already illustrated by the last plot in Fig.~\ref{fig:fatt2}: the momentum distribution 
corresponding to $z=t=2\,\trel$ is peaked at small $p\sim T$, yet it strongly 
deviates from a thermal distribution.

\begin{figure}[h]
\begin{center}
\includegraphics[width=0.3\textwidth]{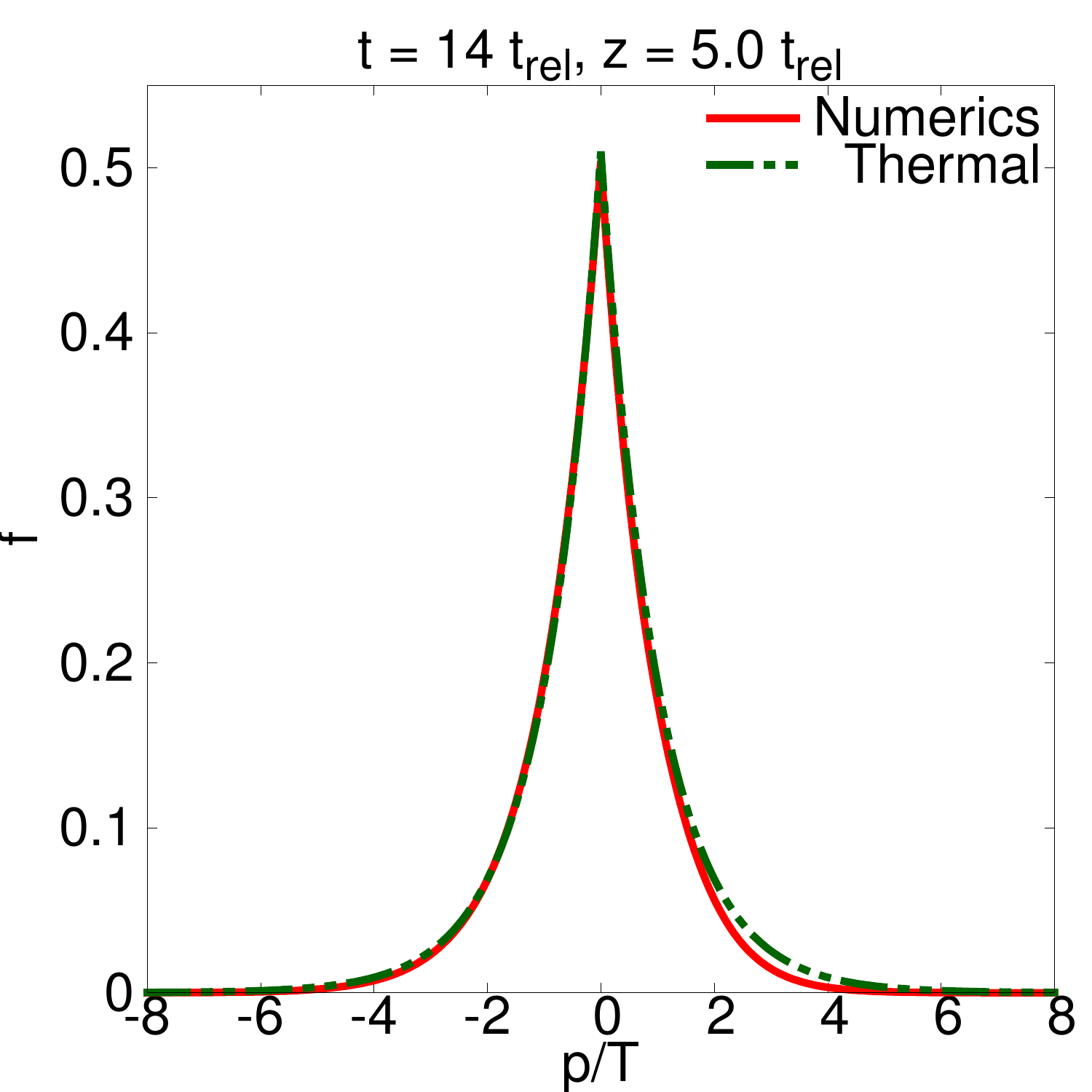}
\end{center}
\caption{The gluon distribution $f(t,z,p)$ for $E=25\,T$ is plotted as a function of $p$ for 
$t=14.2\,\trel\simeq 3\tbr(E)$ and $z=\tbr(E)=5\,\trel$ and compared to a thermal distribution
with the same normalization at $p=0$.}
\label{fig:ft14z5}
\end{figure}

To better distinguish between thermal and non-thermal
(soft) gluons at late times, we have exhibited in Fig.~\ref{fig:fEz}
the $z$--distribution of the energy and number densities, defined as
\be
\varepsilon(t,z)=
\int\rmd p\,|p|\,f(t,z,p)\,,\qquad n(t,z)=
\int\rmd p\,f(t,z,p)\,.\ee
As visible in these plots, even for times as large as $t=1.5\,\tbr(E)$, where we know (say, from 
Figs. \ref{fig:eEvol} (f) and (g)) that the energy is preponderantly carried by soft quanta with
$p\sim T$, the energy distribution is still strongly peaked at $z=t$, meaning that most of these
soft gluons are {\em not} thermal: they have been emitted at late stages 
and did not have the time to thermalize. This finding is in agreement with the discussion 
following \eqn{eq:flow}, where we noticed
that the flux of soft gluons is largest towards the late stages of the cascade.  The situation changes
at the larger time $t=3\,\tbr(E)$, that we can here access only for the jet with $E = 25\,T$ 
(by looking at $t = 14.2\,\trel\simeq 3\,\tbr(25)$). In that case, we see that both densities, 
$\varepsilon(t,z)$ and $n(t,z)$, peak well behind the light-cone --- 
in particular, the number distribution peaks around $z\simeq\tbr(E)=5\,\trel$, 
in agreement with \eqn{eq:Fsoldiff}. This strongly indicates that the entire gluon distribution
produced by the jet has thermalized by $t=3\,\tbr(E)$: the jet is {\em fully quenched}. 
This conclusion can be also checked by plotting
the distribution $f(t,z,p)$ as a function of $p$ for $t=3\,\tbr(E)$ and, say, $z=\tbr(E)$: this is 
shown in Fig.~\ref{fig:ft14z5} which indeed features an almost perfect thermal 
distribution. The distribution of such a fully quenched jet 
in longitudinal phase-space is illustrated by Figs.~\ref{fig:fEvol} (h) and \ref{fig:eEvol} (h).
Clearly, this is very similar to the late-time distributions found in Sect.~\ref{sec:analytic},
cf. Figs.~\ref{Gpz} and \ref{fig:sourced}: a distribution symmetric in $z$ which extends
via diffusion.


\subsubsection{Energy loss towards the medium}
\label{sec:eloss}

Given our numerical results, as presented in the previous subsection, it is furthermore interesting to 
use them to extract the energy lost by the jet towards the medium. A priori, this involves
two components: the energy dissipated into the medium via the drag force
(physically, this is the energy transferred to the plasma constituents through elastic collisions)
and the energy taken away by the gluons from the jet which have reached a thermal distribution in
momentum (since such gluons cannot be distinguished from the medium constituents anymore).  As noticed
at the end of Sect.~\ref{sec:source}, the drag component can be minimized by choosing
$p_*=T$, so it should be enough to compute the energy carried by the thermalized part of the
gluon distribution. Still, to avoid any uncertainty concerning the contribution of the drag,
it is preferable to compute the energy loss as the {\em difference} 
between the original energy $E$ of the LP and the energy carried by the jet constituents which
have {\em not} thermalized. This definition too is a bit ambiguous though, because the distinction between 
thermal and non-thermal gluons is not really sharp, as already noticed. Yet, we have seen
that the gluons in the tail of the distribution at $z\lesssim t-t_{\rm rel}$ are approximately thermal, 
whereas those which belong to the front ($z> t -\trel$) are still far away from thermal equilibrium
--- at least for not too late time, $t\lesssim \tbr(E)$, when the front still exist (see e.g. Fig.~\ref{fig:fatt2}).
This observation motivates the following  definition for the energy loss via thermalization:
\begin{eqnarray}
\Delta E_{\rm ther}(t)\,=\,
E-\int_{t-t_{\rm rel}}^\infty \rmd z \int_{p_*}^\infty \rmd p\, p \, f(t,z,p)\,. \label{eq:eLoss}
\end{eqnarray}
For sufficiently large times $t\gg\tbr(E)$, all the gluons lie $z< t-t_{\rm rel}$ (the front disappears) and
$\Delta E_{\rm ther}\simeq E$. But the most interesting situation in view of the phenomenology at the LHC, 
is that where the medium size $L$ is small relative to the branching time $\tbr(E)$ for the LP, hence
$\Delta E_{\rm ther}$ is small compared to $E$.

\begin{figure}[t]
\begin{center}
\includegraphics[width=0.5\textwidth]{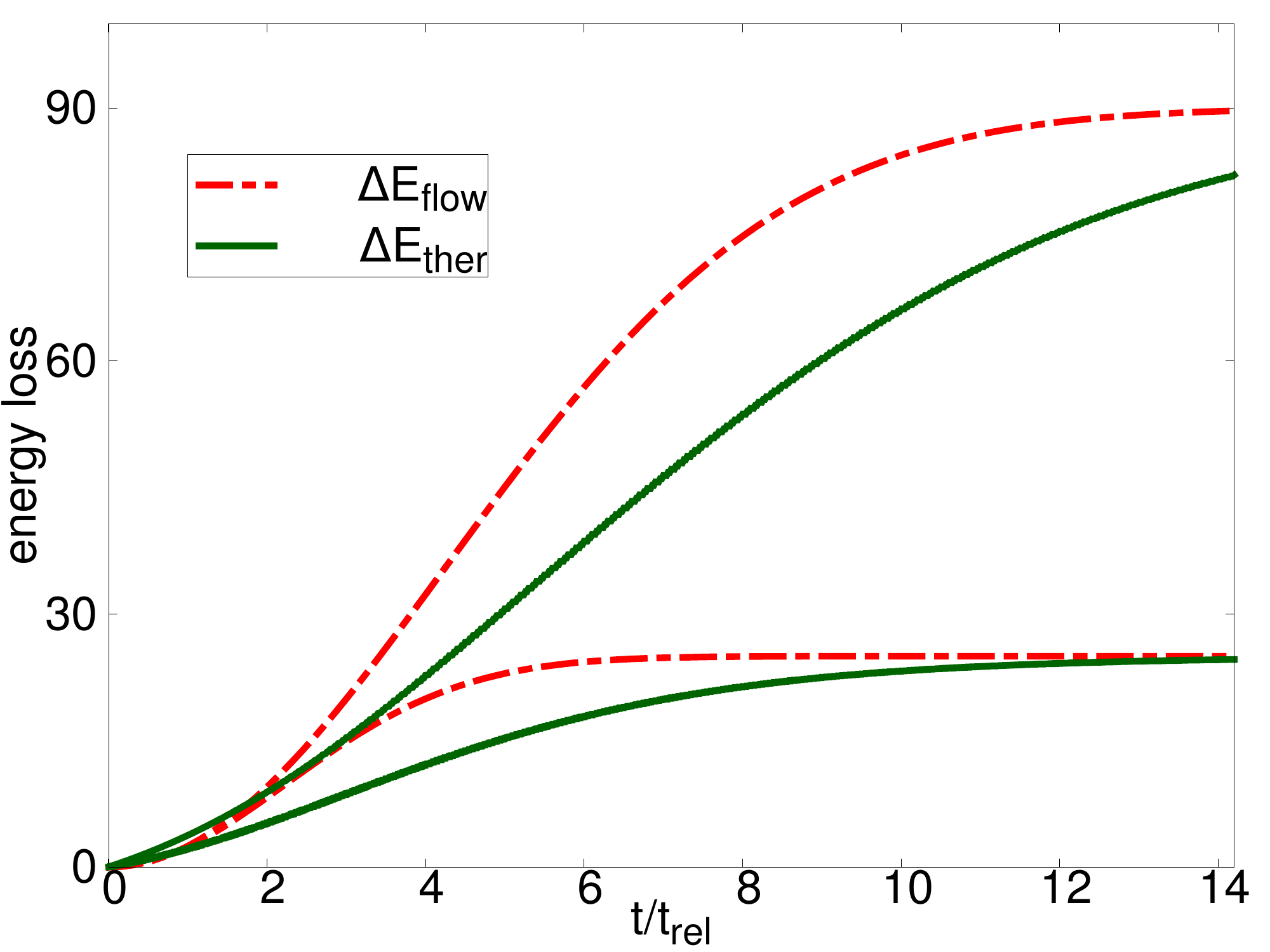}
\end{center}
\caption{The energy loss towards the medium $\Delta E_{\rm ther}(t)$ (in units of $T$),  
\eqn{eq:eLoss}, plotted as a function $t$ for 
$E=25\,T$ and $E=90\,T$. The result is compared to the flow energy \eqref{eq:flow} which
applies to an ideal cascade.}\label{fig:eLoss}
\end{figure}

In Fig.~\ref{fig:eLoss} we present our numerical results for $\Delta E_{\rm ther}(t)$ as a function of $t$
(or, equivalently, the medium size) for the two energies of interest, $E=25\,T$
and $E=90\,T$. For comparison, we also show the corresponding prediction $\Delta E_{\rm flow}(t)$ of
\eqn{eq:flow}; this would be the energy transferred to the medium in the ideal case where
the plasma acts as a perfect absorber for the gluons with $p\lesssim T$ (without affecting the
branching dynamics at $p > T$). Not surprisingly, the energy loss $\Delta E_{\rm ther}(t)$
for the `physical' cascade remains significantly lower than the `ideal' expectation 
$\Delta E_{\rm flow}(t)$ for all times $t\lesssim \tbr(E)$, that is, so long as the gluon cascade has not
fully thermalized. (For large times $t\gg\tbr(E)$, these two quantities approach to each other, as they 
both converge towards the total energy $E$, as they should.) 
The main reason for this discrepancy is the fact that the medium is {\em not} a perfect sink: there is 
a delay in the thermalization of the soft gluons produced via branchings and, as a result,
gluons with $p\sim T$ can still propagate at the speed of light over a time interval $\Delta t\gtrsim \trel$ 
after their production. Since the production rate increases with time, cf.  
\eqn{eq:flow}, it is natural that the difference between $\Delta E_{\rm flow}(t)$ and 
$\Delta E_{\rm ther}(t)$ increases as well so long as the LP
still exists, i.e. for $t\lesssim \tbr(E)$. This trend is indeed visible in Fig.~\ref{fig:eLoss}.

This being said, the energy loss $\Delta E_{\rm ther}$ 
that we have numerically found is significantly large. By inspection of Fig.~\ref{fig:eLoss}, we see that 
$\Delta E_{\rm ther}(L)\simeq 30\,T$ ($=15$\,GeV)
for a jet with $E=90\,T$ ($=45$\,GeV) and for a medium size $L=5\,\trel$ ($=5$\,fm). Furthermore,  as also
visible in Fig.~\ref{fig:eLoss}, the energy loss rises quite fast with time and hence with the medium size $L$~:
at small times $t\ll \tbr(E)$, one roughly has $\Delta E_{\rm ther}(t)\propto t^2$, in agreement with
\eqn{DEflow}, whereas for larger times $t\gtrsim\tbr(E)$, $\Delta E_{\rm ther}(t)$ approaches the total
energy $E$ of the LP: the jet is `fully quenched'.

\begin{figure}[t]
\begin{center}
\includegraphics[width=0.45\textwidth]{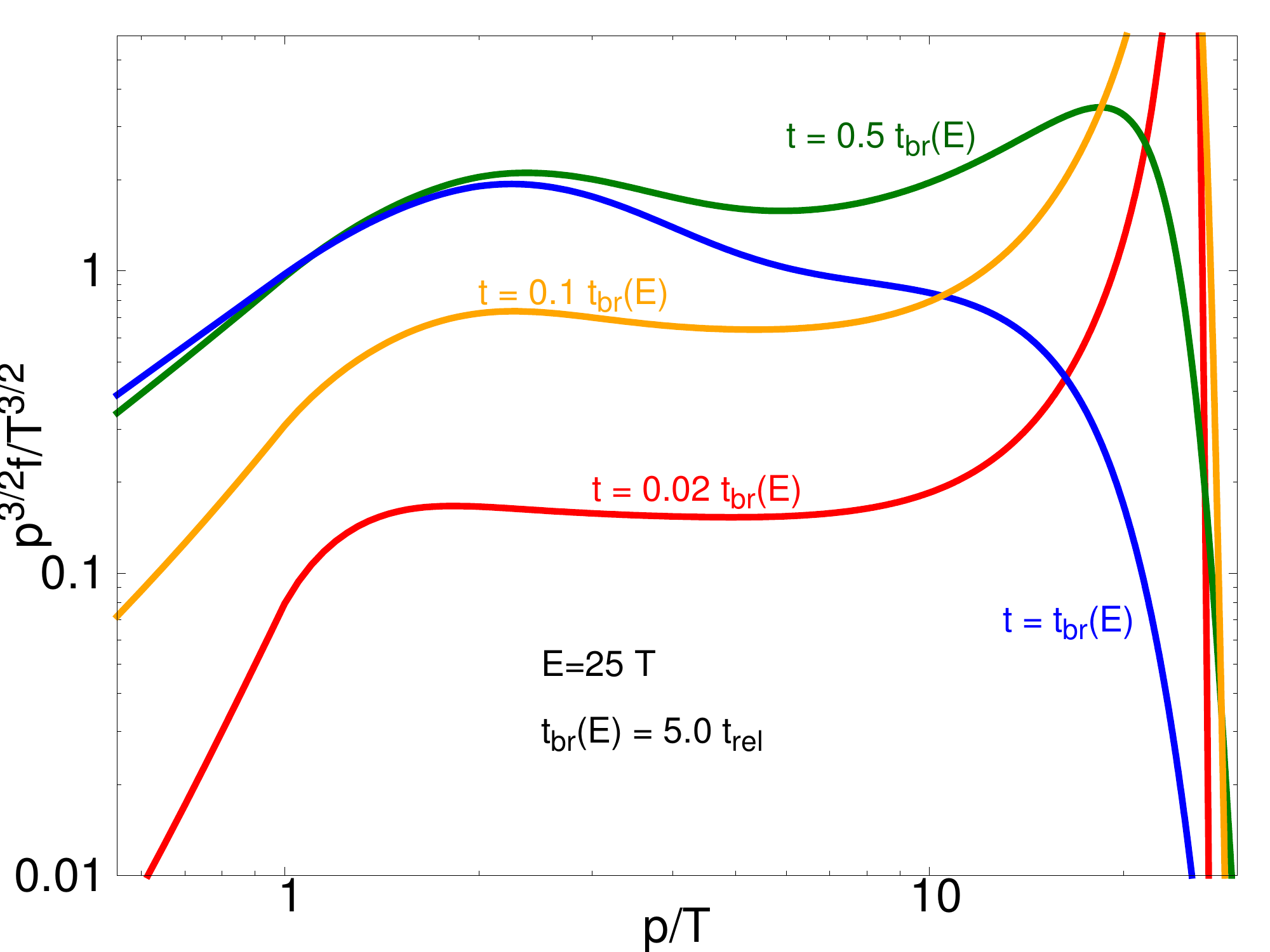}\hspace{0.05\textwidth}\includegraphics[width=0.45\textwidth]{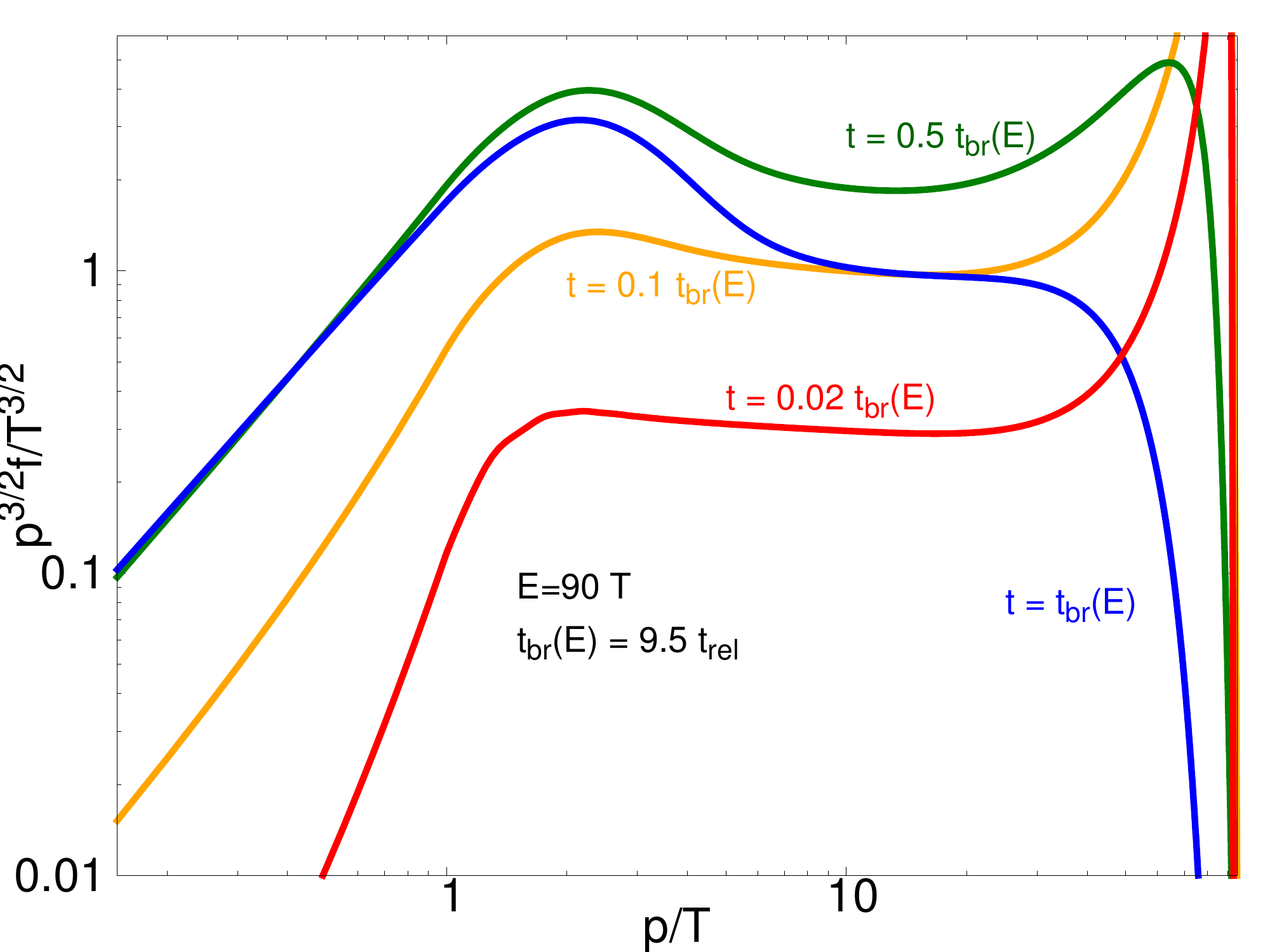}
\caption{The gluon distribution in momentum at $t=z$ for two energies, $E=25\,T$ and $E=90\,T$, and
for 4 values of time, which are now chosen to be the same in units of $\tbr(E)$ for both energies.
The figures show a rather broad window of approximate scaling behavior, $f\propto {1}/{p^{3/2}}$,
at not too large times $t\lesssim \tbr(E)$.
}\label{fig:deviScal}
\end{center}
\end{figure}

Whereas the results in Fig.~\ref{fig:eLoss} point out towards a failure of the hypothesis of a `perfect
sink', this failure remains quite mild, especially at early times $t\ll\tbr(E)$.
This is already suggested by the qualitative similarity between the gluon distribution produced by 
a `real' jet, as shown in Fig.~\ref{fig:fEvol}, and that generated by an ideal gluon cascade, cf. 
Fig.~\ref{fig:sourced} and Fig.~\ref{Spz}. To have a more quantitative test in that sense, we have
studied the gluon spectrum near the front of the jet, that is, the distribution $f(t,z,p)$ produced by the
kinetic equation at $z=t$. At small times $t\lesssim 0.5\tbr(E)$,  the numerical results in Fig. \ref{fig:deviScal}
exhibit a relatively wide window at $T < p \ll E$ where the front distribution function shows 
the same scaling behavior,  $f(z=t, p)\propto {1}/{p^{3/2}}$, as 
the ideal branching process (compare to the curved `w/o cutoff' in Fig.~\ref{fig:spatialHomo}).  
As repeatedly stressed, this scaling law is a hallmark of wave turbulence. 
Fig. \ref{fig:deviScal} should be contrasted to the corresponding results for the  spectrum
$f(t,p)=\int\rmd z f(t,z,p)$ (the curves denoted as `full' in Fig.~\ref{fig:spatialHomo}), which 
show no scaling window at all. So, in this respect at least, the gluon spectrum is potentially misleading, as
anticipated in Sect.~\ref{sec:spectrum}.

More generally, the ensemble of studies that we have performed in this paper demonstrate that the 
detailed phase-space distribution is better suited than the gluon spectrum
for understanding the in-medium evolution of the jets.

\section{Conclusions and perspectives}

In this paper, we have presented a first study of the thermalization of the soft components 
of the gluon cascades generated via multiple branchings by an energetic parton which propagates 
through a weakly-coupled quark-gluon plasma. Our overall picture is rather simple and physically 
motivated, and in our opinion it is also quite robust: indeed, this picture is almost an immediate
consequence of the strong separation of scales between the characteristic time for the medium-induced 
branchings of hard gluons and, respectively, the relaxation time for the thermalization of soft gluons.
In trying to establish this picture beyond parameter estimates, we met with several difficulties
and subtle points, for which we proposed at least partial solutions.

A major difficulty is the overall complexity of the problem, that we have tried to circumvent
via suitable approximations, notably by carefully
separating the gluons from the jet from those in the medium and by projecting the dynamics onto
the one-dimensional, longitudinal, phase-space.  These approximations are fully justified for the
sufficiently hard gluons in the cascades, with momenta $p\gg T$, which control the
dynamics of multiple branchings. On the other hand, these approximations becomes less justified
when moving to the softer gluons with $p\lesssim T$, where they are at most qualitatively right.
(But we have tried to carefully argue that they correctly reproduce the relevant time
scales to parametric accuracy.) A particularly subtle approximation refers to the 
branching dynamics near the lower end of the cascades, at $p\sim T$, where we expect the cascade
to terminate, on physical grounds. In our calculations, we have simply cut off the branching process
at a scale $p_*\sim T$ and found that, thanks to the smearing effect of the elastic collisions,
the results are not very sensitive to the precise value of this cutoff. (We have indeed checked
that our numerical results remain qualitatively and even semi-quantitatively unchanged when varying
this cutoff by a factor of 2 around its central value.)
But it would be of course important, both conceptually and phenomenologically, 
to have a dynamical implementation of this cutoff, which in turn requires a consistent treatment
of the full gluon distribution at soft momenta, including the inherent non-linear effects. 

This discussion points towards the many `technical' limitations in our approach, with potential
physical consequences, which will be hopefully lifted by more detailed, future, analyses.
In principle, the theoretical framework for such studies is well defined: this is set by the
general kinetic equations alluded to in Sect.~\ref{sec:kin}, that can be found
in the literature \cite{Baier:2000sb,Arnold:2002zm}. A main difficulty as compared to previous
numerical studies of such equations in the literature \cite{Schenke:2009gb,Kurkela:2014tea,Kurkela:2015qoa},
is that fact that, for the jet problem at hand,
one needs to explicitly deal with the strong spatial inhomogeneity introduced by the
hard components of the jet.

Another interesting direction of research refers to a better understanding of the  
implications of the present picture for the phenomenology of jets 
in heavy ion collisions at RHIC and the LHC. Our first estimates for the
energy loss via thermalization, which are of course very raw and must be taken with a grain of salt,
are quite encouraging in that sense. In our opinion, it makes sense to compare, at least
qualitatively, the quantity $\Delta E_{\rm ther}$ introduced in \eqn{eq:eLoss}
(the energy carried away by the thermalized gluons in the tail of the jet) with the energy imbalance
at large angles, as measured in the context of the di--jet asymmetry. The
detailed analyses of the corresponding data, 
notably by the CMS collaboration \cite{Chatrchyan:2011sx,Gulhan:2014},
demonstrate that the energy imbalance is carried by an excess of soft hadrons
($p_T\lesssim 2$~GeV) propagating at large angles. It looks natural to associate 
these soft hadrons with the `thermalized gluons' in our current set-up. If so, it is interesting to notice that
our estimates for $\Delta E_{\rm ther}$ in Fig.~\ref{fig:eLoss} are in the ballpark of 10 to 20 GeV, a value
which is not unreasonable for the phenomenology alluded to above. But of course further studies,
to remove some of our theoretical uncertainties and to better defines experimental observables,
are still needed before aiming at a detailed comparison with the phenomenology.

\section*{Acknowledgments}
We would like to thank Al Mueller for inspiring discussions during the early stages of this work and
for a careful reading of the manuscript.
This work is supported by the European Research Council 
under the Advanced Investigator Grant ERC-AD-267258 and by the 
Agence Nationale de la Recherche project \# 11-BS04-015-01. 


\providecommand{\href}[2]{#2}\begingroup\raggedright\endgroup

\end{document}